\documentclass[fleqn,usenatbib]{mnras}

\usepackage{newtxtext,newtxmath}

\usepackage[T1]{fontenc}
\usepackage{caption}
\usepackage{placeins}
\usepackage{amsmath}
\usepackage{bm}

\DeclareRobustCommand{\VAN}[3]{#2}
\let\VANthebibliography\thebibliography
\def\thebibliography{\DeclareRobustCommand{\VAN}[3]{##3}\VANthebibliography}

\usepackage{graphicx}	
\usepackage{amsmath}	
\usepackage{booktabs}
\usepackage{subcaption}  

\newcommand{\lya}{Ly$\alpha$}

\title[$z\sim7$ Protoclusters and Environmental Effects]{Reionization in Protocluster Environments at $z>7$ with JWST/NIRSpec}

\author[Li et al.]{
Qiong Li\thanks{ qiong.li@manchester.ac.uk}$^{1}$,
Christopher J. Conselice$^{1}$, 
Duncan Austin$^{1}$,
Tom Harvey$^{1}$,
Nathan Adams$^{1}$,
\newauthor
Vadim Rusakov$^{1}$,
Lewi Westcott$^{1}$
\\
$^{1}$ Jodrell Bank Centre for Astrophysics, University of Manchester, Oxford Road, Manchester UK
}

\date{Accepted 2026 February 12. Received 2025 December 22; in original form 2025 August 6}

\pubyear{\the\year{}}

\begin{document}
\label{firstpage}
\pagerange{\pageref{firstpage}--\pageref{lastpage}}
\maketitle

\begin{abstract}
Understanding the role of high-redshift protoclusters in cosmic reionization is essential to unveiling the early stages of structure formation. Using deep imaging and spectroscopy from the \textit{James Webb Space Telescope} (\textit{JWST}) JADES Deep Survey in GOODS-South, we identify two prominent protoclusters at \( z > 7 \) and investigate their environmental properties in comparison to field galaxies. Protocluster members exhibit systematically higher ionizing photon production efficiency ($\xi_{\text{ion}}$) and inflated [O\,\textsc{iii}]/H$\beta$ ratios at fixed UV magnitude or stellar mass, likely driven by young, metal-poor stellar populations and intense star formation. Despite these properties, their Ly$\alpha$ emission is weak or absent, and they show high proximate neutral hydrogen column densities, suggesting insufficient ionizing output to maintain ionized bubbles. We also find that a strong Ly$\alpha$ emitter (LAE), JADES-GS-z7-LA, may lie within the same ionized region as one protocluster. Although their Lyman continuum escape fractions ($f_{\mathrm{esc}} \sim 0.1$) are comparable to those of LAEs, individual protocluster galaxies are faint ($M_{\mathrm{UV}} > -19$) and low-mass ($\log(M_*/M_\odot) \sim 8.5$). The enhanced number density within protoclusters boosts the local UV luminosity density by nearly 1~dex. The surrounding gas remains largely neutral, suggesting that reionization was highly patchy and modulated by environment. The protocluster galaxies likely host ionization-bounded nebulae with holes, suppressing Ly$\alpha$ visibility, in contrast to field galaxies that are more consistent with density-bounded nebulae.
\end{abstract}




\section{Introduction}

The assembly and evolution of high-redshift protoclusters play a crucial role in shaping the large-scale structure of the Universe and driving the early stages of cosmic reionization. Protoclusters, defined as the progenitors of present-day galaxy clusters, are expected to host dense populations of star-forming galaxies embedded in overdense regions of the cosmic web \citep{Overzier2016, Chiang2017}. These structures are predicted to collapse into massive galaxy clusters by $z=0$, making them prime laboratories for studying galaxy formation in the early Universe. The advent of \textit{JWST} has provided unprecedented access to high-redshift protoclusters at $z>7$, revealing new insights into their stellar populations, ionization states, and the role of their local environments in shaping their properties \citep{Endsley2022, Morishita2023, Whitler2023}.

A key challenge in understanding this epoch is to determine how galaxies in different environments contributed to reionization and how the surrounding neutral hydrogen affected their evolution. Observations have shown that galaxies in overdense regions tend to exhibit distinct characteristics compared to their field counterparts, with enhanced star formation activity, greater dust content, and differences in nebular line strengths \citep{Harikane2019, Laporte2021,Qiong2024}. These variations may arise from the combined effects of galaxy interactions, feedback processes, and the availability of cold gas for star formation. Furthermore, the presence of a significant neutral hydrogen fraction in these environments can suppress Ly$\alpha$ visibility, potentially providing constraints on the timeline of reionization \citep{Mason2018, Fuller2020, Jung2022}.

Observations from \textit{JWST}, ALMA, and deep ground-based surveys have revealed an increasing number of bright, high-equivalent-width LAEs (Lyman-alpha emitters) at $z>7$, suggesting that certain regions of the Universe were already significantly ionized by these epochs. For instance, spectroscopic confirmations of LAEs at $z=7.6$ in the EGS field \citep{Jung2020} and $z=8.68$ in the CANDELS survey \citep{Zitrin2015} indicate that local ionized regions must have formed around these galaxies, enabling Ly$\alpha$ photons to escape. Similar detections have been reported at $z\sim7.5-9.5$ in the JWST/PASSAGE Survey \citep{Runnholm2025}, indicating that reionization proceeded inhomogeneously, with certain regions ionizing earlier than others. Recent \textit{JWST} studies have further emphasized the importance of environment in regulating Ly$\alpha$ transmission by, for example, identifying an extremely high-equivalent-width LAE within an ionized region at $z\simeq7.3$ in JADES GOODS-S \citep{Saxena2023}, mapping the environments of reionisation-era LAEs with JADES and FRESCO and showing that many coincide with large-scale galaxy overdensities, including a prominent structure at $z\sim7.3$ in GOODS-S \citep{Witstok2024,Helton2023,Meyer2024}, and using larger Ly$\alpha$ samples across multiple \textit{JWST} fields to demonstrate that Ly$\alpha$ visibility is enhanced in overdense regions and can be used to infer the sizes of ionized bubbles during reionization \citep{Endsley2022,Tang2023}.

The detection of these high-redshift LAEs and their inferred ionized bubbles has profound implications for cosmology. The size and clustering of ionized regions provide constraints on the timing and spatial structure of reionization, helping to determine whether early galaxies alone could have driven this transition. Measurements of Ly$\alpha$ damping wings in quasar spectra indicate that reionization was still incomplete at $z\sim7.5$ \citep{Greig2017, Davies2018}, while observations of LAEs suggest that local reionized bubbles existed around star-forming galaxies even at $z\sim8.5$ \citep{Mason2019, Witstok2024}. These ionized bubbles are thought to have formed through clustered star formation, where groups of galaxies jointly ionized their surroundings. However, the precise mechanisms governing the formation and growth of these bubbles remain unclear. Understanding how environmental factors influence reionization at these scales is thus essential for refining theoretical models and interpreting upcoming \textit{JWST} observations.

While some overdensities at $z>7$ host strong LAEs and early ionized bubbles (e.g. \citealt{Whitler2024}), other protoclusters appear to lack detectable Ly$\alpha$ emission (e.g. \citealt{Morishita2023}), suggesting a more complex interplay between neutral gas and galaxy properties. Neutral hydrogen plays a fundamental role in shaping the evolution of early galaxies, influencing both their internal star formation and their visibility in Ly$\alpha$. Studies of damped Lyman-$\alpha$ absorbers (DLAs) and Lyman-limit systems (LLSs) suggest that significant amounts of neutral hydrogen persist in the circumgalactic medium (CGM) of high-redshift galaxies, particularly in dense regions where large-scale gravitational collapse has funneled gas into star-forming halos \citep{Fan2006,Stern2021}. The presence of these gas reservoirs complicates the interpretation of Ly$\alpha$ observations, as the attenuation of Ly$\alpha$ photons by neutral gas could be environment-dependent. Theoretical and observational studies indicate that neutral gas reservoirs regulate metal enrichment, star formation efficiency, and feedback processes in young galaxies \citep{Kakiichi2018, Hutter2021}. Recent work by \citet{Harikane2025} emphasizes that massive neutral gas reservoirs are essential for sustaining the formation of early massive galaxies, with cold, metal-poor gas flows providing the raw material for the first major star formation episodes. The presence of such gas reservoirs in high-redshift overdensities could lead to longer timescales for gas ionization, delaying the emergence of observable Ly$\alpha$ emission.


This study investigates the environmental dependence of Ly$\alpha$ visibility in high-redshift protoclusters by examining two prominent overdensities in the JADES field at $z\sim7.2$. We compare the physical and ionization properties of galaxies in these overdense regions with those in the field, aiming to determine whether local overdensities systematically influence Ly$\alpha$ emission and neutral gas content. 
Our goal is to assess whether dense regions are delayed or advanced sites of reionization, addressing a key question in the spatial progression of cosmic reionization.

The structure of this paper is as follows. Section~\ref{sec:Observations} describes the observational data and selection criteria for protocluster and control galaxy samples. Section~\ref{sec:z7pc} outlines the methodology used to measure local densities, characterize ionized environments, and identify neutral hydrogen reservoirs. In Section~\ref{sec:result}, we present a detailed comparison between protocluster and field galaxies, analyzing their Ly$\alpha$ equivalent widths, UV magnitudes, star formation histories and neutral hydrogen.  
Section~\ref{sec:compare}, we place our findings in the context of reionization models, discussing how environmental factors modulate Ly$\alpha$ escape and whether overdense regions effectively drive reionization. 
Finally, we summarize our conclusions in Section~\ref{sec:conclusion}.

Throughout this work, we assume a flat $\Lambda$CDM cosmology with $H_0 = 70$ km s$^{-1}$ Mpc$^{-1}$, $\Omega_m = 0.3$, and $\Omega_\Lambda = 0.7$. All magnitudes are given in the AB system \citep{Oke1983}.

\section{Observations and Data Reduction}\label{sec:Observations}

The \textit{JWST} Advanced Deep Extragalactic Survey \citep[JADES;][]{Rieke2023,Bunker2023, DEugenio2025} provides one of the deepest extragalactic observations to date, covering the GOODS-S and GOODS-N fields. In this study, we focus on the GOODS-S region, specifically the publicly available data covering the `DEEP' subregion (PI: Eisenstein, N. Lützgendorf, ID:1180, 1210). The observations span a spatial coverage of approximately 24.4–25.8 arcmin\(^2\) and utilize nine NIRCam filters: F090W, F115W, F150W, F200W, F277W, F335M, F356W, F410M, and F444W. The observations were conducted using a minimum of six dithered exposures per pointing, ensuring optimal cosmic ray rejection and enhanced spatial resolution. The total integration times range from 14 ks in the bluer bands to 60 ks in the deeper exposures. The 5\(\sigma\) depth of these data spans 29.58 to 30.21 AB mag, with F277W being the deepest. We also use HST F606W and F814W GOODS-S mosaic (v2.5) data from the Hubble Legacy Fields team \citep{Illingworth2013,Whitaker2019}, which we realign to match the WCS of the JADES imaging.

The raw imaging data were processed using the \textit{JWST} calibration pipeline \citep{Bushouse2022}, incorporating several key steps. Initially, the Stage 1 pipeline applied detector-level corrections, including bias subtraction, dark current removal, and non-linearity corrections. The Stage 2 pipeline then performed flat-fielding and flux calibration, using the latest in-flight reference files (pmap v1364). To mitigate the impact of residual background variations, a two-step sky subtraction method was applied: first, a constant background level was removed from individual exposures, followed by a more refined 2D background modeling using \texttt{photutils} \citep{Bradley2022}. The final mosaic construction was performed using the Stage 3 pipeline, aligning all exposures to a common astrometric frame based on \textit{Gaia} DR3 \citep{Gaia2023} and drizzling the images to a 0.03 arcsec/pixel scale.

The photometric catalog was derived using \texttt{SExtractor} \citep{Bertin1996}, operating in dual-image mode, with a weighted stack of the three reddest wide-band filters (F277W, F356W, and F444W) used for source detection. Aperture photometry was extracted using circular apertures of 0.32 arcsec diameter, with PSF corrections applied to account for flux losses based on simulated \textit{WebbPSF} models \citep{Perrin2012}. To improve photometric accuracy, we estimate and subtract the local sky background within the $32 \times 32$ pixel regions, where the real sources are masked. The final photometric errors were computed using the normalized median absolute deviation (NMAD) of the empty-sky apertures, ensuring robust uncertainty estimates. For a more in-depth description of the data reduction process and catalogs, which were produced by the {\tt galfind} software\footnote{\url{https://github.com/duncanaustin98/galfind}}, refer to \citet{Adams2024,conselice2024epochs}.

We also incorporate publicly available JADES NIRSpec spectroscopy \citep{Bunker2023, ferruit2022near}, obtained from the DR1+DR3 JADES data release, with observations targeting the GOODS-S field (PI: Eisenstein, N. Lützgendorf, ID:1180, 1210). The spectroscopy was acquired using four disperser/filter configurations: G140M/F070LP, G235M/F170LP, G395M/F290LP, and G395H/F290LP. These configurations provide wavelength coverage spanning \(0.70 - 1.27 \, \mu\)m, \(1.66 - 3.07 \, \mu\)m, \(2.87 - 5.10 \, \mu\)m, and \(2.87 - 5.14 \, \mu\)m, respectively, with resolving powers of \( R \approx 1000 \) for the medium-resolution gratings and up to \( R \approx 2700 \) for the high-resolution G395H grating. 
In this study, we primarily utilize PRISM/CLEAR spectroscopy, which provides a continuous wavelength coverage from \(0.6 \, \mu\)m to \(5.3 \, \mu\)m at a spectral resolution of \( R \approx 30 - 330\), allowing for comprehensive continuum and emission line analysis.

The NIRSpec spectroscopic data reduction followed a multi-stage process using the \textit{JWST} official pipeline, complemented by additional post-processing steps. The Stage 1 reduction applied detector-level corrections, including bias subtraction, dark current removal, and cosmic ray flagging. The Stage 2 pipeline performed wavelength calibration, flat-fielding, and initial flux calibration, leveraging the most recent in-flight calibration files. The Stage 3 pipeline extracted and rectified the two-dimensional spectra, correcting for spectral trace distortions using calibration reference files. Background subtraction was achieved through a combination of nod-subtraction for individual sources and global sky-subtraction using empty-sky regions. 

For one-dimensional spectral extraction, optimal weighting techniques following \citet{Horne1986} were used, improving the signal-to-noise ratio for faint sources. Flux calibration was performed using standard reference stars observed within the same program, ensuring relative spectrophotometric accuracy. To account for telluric absorption residuals, the correction based on a median-stacked sky spectrum from multiple sources was applied. 

\begin{figure}
    \includegraphics[width=8.5cm]{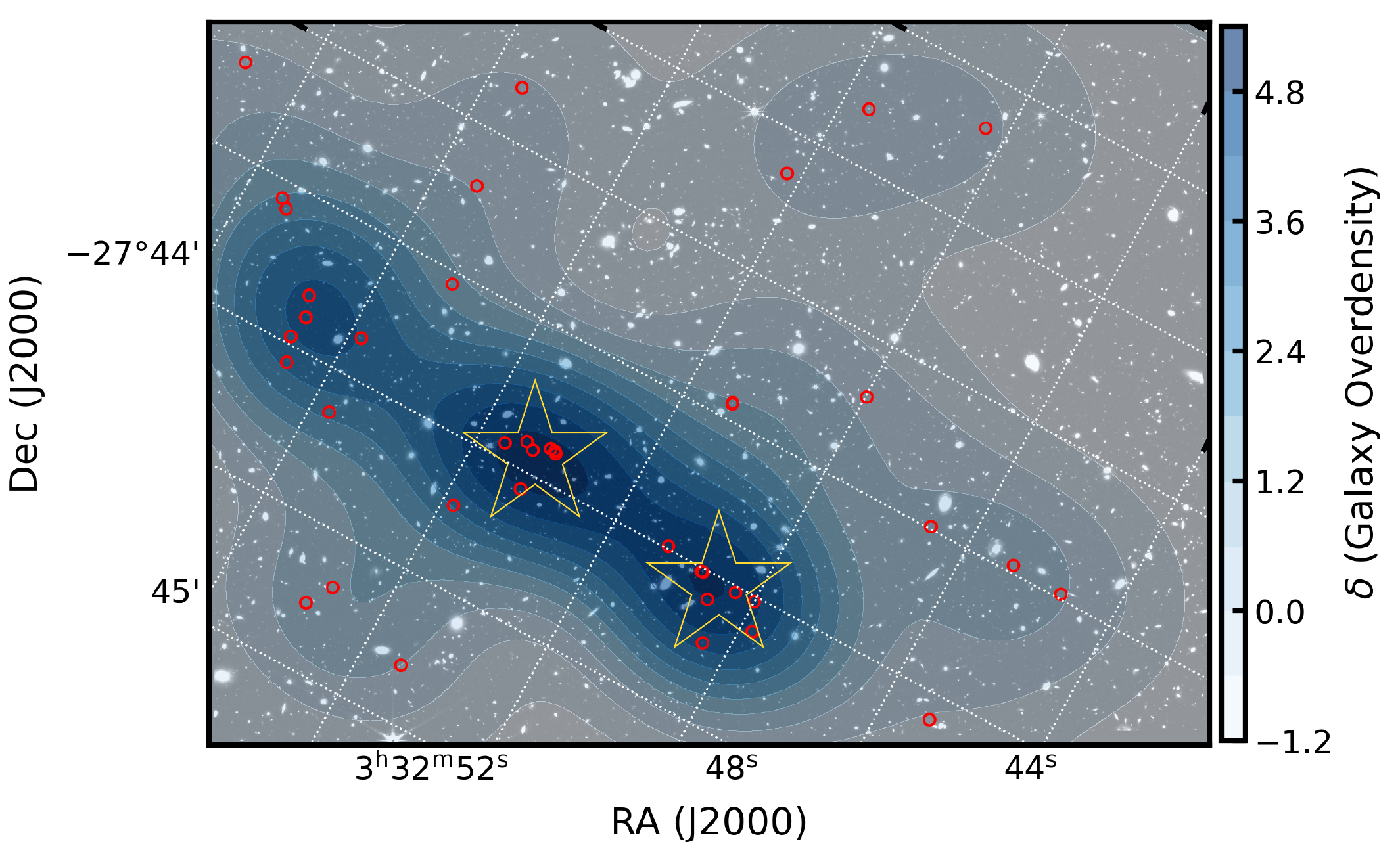}
    \caption{Projected galaxy overdensity map at $z=7.15-7.45$ in the observed field. The color scale represents the overdensity parameter $\delta$, with darker shades indicating higher local galaxy densities. Red points are photometric individual galaxy positions, while gold star markers highlight the most significant overdense regions, corresponding to the identified protocluster candidates.}
    \label{fig:density_maps}
\end{figure}

\section{High-redshift Protocluster Identification at \texorpdfstring{$z>7$}{z>7}}
\label{sec:z7pc}

We present the identification of high-redshift protoclusters using a local density-based approach. This method enables the detection of structures that were missed, particularly for $z > 7$ where the limited number of galaxies poses significant challenges. In the JADES field, we identify two prominent protoclusters at $z \sim 7.15 - 7.45$, 15 photometric galaxies are associated with these overdensities. They are initially selected by photometric density mapping and later confirmed spectroscopically.

\subsection{Photometric SED Fitting}

To estimate photometric redshifts, we use the {\tt EAZY-py} code \citep{brammer2008eazy} with a combination of 12 default templates and 6 additional templates from \citet{Larson2023}. These include young (ages $10^6$–$10^7$ yr), low-metallicity (5\% $Z_\odot$) stellar populations with strong emission lines and blue UV slopes, characteristics commonly observed in $z>6$ galaxies \citep{Finkelstein2022,Nanayakkara2022,cullen2023ultraviolet}. The \citet{Larson2023} templates are generated using BPASS v2.2.1 and processed with {\tt CLOUDY} v17.0 (log $U = -2$) to include nebular emission. We also apply IGM attenuation from \citet{Madau1995}. A minimum 10\% photometric uncertainty is assumed to account for potential systematics in photometry and model mismatches.

Redshift probability distributions derived from \texttt{EAZY} are then used as inputs for SED fitting with the Bayesian Analysis of Galaxies for Physical Inference and Parameter Estimation Software (\texttt{BAGPIPES}; \citealt{Carnall2018,Carnall2023}). We adopt \citealt{Bruzual2003} stellar models with a Kroupa IMF and assume a lognormal star formation history (SFH), parameterized by the time of peak star formation and the SFH width (FWHM), each with priors spanning [0.01, 10] Gyr. Star formation is truncated before the cosmic age at the given redshift, as required by the model. Nebular emission and continuum are modeled using {\tt CLOUDY}-based templates \citep{2013RMxAA..49..137F}, and we adopt the \citet{Calzetti} dust attenuation curve. We assume log-uniform priors for dust attenuation ($A_V \in [0.0001, 10]$), gas-phase metallicity ($Z/Z_\odot \in [10^{-3}, 3]$), and ionization parameter (log$_{10} U \in [-3, -1]$), consistent with expectations for young, metal-poor galaxies at high redshift. All these settings are consistent with the configuration used in previous papers of the EPOCHS series (e.g. \citealt{Harvey2025, Austin2025}).

\subsection{Spectroscopic Fitting with BAGPIPES}\label{sec:spec_bagpipes}
To further confirm the redshifts of protocluster member galaxies and derive their physical properties, we use \texttt{BAGPIPES} to conduct spectro-photometric fitting of 1D prims spectra released publicly by the JADES collaboration.

The first step in our analysis involves extracting 1D prism spectra from the JADES public release dataset.
We use a non-parametric star formation history (SFH) model using a continuity prior \citep{leja2019measure}, allowing for a flexible description of the galaxy’s star formation activity over time. The SFH is defined using six logarithmically spaced time bins, where the star formation rate (SFR) in each bin is independently constrained by the observed spectrum. To account for the effects of stellar and nebular emission, we model the stellar metallicity with a uniform prior in the range $Z/Z_{\odot} \in [0, 3]$, while nebular emission is included using \textsc{Cloudy}-based templates \citep{Gutkin2016}, with the ionization parameter varying between $\log U = -3$ and $-1$. Dust attenuation is treated using the \citet{Calzetti} attenuation law, and we explore alternative parameterizations such as the flexible \citet{Salim2018} model, which introduces a power-law modification to the attenuation curve. The spectral resolution is incorporated using a wavelength-dependent instrumental response function, allowing us to account for variations in the spectral resolving power across different wavelength ranges.

The fitting procedure is performed using the \texttt{multinest} nested sampling algorithm \citep{Feroz2009}, ensuring efficient exploration of the parameter space. For each galaxy, we sample 500 posterior realizations to estimate the full probability distribution of physical parameters, including stellar mass, SFR, dust attenuation, and metallicity. The resulting best-fit model spectra are saved alongside their 16th, 50th, and 84th percentile uncertainty envelopes, providing a robust characterization of the uncertainties associated with each parameter.

\subsection{Sample Selection and Overdensity Estimation}
To study the local environments of high-$z$ galaxies, we use the k-dimensional tree (KDTree) data structure to efficiently search for nearest neighbors. This enables us to analyze the properties of the galaxy as a function of the local environment (i.e., the number of neighbors or the local density).

To ensure a reliable measurement of the galaxy environment, we use the nearest neighbor method to define local density and identify potential galaxy group members, following the approach of \citet{Lopes2016}. Specifically, for each galaxy in our dataset, we compute its projected distance, $d_n$, to the $n$th nearest neighboring galaxy, while applying a maximum velocity offset constraint to mitigate contamination. The velocity offset mask is defined as:
\begin{equation}
    \mathcal{M}^v(z) =
    \begin{cases}
    1 & \text{if } \Delta z < 0.1  \\
    0 & \text{otherwise}
    \end{cases}.
\end{equation}
This selection of $\Delta z < 0.1$ ensures consistency with our photometric redshift uncertainty constraints and effectively captures the entirety of overdensity regions, also as suggested by \citet{Chiang2013} and \citet{Muldrew2015}.

The local galaxy density, $\Sigma_n(z)$, is defined as:
\begin{equation}
\Sigma_n (z) = \frac{n}{\pi d_n^2},
\end{equation}
where $d_n$ is the projected distance in Mpc to the $n$th nearest neighbor, and $\Sigma_n$ is expressed in units of galaxies per Mpc$^2$. The parameter $n$, representing the rank of the density-defining neighbor, must be carefully chosen to ensure sensitivity to environmental variations. A common choice in the literature, adopted here, is $n = 5$ \citep{Lopes2016, Santos2014}, as this value remains smaller than the number of galaxies typically found in clusters while ensuring a robust local density estimate.

The selection criteria for high-redshift cluster galaxies at $z>7$ are consistent with previous studies (e.g., \citealt{Qiong2024}), requiring galaxies to lie at redshifts $z > 7$ and to reside in the upper quartile of the $\Sigma_5$ surface density distribution. Based on these criteria, we identified two notable protoclusters in the `DEEP' region of the JADES GOODS-South field. The corresponding density maps, constructed from photometric data, are shown in Figure~\ref{fig:density_maps}.

We identify two nearby protocluster candidates in the JADES--GS field that are closely associated both spatially and in redshift. Given their proximity, we primarily describe their {combined} properties below. The two overdensities are traced by a total of $N_{\rm phot}=15$ photometrically selected galaxies, with $7$ and $8$ members in each structure, respectively. In both cases, $5$ galaxies are spectroscopically confirmed. The two protocluster candidates have galaxy overdensity significances of $\delta_{\rm gal}=5.0$ and $\delta_{\rm gal}=5.3$. The photometric members span a similar redshift range, $z_{\rm phot}\simeq7.13$--$7.41$. The typical photometric-redshift uncertainties of the JADES catalog, $\Delta z\sim0.14$. The combined structure extends over an angular diameter of $\sim1.8'$, corresponding to a comoving size of $\sim4.8$~cMpc at $z\sim7$. In contrast, the spectroscopically confirmed galaxies at the overdensity peak occupy a much narrower redshift interval of $z_{\rm spec}=7.24$--$7.28$.

Given their small angular separation and overlapping redshift distributions, it is plausible that the two overdensities trace a single extended large-scale structure, rather than two completely independent systems. We further examine the spatial relation between the protocluster candidates and the Ly$\alpha$ emitter reported by \citep{Saxena2023}. The projected distance between JADES--GS--z7--LA and the centres of the protocluster candidates is $\sim40''$, corresponding to $\sim0.2$~physical~Mpc at $z\sim7$.

\subsection{Spectroscopic Confirmation and Validation}

In \cite{Qiong2024}, we conducted a purely photometric-redshift–based search for overdensities across $z = 4-10$, which yielded several candidate structures but without requiring spectroscopic confirmation. In this work, our analysis is restricted to $z > 7$ and aims to identify overdense regions that are supported by spectroscopic redshifts from JADES/NIRSpec.

We examined all galaxies with $z > 7$ within the JADES NIRCam footprint, including both spectroscopic and high-quality photometric redshifts. Among these, only two overdensities contain multiple spectroscopically confirmed members. These are the only two spectroscopically validated structures at $z > 7$ identified in the current dataset. Additional photometric overdensity candidates exist but lack the required spectroscopic follow-up and are therefore not included in our analysis.

To validate the identified protoclusters, we reviewed available spectroscopic data. Remarkably, 10 out of 15 galaxies in the candidate sample have spectroscopic confirmations. To ensure the robustness of the protocluster membership, we conducted the following additional checks:
\begin{enumerate}
    \item Investigated all potential member galaxies within $\Delta z < 0.1$ and $r_p < 100$~kpc.
    \item Cross-referenced the results with previously identified protoclusters reported in the literature, including spectroscopically confirmed galaxy overdensities in GOODS-N and GOODS-S \citep{Helton2023}. 
\end{enumerate}

\begin{figure*}
\centering
 \includegraphics[width=0.95\textwidth]{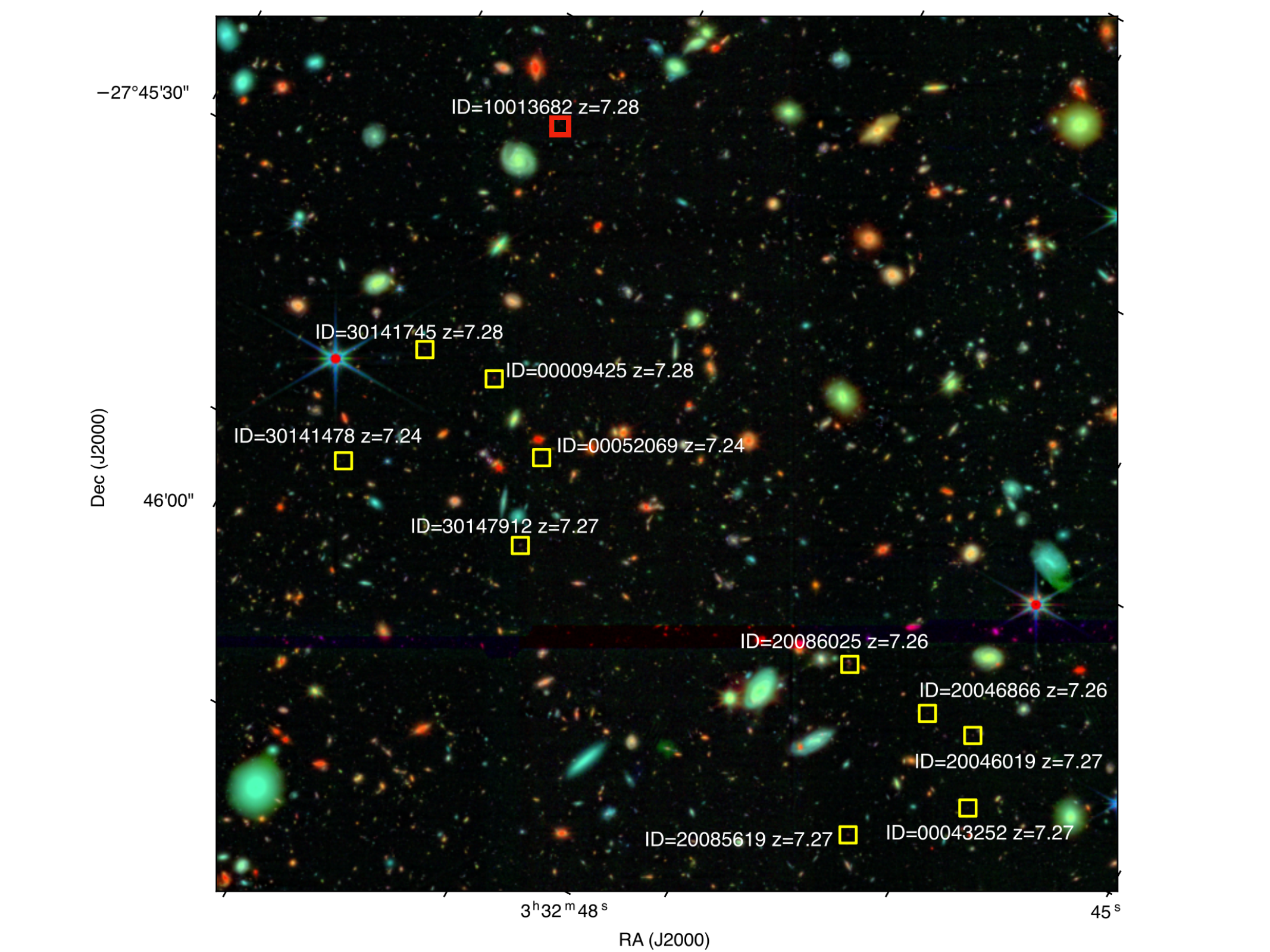} 
    \caption{
    JWST/NIRCam RGB images of two protoclusters at $z\sim7.2$, constructed using the F115W (blue), F200W (green), and F444W (red) filters. The yellow squares indicate spectroscopically confirmed members of the overdensity, with their respective IDs and redshifts labeled. These protoclusters exhibit significant overdensities of star-forming galaxies. The red square marks the location of a bright LAE (JADES-GS-z7-LA) located $\sim40''$ away from the center of overdensity.}
    \label{fig:protocluster_members}
\end{figure*}

Our analysis confirmed the presence of two high-redshift protoclusters in the JADES field. The first structure comprises 5/7 spectroscopically confirmed member galaxies, while the second includes 5/8 such members. The spatial distribution of these galaxies is shown in Figure~\ref{fig:protocluster_members}.

In addition to the JADES DR1+DR3 spectroscopy used in our primary analysis, we have also cross-checked the FRESCO spectroscopic catalogue in the relevant region \citep{Helton2023}. We find one additional galaxy with a secure FRESCO redshift ($z=7.272$) that is spatially coincident with the same overdensity. While this object is not included in our formal analysis to preserve a homogeneous spectroscopic sample, its presence provides independent support for the reality of the structure.

For each galaxy we derive an individual $3\sigma$ upper limit on the 
rest-frame Ly$\alpha$ equivalent width. We first fit a power-law 
continuum to the rest-frame spectrum while masking the Ly$\alpha$ 
region and all strong spectral features. The noise level around 
Ly$\alpha$ is measured in an adaptive $\pm 20$--$50$\,\AA\ window, 
and a Gaussian line profile with ${\rm FWHM}=300~{\rm km\,s^{-1}}$ 
is assumed to convert the noise into a $3\sigma$ flux limit. Across the ten $z>7$ galaxies in the overdense region, we obtain 
${\rm EW}_{0,\,3\sigma}$ limits spanning 
$\approx 6$--$23$\,\AA, with a mean of 
$\langle{\rm EW}_{0,\,3\sigma}\rangle \approx 15$\,\AA.
These values are consistent with the Ly$\alpha$ non-detection limits 
reported in recent JWST studies (e.g., \citealt{Tang2023}).
Upon further examination of the spectra, we find that almost all protocluster galaxies exhibit weak or no detectable Ly$\alpha$ emission ($<3\sigma$ detections), consistent with the suppression expected from a largely neutral intergalactic medium at high redshift. Notably, 6/10 galaxies show strong nebular emission lines, with both [O\,\textsc{iii}] and H$\beta$ detected at $>5\sigma$ significance, indicating a highly ionized and actively star-forming interstellar medium.

Our sample consists of 10 spectroscopically confirmed galaxies at $z\sim7.2$, forming a compact overdensity within a projected scale of $\sim30$ arcseconds. As noted in Section~\ref{sec:spec_bagpipes}, the redshifts used in our overdensity analysis are obtained from our Bagpipes SED-fitting procedure. The stellar masses, SFRs and the specific star formation rate (sSFR) are extracted from spectral energy distribution fitting, using a median-likelihood approach. The UV magnitudes are measured in the rest-frame 1450–1550 Å window, avoiding contamination from strong emission lines. 

The full set of 1D and 2D spectra for all spectroscopically confirmed protocluster members is presented in Appendix~\ref{appendix:spectra}, and their derived physical properties and best-fit spectral energy distributions are summarized in Appendix~\ref{appendix:properties}.

The identification of the two protoclusters relies primarily on the high-quality JADES photometric redshifts. The median photometric-redshift uncertainty from our SED fitting is $\Delta z \simeq 0.14$. The outlier fraction, defined as the fraction of photometric redshifts that deviate from the spectroscopic redshift by more than 15\% in $(1+z)$, is very low at $\eta = 2.9\%$. Here, we chose galaxies within a $\Delta z<0.1$ redshift cube, in order to better match our photometric redshift uncertainty constraints. We tested thresholds of $\Delta z<0.15$ and $\Delta z<0.2$ and found that the local galaxy density values for a few galaxies do change with these larger thresholds.  Additionally, according to the predictions by \citet{Chiang2013} and \citet{Muldrew2015}, the effective diameter of protoclusters, defined as 2 $\times$ the effective radius (2Re), is about 20 cMpc at $z = 5$. Hence, selecting galaxies with $\Delta z<0.1$ conveniently covers the entire overdensity region. Tests in which galaxy redshifts are perturbed within their photometric uncertainties show that the significance of both structures remain unchanged. The photometric-redshift errors therefore primarily smooth the redshift distribution rather than altering the identification of the overdense peaks.

The spectroscopic sample, although incomplete owing to the NIRSpec target-allocation strategy and flux limits, provides an independent validation of the photometric structures. This is also the reason why we rely on the photometric sample for the initial identification of candidate members, as the spectroscopic coverage is sparse and non-uniform across the field. In fact, several additional high-redshift overdensities identified from photometric redshifts are not pursued further in this work precisely because of the lack of corresponding spectroscopic coverage, as discussed in our previous analysis \citep{Qiong2024}. The confirmed NIRSpec redshifts nevertheless cluster within the same two redshift peaks identified photometrically, supporting the robustness of the protocluster membership. These considerations demonstrate that the observed overdensities and their inferred physical scales are not driven by selection bias, but remain stable against both photometric-redshift uncertainties and spectroscopic incompleteness.

\section{Results}\label{sec:result}
\subsection{Stacked Spectra and \texorpdfstring{Ly$\alpha$}{Lyalpha} Visibility in Dense Environments}\label{sec:stacked}

The study of high-redshift galaxy populations in different environments provides crucial insights into the early stages of galaxy formation and the role of star-forming galaxies (SFGs) in cosmic reionization. Some of these galaxies exhibit a significant decrease in flux redward of the Ly$\alpha$ line. This feature may suggest the presence of substantial reservoirs of neutral, cold gas. In this section, we stack the spectra of three sets of high-redshift galaxies in different environment, aiming to characterize their spectral properties, and compare the prevalence of these features in protocluster galaxies to those observed in field galaxies. 

\subsubsection{Stacked spectrum}

To stack the spectrum, we ues the spectroscopic data retrieved from the publicly available JADES datasets. The sample includes confirmed protocluster galaxies, as well as comparison samples from different environments, categorized as cluster or field galaxies \citep{Bunker2023}.

Each galaxy spectrum is corrected for redshift, calibrated, and normalized before stacking. The normalization is performed using the mean flux in the rest-frame wavelength range of 1500–2500 Å. This ensures that all spectra contribute equally to the final stacked spectrum, minimizing biases from variations in intrinsic brightness.

The stacking procedure follows a weighted averaging method. For each galaxy, the spectrum is interpolated onto a common rest-frame wavelength grid spanning from 800 Å to 6300 Å, with a resolution of 5000 grid points. A weighted mean is computed using inverse variance weighting, ensuring that spectra with higher SNR contribute more significantly. The stacked spectrum is then used to measure emission line fluxes and continuum properties. 
To assess whether the stacked spectra are dominated by a small number of bright or high signal-to-noise galaxies, we performed a test of the stacking methodology. In addition to our fiducial inverse-variance stacking, we constructed alternative stacks using equal weighting, while keeping all other steps identical, including rest-frame interpolation, spectral binning, continuum fitting, and the definition of the Ly$\alpha$ equivalent width upper limit. Using a 20~\AA\ rest-frame window to estimate the 3$\sigma$ Ly$\alpha$ equivalent width limit, we obtain $EW_{\mathrm{Ly}\alpha}^{3\sigma} = 9.56$~\AA\ for the inverse-variance stack and $11.02$~\AA\ for the equal-weight stack, differing by only $1.46$~\AA. This difference is well within the statistical uncertainties and does not affect our conclusions. We further quantify the contribution of individual galaxies to the inverse-variance stack and find that the most heavily weighted galaxy contributes 22\% of the total weight, while the top three galaxies together contribute 49\%. This indicates that the stacked spectra are not dominated by a single high-S/N source. We therefore conclude that our adopted inverse-variance stacking does not introduce significant bias and adopt it throughout the analysis.

To remove contamination from strong emission lines, a masking procedure is applied. A set of known strong spectral lines, including Ly$\alpha$, C\,\textsc{iv}, He\,\textsc{ii}, and [O\,\textsc{iii}], are masked within a $\pm50$\,\AA\ window. For some broad lines, the masking window is widened accordingly. Additionally, wavelengths below 1200 Å are excluded due to potential noise contamination from the instrument's sensitivity limits. The continuum is then fitted using a power-law function of the form ($
    f_{\lambda} = a\, \lambda^{\beta},
$)
where \( a \) and \( \beta \) are free parameters obtained through a non-linear least-squares fit.

\begin{figure*}
    \centering
    \includegraphics[width=0.85\textwidth]{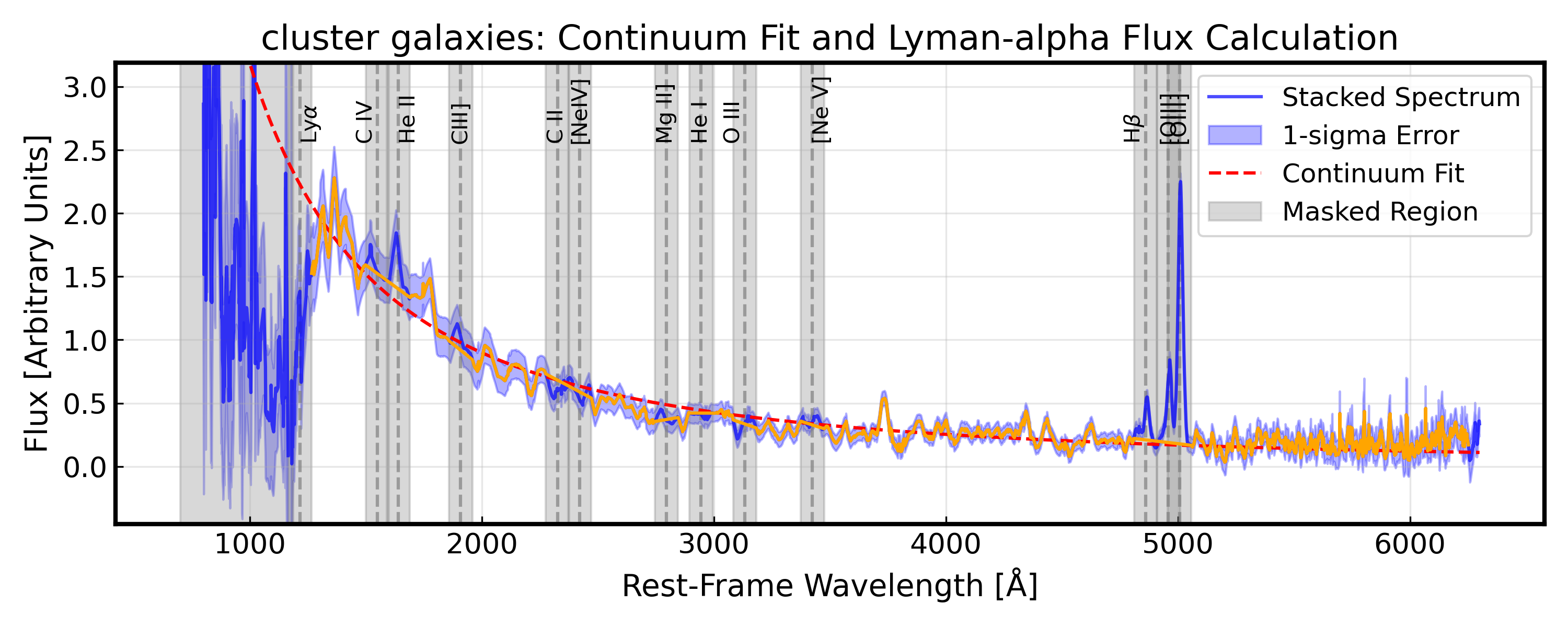}
    \caption{Stacked spectrum of 10 protocluster galaxies at $z\sim7.2$. The shaded regions indicate the masked spectral lines. The orange line indicates regions of the spectrum for continuum fitting, where strong emission lines have been masked out. The red dashed line represents the fitted continuum.}
    \label{fig:stacked_spectrum}
\end{figure*}

Figure \ref{fig:stacked_spectrum} presents the stacked spectrum of the protocluster galaxies. The continuum is well-fitted by a power-law model, with the best-fit parameters providing an estimate of the underlying stellar population's spectral energy distribution. The masked spectral lines are indicated by shaded regions. After subtracting the fitted continuum, the Ly$\alpha$ flux is calculated by integrating the flux within a $\pm20$ Å window centered at 1215.67 Å in the rest frame. The equivalent width (EW) of Ly$\alpha$ is determined using the continuum flux density at 1215.67 Å. Propagating errors are from both the flux measurement and the continuum fit.

To assess whether differences in the intrinsic properties of the galaxies could bias the stacking results, we examine the UV luminosities and stellar masses of the protocluster and field samples used in the continuum stacks. The protocluster members occupy a relatively narrow range of $M_{\rm UV} \sim -17$ to $-19$, while the field samples span a broader luminosity distribution. Similarly, the protocluster galaxies predominantly lie in an intermediate stellar-mass regime ($\log M_* \sim 7.5$--$8.5$), whereas the field samples include both lower- and higher-mass systems. These variations are modest and fall within the typical diversity of $z\sim7$ star-forming galaxies, and therefore are not expected to drive systematic differences in the stacked continua.

\subsubsection{Control Sample Selection and Comparison with Protocluster Galaxies}\label{sec:control_sample}

To robustly assess the impact of environment on galaxy properties at high redshift, we construct two control samples based on their local density measurements. The first control sample consists of galaxies identified as residing in strictly underdense environments. These galaxies were manually inspected to confirm their isolation, with local density values ($\Sigma_5$) falling within the lowest quartile of the distribution. To ensure a sufficient sample size, the redshift range was slightly extended to $z=7.0-9.5$, resulting in a final selection of 35 galaxies.

\begin{figure}
    \centering
    \includegraphics[width=0.5\textwidth]{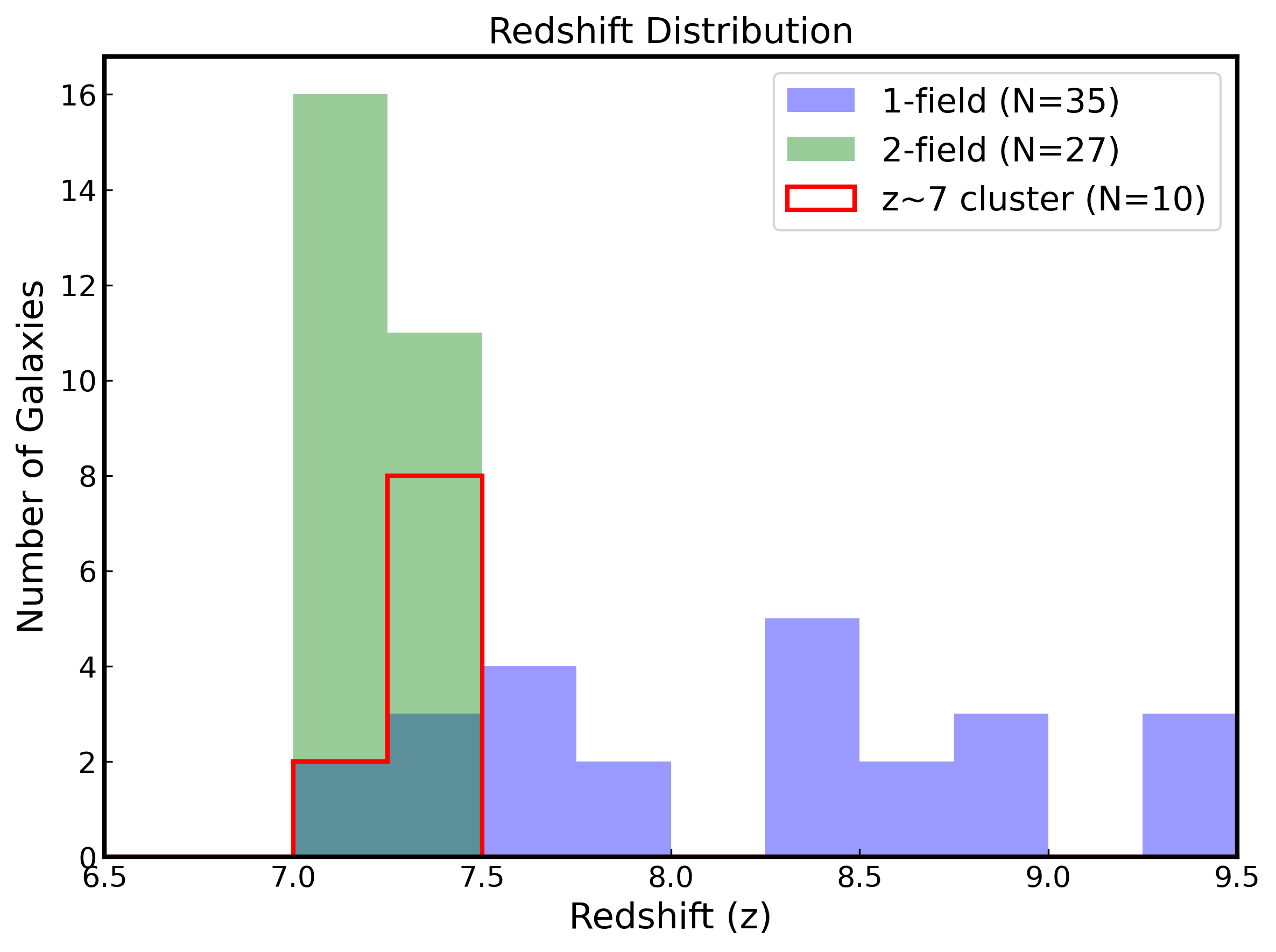}
    \caption{Redshift distribution of the \(z \sim 7\) protocluster and field galaxy control samples. The protocluster exhibits a strong overdensity near \(z \sim 7.1\).}
    \label{fig:redshift}
\end{figure}

The second control sample was designed to match the redshift distribution of galaxies within the identified protocluster while ensuring that none of the selected galaxies resided in highly overdense regions. Unlike the strictly underdense sample, this selection did not impose a stringent isolation criterion but instead focused on excluding galaxies from the highest-density regions. This resulted in a final sample of 27 galaxies within the redshift range $z=7.1-7.4$, allowing for a direct comparison with the protocluster galaxies at matching redshifts. Figure~\ref{fig:redshift} presents the redshift distribution of the three samples. 


We checked all spectra and excluded sources with poor data quality to ensure reliable analysis. The stacked spectra for the two control samples are shown in Figure~\ref{fig:stacked_spectra_control} in the Appendix. The top panel corresponds to the strict field sample, while the bottom panel shows the broader control sample.
Both the control samples and the $z \sim 7$ protocluster galaxies were analyzed using the same \texttt{BAGPIPES} spectral fitting settings \citep{Carnall2018}. The fitting procedure used the same assumptions regarding star formation histories, dust attenuation laws, and metallicity priors to ensure a consistent comparison across different environments. For both control samples and the protocluster galaxies, the measured Ly$\alpha$ fluxes and equivalent widths (EWs) remain below the $3\sigma$ detection threshold, indicating a lack of strong Ly$\alpha$ emission across all environments at $z>7$.

\subsection{Ionizing Photon Production in Protoclusters and Field Galaxies}

\begin{figure*}
    \centering
    \includegraphics[width=0.48\textwidth]{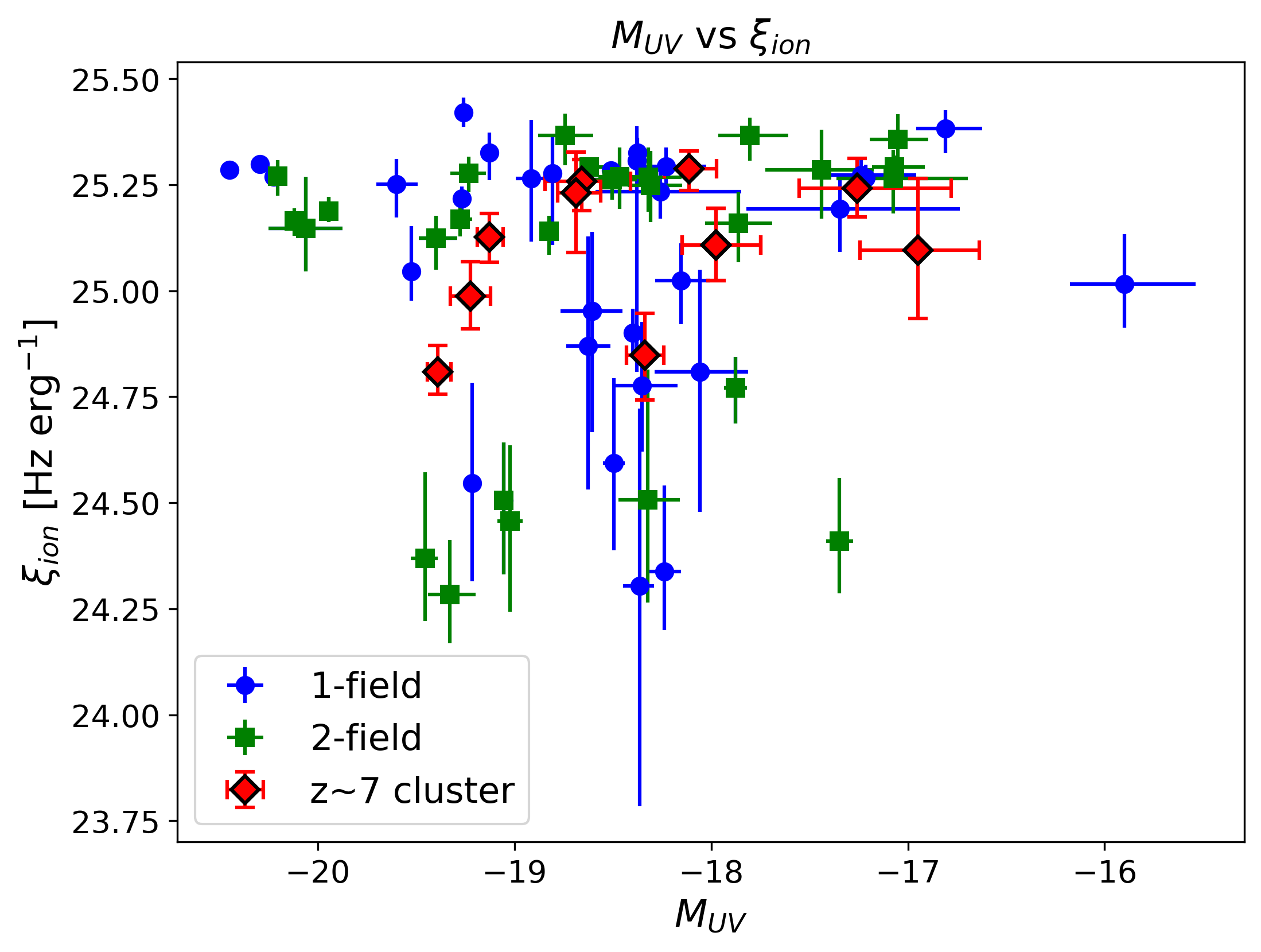}
    \includegraphics[width=0.48\textwidth]{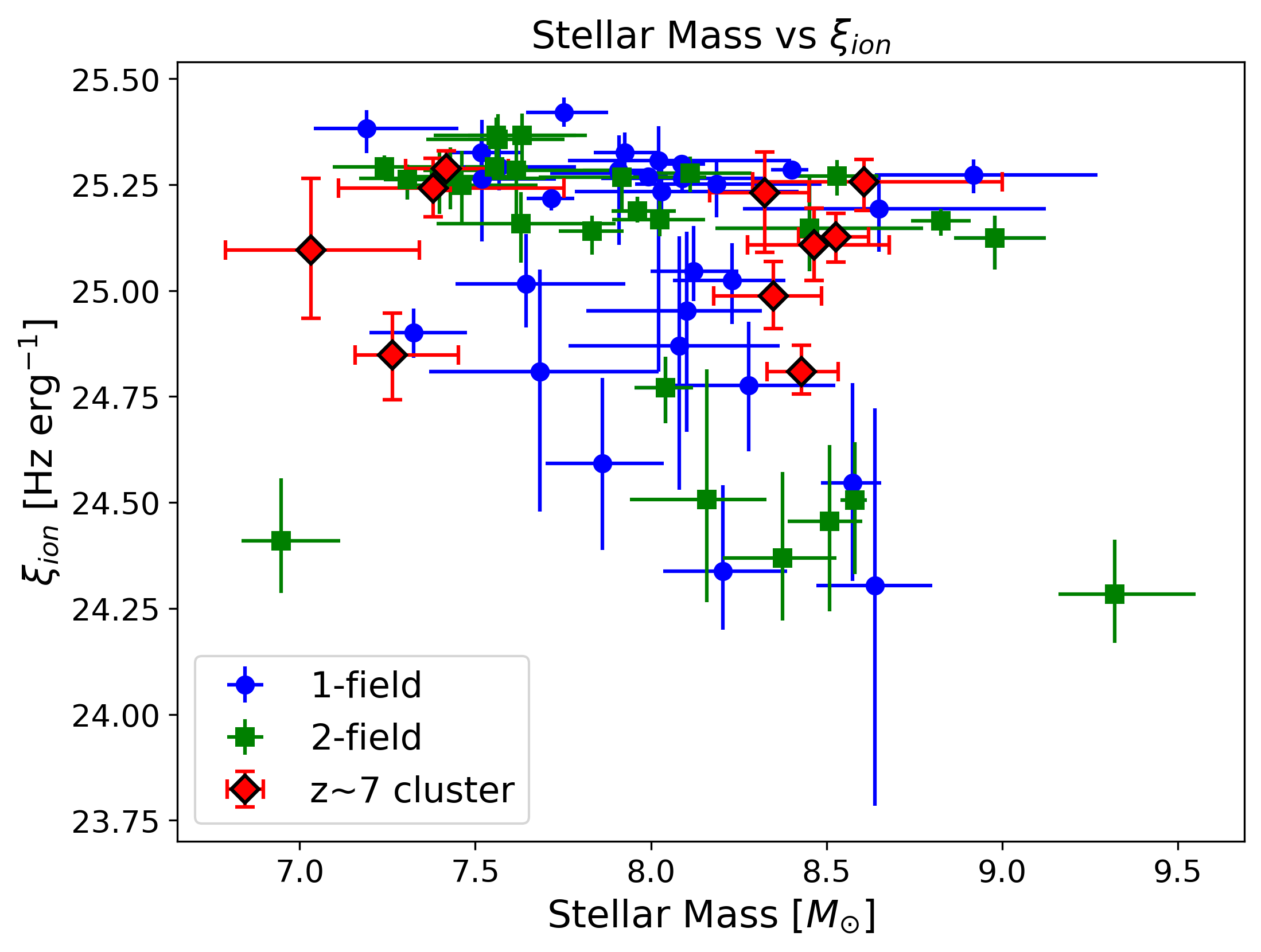}
    \caption{
    Left: Comparison of \(\xi_{\text{ion}}\) as a function of \(M_{\text{UV}}\) for our protocluster and field samples. 
    Right: Stellar mass vs. \(\xi_{\text{ion}}\). 
    In both panels, protocluster galaxies exhibit systematically higher \(\xi_{\text{ion}}\) values compared to field galaxies at fixed \(M_{\text{UV}}\) or stellar mass. 
    The labels “1-field” and “2-field” refer to the two control samples described in Section~\ref{sec:control_sample}. `1-field' is the strictly underdense sample selected from the lowest quartile of the local density distribution across the full redshift range, 
    while `2-field' represents the redshift-matched control sample selected to avoid overdense regions at $z = 7.1-7.4$.}
    \label{fig:MUV_xi_ion}
\end{figure*}

\begin{figure*}
    \centering
    \includegraphics[width=0.48\textwidth]{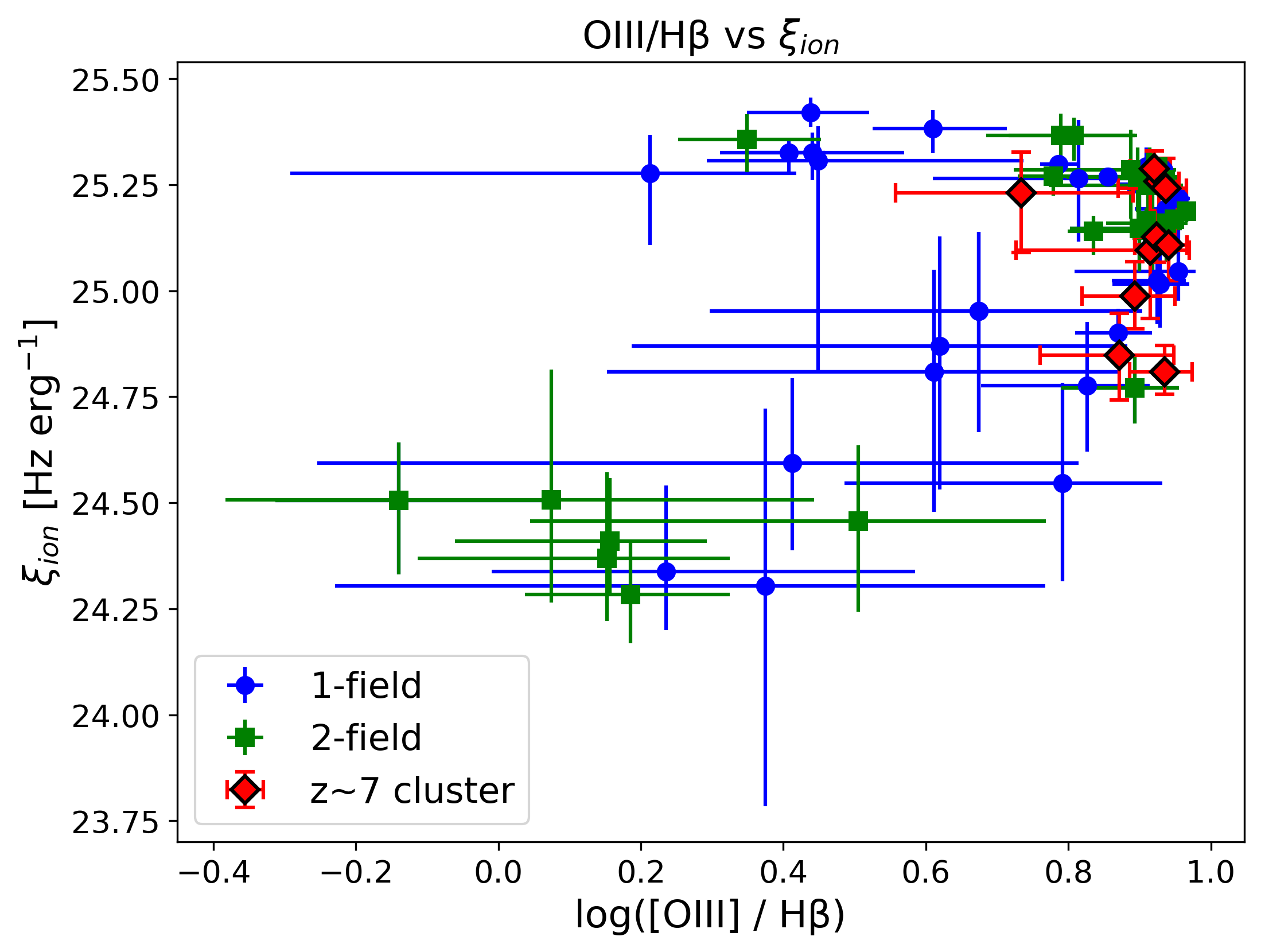}
    \includegraphics[width=0.48\textwidth]{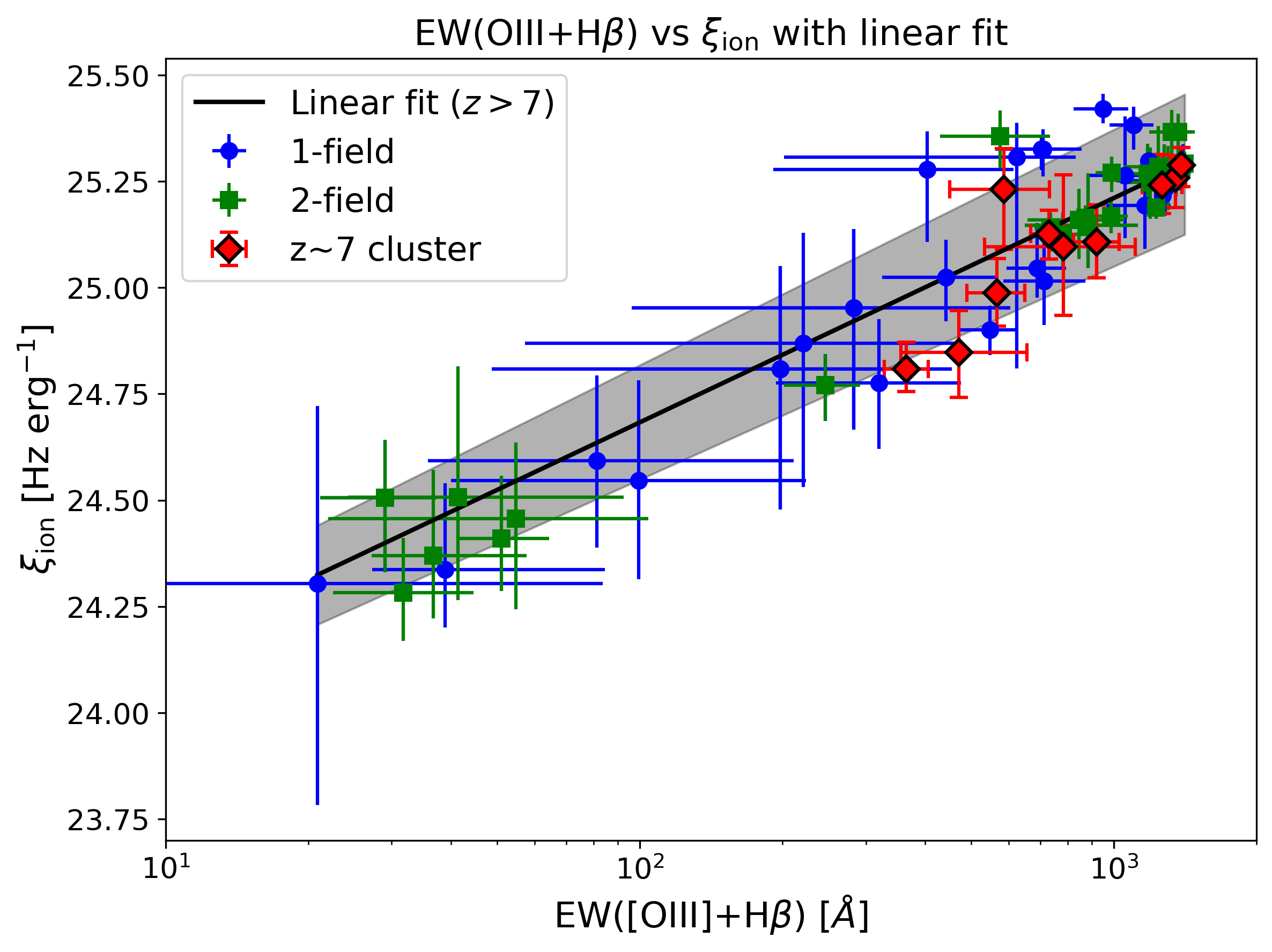}
    \caption{
    Left: The relationship between the nebular line ratio \([\text{OIII}]/\text{H}\beta\) and \(\xi_{\text{ion}}\) for protocluster and field galaxies. Protocluster galaxies (red diamonds) exhibit systematically higher \([\text{OIII}]/\text{H}\beta\) ratios, suggesting enhanced ionization conditions and possibly lower metallicities compared to the field populations (blue and green points). 
    Right: The correlation between \(\xi_{\text{ion}}\) and the equivalent width of \([\text{OIII}]+\text{H}\beta\). A strong positive trend is evident, with galaxies exhibiting larger equivalent widths also having higher \(\xi_{\text{ion}}\). The logarithmic scale is on the x-axis. 
    }
    \label{fig:xi_ion_oiii_hb}
\end{figure*}

The production and escape of ionizing photons from early galaxies are fundamental parameters in understanding the contribution of high-redshift galaxies to cosmic reionization (e.g., \citealt{Bouwens2016,Begley2025,Bosman2024,Llerena2025}). The intrinsic ionizing photon production rate, \( \dot{N}_{\text{ion}}^{\text{int}} \), represents the number of photons capable of ionizing hydrogen that are produced per second by a galaxy. The ionizing photon production efficiency, \( \xi_{\text{ion}} \), is defined as the number of ionizing photons emitted per unit UV luminosity and serves as a key metric for determining the contribution of galaxies to the ionizing background (e.g., \citealt{Finkelstein2019,Duncan2015}). 

To quantify these properties, we compute \( \dot{N}_{\text{ion}} \) using Balmer recombination lines and derive \( \xi_{\text{ion}} \) from the UV continuum at 1500 Å. We apply these calculations to both the field galaxies and the protocluster sample at \( z \sim 7 \) using spectral modeling with \texttt{BAGPIPES} \citep{Carnall2018}. Under the assumption of Case B recombination \citep{Osterbrock2006} and standard ISM electron temperatures and densities, the intrinsic ionizing photon production rate can be derived from the intrinsic H$\alpha$ luminosity \( L_{\text{H}\alpha} \), which is related to the number of ionizing photons by:

\begin{equation}
    \dot{N}_{\text{ion, case B}}^{\text{int}} = \frac{L_{\text{H}\alpha}}{1.36 \times 10^{-12} \, \text{erg/photon}}
\end{equation}

where the conversion factor \( 1.36 \times 10^{-12} \) erg/photon accounts for the fraction of ionizing photons that lead to H$\alpha$ emission. The H$\alpha$ luminosity is obtained from the fitted H$\alpha$ flux \( f_{\text{H}\alpha} \) from \texttt{BAGPIPES}, corrected for dust attenuation. 

The ionizing photon production efficiency is defined as:

\begin{equation}
    \xi_{\text{ion, case B}} = \frac{\dot{N}_{\text{ion, case B}}^{\text{int}}}{L_{\text{UV,1500}}}
\end{equation}

where \( L_{\text{UV,1500}} \) is the dust-corrected UV luminosity at 1500 Å. This is calculated by integrating the rest-frame UV spectrum over a 100 Å window centered at 1500 Å and applying a dust correction. 

Figure~\ref{fig:MUV_xi_ion} (a) shows the relationship between absolute UV magnitude (\(M_{\text{UV}}\)) and \(\xi_{\text{ion}}\) for our sample. The protocluster galaxies (red diamonds) are clustered around \(M_{\text{UV}} \sim -17\) to \(-19\), while the field samples extend to fainter and brighter magnitudes. Notably, the protocluster galaxies exhibit systematically higher \(\xi_{\text{ion}}\) values compared to field galaxies at similar \(M_{\text{UV}}\). Figure~\ref{fig:MUV_xi_ion} (b) shows the dependence of \(\xi_{\text{ion}}\) on stellar mass. While field galaxies span a broad mass range, protocluster members predominantly occupy an intermediate-mass regime (\(\log(M_*) \sim 7.5 - 8.5\)). At fixed stellar mass, protocluster galaxies tend to exhibit slightly higher \(\xi_{\text{ion}}\). This suggests that galaxies in overdense regions may experience conditions that enhance the production efficiency of ionizing photons. Possible explanations include variations in stellar population properties, such as younger, more metal-poor stellar populations, or differences in feedback mechanisms that regulate nebular escape fractions \citep{Bouwens2016,Endsley2023}.

Figure~\ref{fig:xi_ion_oiii_hb} (a) presents the relationship between \(\xi_{\text{ion}}\) and the [O\,\textsc{iii}]/H\(\beta\) line ratio. The protocluster galaxies exhibit higher [O\,\textsc{iii}]/H\(\beta\) values, indicative of high-excitation conditions likely driven by harder ionizing spectra and enhanced ionization parameters. Such high ratios are characteristic of galaxies hosting young, massive stellar populations and compact star-forming regions, where radiation fields are both intense and hard. While high [O\,\textsc{iii}]/H\(\beta\) ratios are often associated with low gas-phase metallicities, they are also influenced by ionization parameter and starburst age. The alignment of protocluster galaxies along this trend suggests that environmental factors may accelerate or intensify these ionizing conditions \citep[e.g.,][]{Nakajima2016,Tang2019}. In Figure~\ref{fig:xi_ion_oiii_hb} (b), we investigate the relationship between the equivalent width of the [O\,\textsc{iii}]+H\(\beta\)  and \(\xi_{\text{ion}}\). A clear positive correlation is observed, which galaxies with higher \(\xi_{\text{ion}}\) also exhibit stronger nebular emission. We fit \(\xi_{\text{ion}}\) as a
function of EW([O\,\textsc{iii}]+H\(\beta\)) to obtain the following relation:
\begin{equation}
\begin{aligned}
\log \left( \frac{\xi_{\mathrm{ion}}}{\mathrm{Hz\,erg}^{-1}} \right) 
&= (0.53 \pm 0.03) \times \log \left( \frac{EW([\mathrm{O\,III}]+\mathrm{H}\beta)}{\text{\AA}} \right) \\
&\quad + (23.63 \pm 0.08)
\end{aligned}
\end{equation}
This trend is consistent with expectations from photoionization models, where young, metal-poor stellar populations produce harder radiation fields, resulting in both higher ionizing photon production efficiency and stronger emission lines \citep{Stark2017}. Observationally, extreme emission line galaxies (EELGs) at both intermediate and high redshifts show a similar behavior, with high [O\,\textsc{iii}]+H\(\beta\) equivalent widths typically corresponding to low metallicities and high sSFRs \citep[e.g.,][]{Tang2019,Endsley2023,Reddy2018,Topping2022}. Notably, the protocluster galaxies in our sample lie systematically along the high-\(\xi_{\text{ion}}\), high-EW sequence, implying that environment can drive more bursty and efficient star formation, possibly preceding chemical enrichment. 

Overall, these results highlight the complex interplay between environment, stellar populations, and ionization conditions in shaping the efficiency of ionizing photon production. The observed differences between protocluster and field galaxies suggest that dense environments may host more extreme ionizing sources, potentially driven by differences in star formation histories, feedback processes, and metallicity evolution. The enhanced \(\xi_{\text{ion}}\) values in protoclusters imply a more efficient production of ionizing photons, which could contribute significantly to cosmic reionization at high redshifts.

\begin{figure*}
    \centering
    \includegraphics[width=0.75\textwidth]{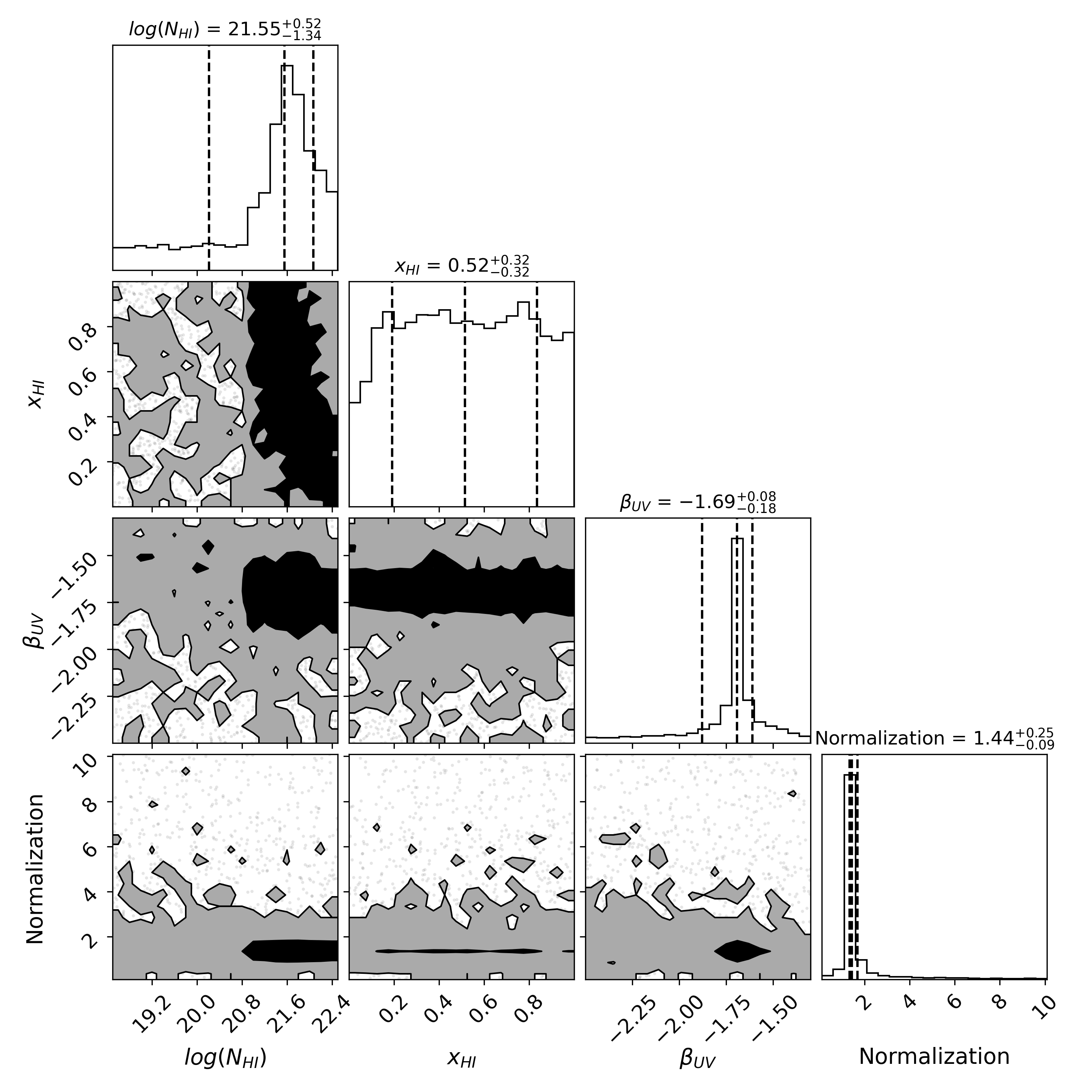}
    \caption{Corner plot of the posterior distributions of the DLA model parameters derived from nested sampling (\texttt{dynesty}) for the cluster galaxies with Ly$\alpha$ emission. Shown are the joint posterior distributions of the neutral hydrogen column density ($\log N_{\mathrm{HI}}$), the neutral fraction ($x_{\mathrm{HI}}$), the UV continuum slope ($\beta_{\mathrm{UV}}$), and the continuum normalization factor. The middle dashed vertical lines in the marginalized panels indicate the median (50th percentile) values, with the other two dashed lines showing the 16th and 84th percentiles. The contours in the joint distributions represent the 68\% and 95\% credible intervals (corresponding to 1$\sigma$ and 2$\sigma$ confidence regions).}
    \label{fig:dla_corner}
\end{figure*}

A potential explanation for these differences lies in the effects of environmental gravitational infall-driven accretion mechanisms. Galaxies residing in dense regions may experience stronger gravitational interactions and gas accretion, leading to bursts of star formation that result in younger, more metal-poor stellar populations. These conditions can increase the hardness of the ionizing spectrum and the efficiency of ionizing photon escape \citep{Mason2018}. Importantly, at fixed stellar mass, galaxies in overdense regions exhibit higher \(\xi_{\text{ion}}\), indicating a true environmental dependence that is not driven by stellar mass. Additionally, differences in gas-phase metallicities between protocluster and field galaxies may play a role in shaping the observed trends, as lower metallicity systems tend to produce harder ionizing spectra and exhibit stronger nebular emission \citep{Nakajima2016}.

The tight correlation between \(\xi_{\text{ion}}\) and nebular line properties further suggests that high-ionization systems are preferentially found in protocluster environments. This implies that ionization conditions in overdense regions may be systematically different from those in the field, with implications for the interpretation of emission-line diagnostics in high-redshift surveys. Future spectroscopic campaigns, particularly with JWST and ground-based facilities, will be crucial in further characterizing these systems and refining our understanding of early galaxy evolution \citep{Endsley2023, Robertson2022}.

Recent JWST observations have extended $\xi_{\rm ion}$ measurements to much fainter galaxies at $z\gtrsim6$, and have reignited discussion of the dependence of $\xi_{\rm ion}$ on UV luminosity. Several studies report a positive $\xi_{\rm ion}$–$M_{\rm UV}$ trend, in which fainter galaxies exhibit higher ionizing efficiencies (e.g. \citealt{Simmonds2024,Llerena2025,Prieto-Lyon2023}). However, more recent analyses suggest that this trend may be biased by the enhanced detectability of faint, bursty systems, and that the underlying distribution is instead flat or even negatively sloped, with brighter galaxies exhibiting comparable or larger $\xi_{\rm ion}$ values (e.g. \citealt{Endsley2023,Begley2025,Simmonds2024,Pahl2025}).

In order to place our measurements in context, we compare our inferred ionizing photon production efficiencies with recent JWST-based studies at similar redshifts. The protocluster and field galaxies in our sample occupy the same broad range of ionizing efficiencies reported in deep JWST surveys, where typical intrinsic values lie between $\log(\xi_{\mathrm{ion}} / \mathrm{Hz,erg^{-1}}),\approx,25.0$--$25.8$ for galaxies at $z\sim6$--9 \citep[e.g.,][]{Begley2025,Endsley2023,Simmonds2024}. Our protocluster galaxies lie toward the upper end of this distribution, whereas field galaxies span its full range. This consistency indicates that the physical conditions governing the ionizing spectra of our galaxies are similar to those found in representative high-redshift populations, while the overdense environment may enhance the ionizing efficiency.

When making such comparisons, it is important to note that different works adopt different methodologies for inferring $\xi_{\mathrm{ion}}$. Some studies derive $\dot{N}{\mathrm{ion}}$ from H\(\alpha\) or inferred star-formation rates rather than H\(\beta\), and UV luminosities may be obtained either from photometry or model spectra (e.g. \citealt{Stefanon2022}). These methodological differences can shift $\xi{\mathrm{ion}}$ by $\sim$0.2--0.4 dex, yet our measurements remain fully consistent with the latest JWST constraints once these systematics are taken into account.

\subsection{Environmental Effects on Neutral Hydrogen}\label{sec:dla}

Ly\(\alpha\) emission and absorption features serve as crucial diagnostics for studying the physical conditions in high-redshift galaxies and their surrounding intergalactic medium. Ly\(\alpha\) radiation, being a resonant line, is highly sensitive to the presence of neutral hydrogen, dust attenuation, and the velocity structure of the gas within galaxies and their circumgalactic medium. The visibility of Ly\(\alpha\) is often suppressed in dense or neutral regions, especially in the presence of damped Lyman-\(\alpha\) absorbers (DLAs), making it essential to accurately model and interpret this feature (\citealt{Ouchi2020, Hayes2023,Heintz2025,Huberty2025}).

In this section we investigate the neutral hydrogen column densities (\(N_{\mathrm{HI}}\)) of galaxies located in proto-clusters at \(z > 7\), comparing their properties under different Ly\(\alpha\) conditions (absence of Ly\(\alpha\) and strong Ly\(\alpha\)) with those of field galaxies.

\begin{figure}
    \centering  \includegraphics[width=0.5\textwidth]{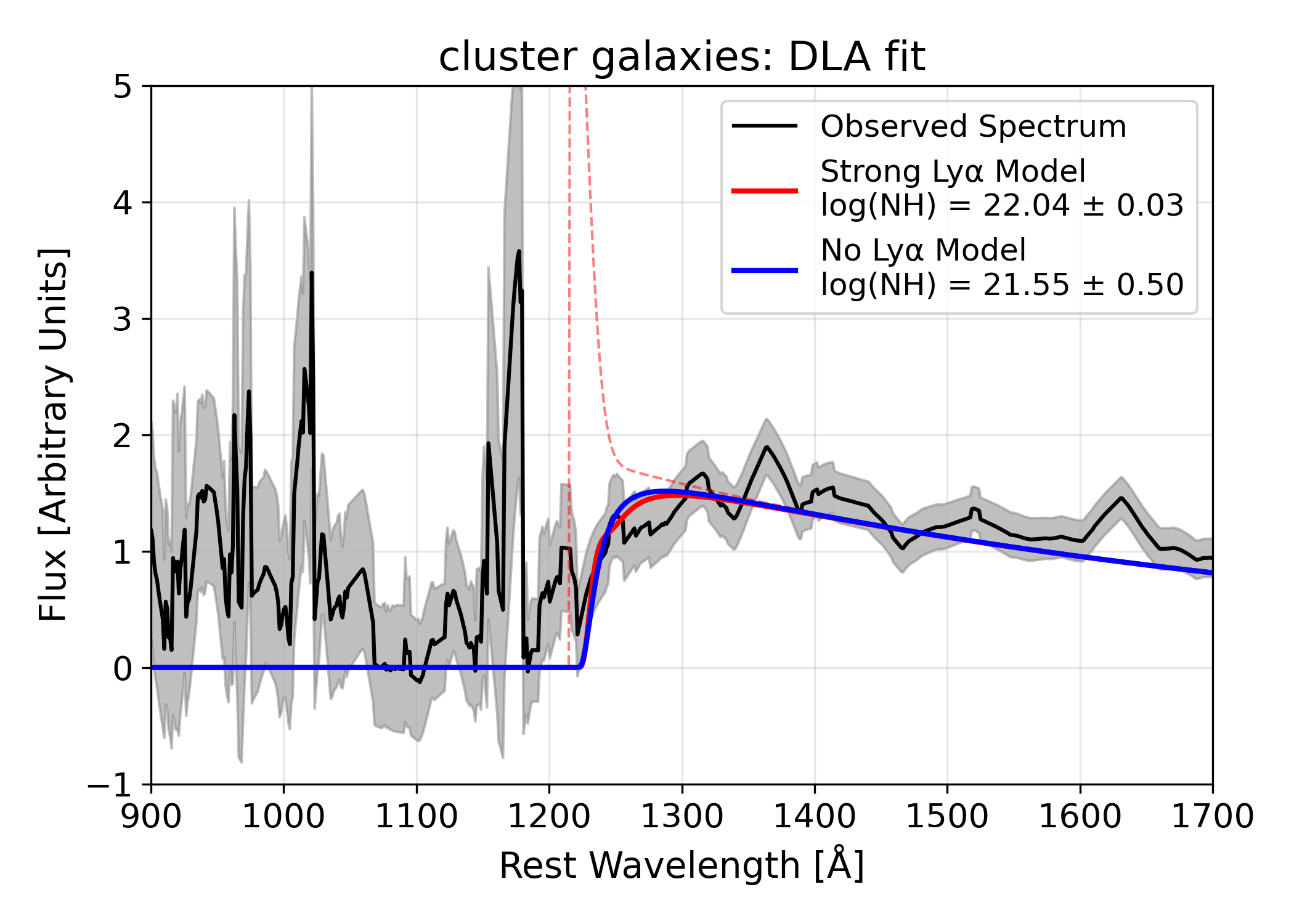}
    \includegraphics[width=0.5\textwidth]{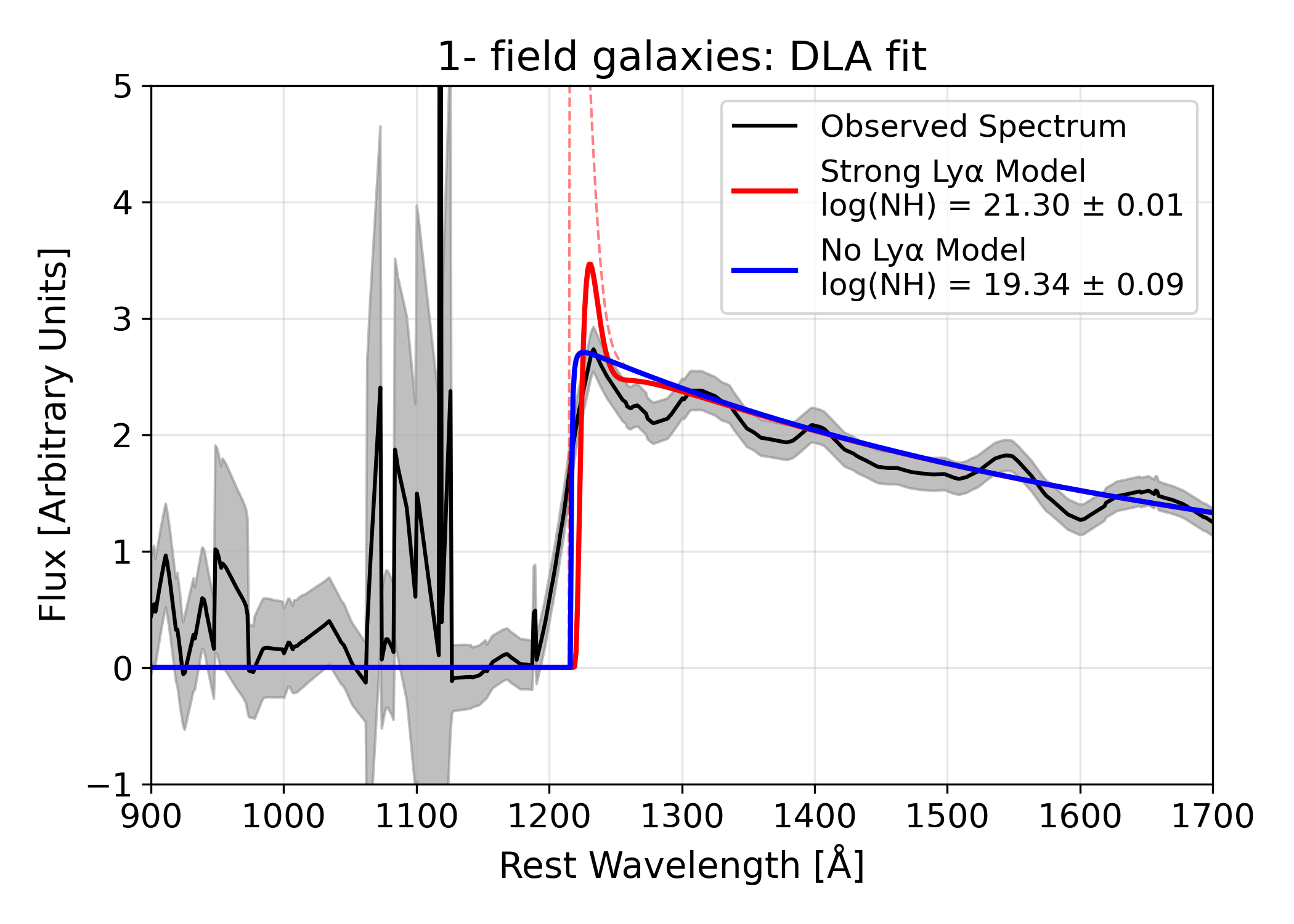}
    \includegraphics[width=0.5\textwidth]{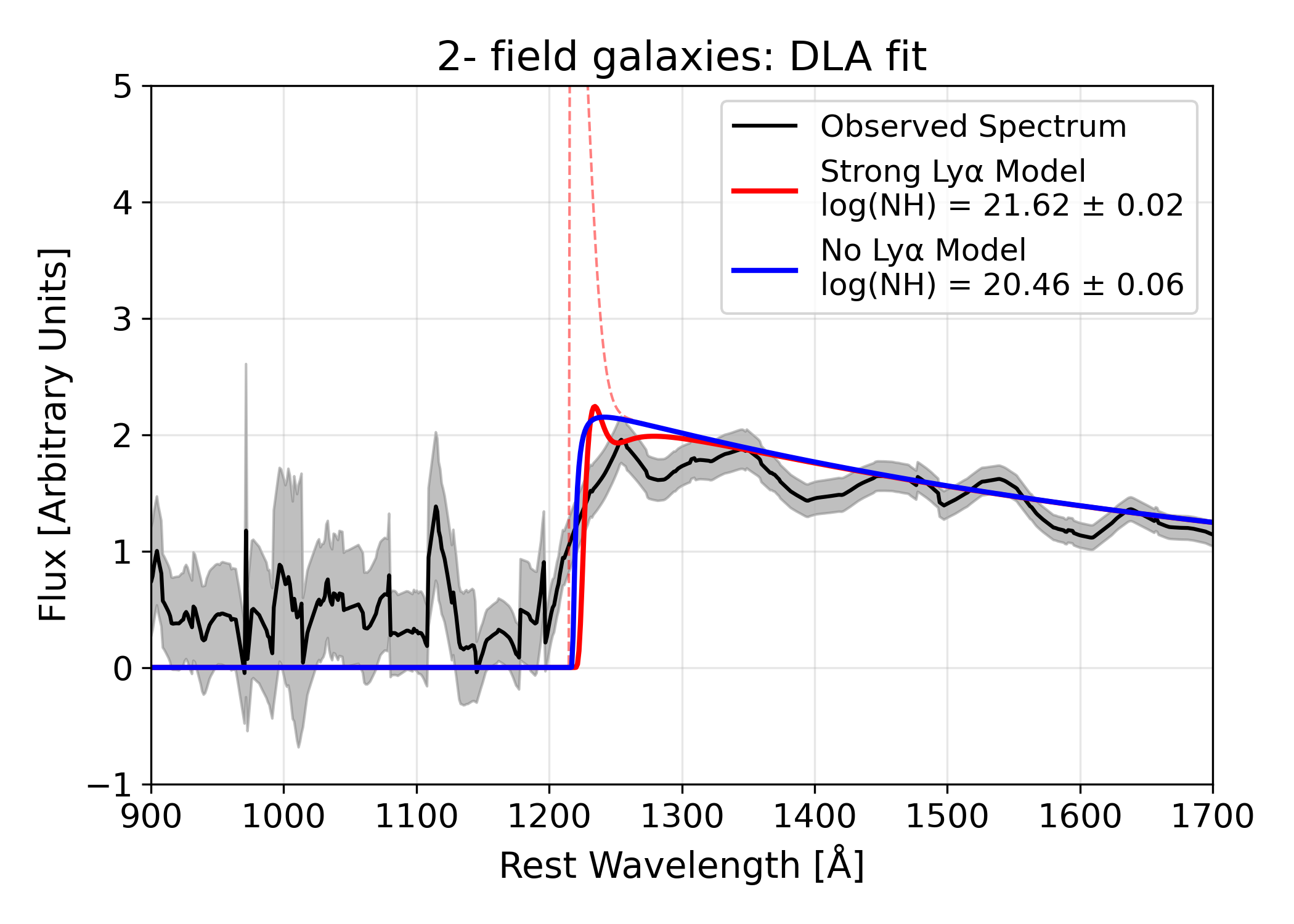}
    \caption{DLA profile fitting to the stacked rest-frame spectra of galaxies in different environments.
    The black curves show the observed spectrum, with gray bands indicating the $1\sigma$ uncertainty.
    The red dashed line indicates the assumed Ly$\alpha$ strength inferred from H$\beta$. The red solid curves show the best-fit models under this assumption, whereas the blue solid curves represent models with suppressed Ly$\alpha$ emission.
    The fitted column densities $N_{\rm HI}$ are indicated in the legend.}
    \label{fig:dla_fitting}
\end{figure}

\subsubsection{Modeling DLA Absorption to Constrain $N_{\mathrm{HI}}$ and UV Continuum Slope}

In the absence of directly measurable intrinsic Ly\(\alpha\), H\(\beta\) emission provides a robust proxy under the assumption of case B recombination. H\(\beta\), being less affected by radiative transfer effects, allows us to estimate the intrinsic Ly\(\alpha\) properties, such as luminosity and line width, which can subsequently be used to analyze scenarios where Ly\(\alpha\) emission is present or absent. Additionally, fitting the DLA feature enables an estimate of the neutral hydrogen column density (\(N_{\rm HI}\)), which is vital for understanding the neutral gas content in these systems (\citealt{Dijkstra2014, Sobral2018}).

To bracket the full range of possible Ly$\alpha$ emission scenarios, we consider two extreme cases: complete absence and maximal intrinsic production of Ly$\alpha$. 
For strongly Ly$\alpha$ scenario, we estimate the expected intrinsic Ly$\alpha$ emission in high-redshift galaxies by modeling the H$\beta$ emission line with a Gaussian profile added to a linear continuum. Under the assumption of case B recombination \citep{Osterbrock2006}, the Ly$\alpha$ luminosity is scaled from H$\beta$,
assuming typical nebular conditions ($T_e = 10^4\,\mathrm{K}, n_e = 100\,\mathrm{cm}^{-3}$). The Ly$\alpha$ line is modeled with the same velocity width as H$\beta$, centered at 1215.67~\AA\ in the rest frame. For the Ly$\alpha$-absent scenario, we set the intrinsic Ly$\alpha$ emission to 0 and directly fit a DLA absorption model. These two cases thus provide physically the upper and lower limits on the possible Ly$\alpha$ signal in high-redshift galaxies.

To assess the impact of neutral hydrogen absorption, we model the absorption feature using both the local neutral hydrogen column (DLA or CGM) and the large–scale intergalactic medium (IGM). The DLA absorption profile is computed using a Voigt profile, following the standard formalism (e.g.\ \citealt{Tepper2006,Dijkstra2014}), where the optical depth $\tau_{\rm DLA}(\lambda)$ is evaluated from the Hjerting (or equivalently Voigt) function for a given neutral hydrogen column density $N_{\rm HI}$. In addition to the local absorber, we include the damping–wing contribution from the neutral IGM using the analytic formulation developed by \citet{McQuinn2008}. $x_{\rm HI}$ is the volume–averaged neutral fraction of the IGM. This is the same physical model widely used in previous analyses of high-redshift damping wings (e.g.\ \citealt{Mortlock2011,Greig2017b,Banados2018}).

The transmission function is calculated as $\exp(-\tau)$. Assuming independent absorption contributions, the total optical depth is written as 
\[
\tau(\lambda) = \exp\left[-\tau_{\rm DLA}(\lambda) - \tau_{\rm IGM}(\lambda)\right],
\]
and the observed stacked spectrum is modeled as a power-law UV continuum multiplied by this transmission along the line of sight:
\[
f^{\rm model}_\lambda(\lambda) =
A \left( \frac{\lambda}{1500\,\mathrm{\AA}} \right)^{\beta_{\rm UV}}
\exp\!\left[ - \tau_{\rm DLA}(\lambda,N_{\rm HI}) - \tau_{\rm IGM}(\lambda,x_{\rm HI}\,) \right],
\]
where $A$ is the continuum normalization, $\beta_{\rm UV}$ is the ultraviolet continuum slope, $x_{\mathrm{HI}}$ is the volume-averaged neutral hydrogen fraction.
Four parameters are jointly fitted: $\log(N_{\mathrm{HI}})$, $x_{\mathrm{HI}}$, $\beta_{\rm UV}$, and normalization. We use the nested sampling algorithm from the \texttt{dynesty} package to explore the posterior distributions. Median values and 68\% credible intervals are derived from equal-weight posterior samples, and are shown as blue lines in the corner plot (Fig.~\ref{fig:dla_corner}). The reported values reflect the true posterior structure rather than the single maximum likelihood point.

For the cluster galaxies, the best-fit DLA model is shown in Figure~\ref{fig:dla_corner}, along with the posterior distributions of the derived quantities: 
\(\log(N_{\mathrm{HI}}/\mathrm{cm}^{-2}) = 21.55^{+0.52}_{-1.34}\), 
\(x_{\mathrm{HI}} = 0.52^{+0.32}_{-0.32}\), and 
\(\beta_{\mathrm{UV}} = -1.69^{+0.08}_{-0.18}\). We also tested a model assuming a fully neutral IGM (\(x_{\mathrm{HI}} = 1\)) and found that the best-fit parameters remained largely consistent. The resulting model is shown in the Appendix~\ref{appendix:dla}. Given the minimal differences, we adopt the fit with \(x_{\mathrm{HI}}\) as a free parameter throughout the main analysis. For the two control samples, we applied the same DLA fitting procedure using \texttt{dynesty}'s nested sampling algorithm. The inferred best-fit parameters for all three samples are summarized in Table~\ref{tab:dla}.

\begin{table}
\centering
\caption{Summary of DLA fitting results for different galaxy stacked spectral samples. The table lists the redshift range, number of galaxies ($N$), and median values with 68\% credible intervals for $\log(N_{\mathrm{HI}}/\mathrm{cm}^{-2})$, neutral hydrogen fraction $x_{\mathrm{HI}}$, and UV continuum slope $\beta_{\mathrm{UV}}$.}
\label{tab:dla}
\begin{tabular}{|l|c|c|c|c|c|}
\hline
Sample & $z$ & $N$ & $\log(N_{\mathrm{HI}})$ & $x_{\mathrm{HI}}$ & $\beta_{\mathrm{UV}}$ \\
\hline
$z\sim7$ cluster   & 7.24--7.28   & 10 & $21.55^{+0.52}_{-1.34}$ & $0.52^{+0.32}_{-0.32}$ & $-1.69^{+0.08}_{-0.18}$ \\
1-field   & 7.24--13.35  & 35 & $19.34^{+0.90}_{-0.28}$ & $0.08^{+0.03}_{-0.07}$ & $-2.18^{+0.17}_{-0.07}$ \\
2-field   & 7.00--7.48   & 27 & $20.46^{+0.64}_{-0.32}$ & $0.19^{+0.49}_{-0.16}$ & $-1.63^{+0.04}_{-0.18}$ \\
\hline
\end{tabular}
\end{table}

\subsubsection{High $N_{\mathrm{HI}}$ in Protocluster Galaxies}

As shown in Table~\ref{tab:dla}, the neutral hydrogen column densities and UV slopes vary significantly across environments.
The measured log\(N_{\mathrm{HI}}\) values for proto-cluster galaxies are found to be \({21.55}^{+0.52}_{-1.34} \, \mathrm{cm^{-2}}\) in the absence of Ly\(\alpha\) and \({22.04}^{+0.22}_{-0.03} \, \mathrm{cm^{-2}}\) when Ly\(\alpha\) emission is assumed to be strong, with Ly$\alpha$ escape fraction $f_{\rm esc}^{\mathrm{Ly}\alpha}=1$. These values are systematically higher than those of field galaxies in both control samples. In Control Sample 1, the log\(N_{\mathrm{HI}}\) values were \({19.34}^{+0.90}_{-0.28} \, \mathrm{cm^{-2}}\) and \({21.30}^{+0.18}_{-0.01} \, \mathrm{cm^{-2}}\) for the no Ly\(\alpha\) and strong Ly\(\alpha\) cases, respectively. Similarly, in Control Sample 2, the log\(N_{\mathrm{HI}}\) values were \({20.46}^{+0.64}_{-0.32} \, \mathrm{cm^{-2}}\) and \({21.62}^{+0.24}_{-0.02} \, \mathrm{cm^{-2}}\) for the no Ly\(\alpha\) and strong Ly\(\alpha\) cases, respectively.

The results indicate that proto-cluster environments at \(z > 7\) are characterized by significantly higher neutral hydrogen column densities compared to field environments. The higher \(N_{\mathrm{H}}\) values in proto-cluster galaxies suggest that dense environments promote the retention of neutral hydrogen reservoirs. 
The higher column densities in proto-clusters are consistent with the expectation that dense environments support higher gas accretion rates and reduced ionization due to shielding effects. These findings are consistent with theoretical models that predict enhanced neutral hydrogen retention in high-redshift proto-clusters due to gravitational potential wells and reduced photoionization from external sources (\citealt{Dijkstra2014, Kakiichi2018}).

Our results align with recent studies that investigate the role of environment in shaping the properties of high-redshift galaxies. For instance, \cite{Sobral2018} found that galaxies in dense environments exhibit enhanced gas accretion and lower ionization fractions, leading to higher \(N_{\mathrm{H}}\). The $N_{\mathrm{H}}$ we derive falls within the range of $10^{20.5} - 10^{22} \, \mathrm{cm^{-2}}$, consistent with the measurements of DLAs in overdense regions reported by \citet{Reddy2023}.

The results also indicate that the presence or absence of Ly$\alpha$ emission strongly affects the inferred $N_{\rm HI}$, reflecting the complex interaction between neutral gas and Ly$\alpha$ radiative transfer in high-redshift galaxies \citep{Dijkstra2014,Sobral2018,Reddy2023,Fudamoto2022,Finkelstein2019}. In both protocluster and field environments, we find that galaxies with weaker Ly$\alpha$ emission tend to show higher neutral hydrogen column densities. This trend is consistent with Ly$\alpha$ radiative transfer models, where resonant scattering makes Ly$\alpha$ photons more difficult to escape in gas-rich systems, leading to weaker observed emission \citep[e.g.,][]{Verhamme2012,Choustikov2024}.

However, the systematically higher \(N_{\mathrm{HI}}\) values in protocluster galaxies suggest that environmental factors such as large-scale overdensities play a key role in regulating Ly\(\alpha\) escape. These dense regions may promote increased gas accretion and reduced ionization due to shielding effects, resulting in more efficient Ly\(\alpha\) trapping \citep[e.g.,][]{Ouchi2020}. Conversely, in systems where Ly\(\alpha\) is absent, we infer lower column densities, which may reflect more ionized or turbulent environments where Ly\(\alpha\) photons are either absorbed or scattered out of the line of sight. Further studies incorporating a larger sample size and cosmological simulations will provide deeper insights into the nature of DLAs and their relationship with Ly\(\alpha\)-emitting galaxies.

In addition to the neutral-hydrogen constraints inferred from the DLA modelling, we also estimate the Ly$\alpha$ equivalent-width upper limit directly from the stacked protocluster spectrum. Using the stacked $1\sigma$ error spectrum around 1216\,\AA\ and integrating the noise over a fixed 20\,\AA\ rest-frame window, we derive a $3\sigma$ Ly$\alpha$ flux upper limit. Dividing this limit by the continuum level inferred from our power-law fit yields a rest-frame equivalent-width upper limit of ${\rm EW}_{0}({\rm Ly}\alpha) < 9.6$\,\AA. This stacked constraint is substantially lower than the Ly$\alpha$ equivalent widths typically observed for galaxies of similar UV luminosity near the end of reionization, which frequently exceed $\sim20-40$\,\AA\ \citep[e.g.][]{Stark2010,Jung2018}. The strongly suppressed Ly$\alpha$ visibility, together with the elevated neutral-hydrogen column densities inferred for protocluster members, suggests that this overdense region retains significant reservoirs of neutral gas and is not yet permeated by a large ionized bubble.

\section{Discussion}\label{sec:discussion}

\subsection{JADES-GS-z7-LA and and Its Connection to these Protoclusters}

\subsubsection{A Prominent LAE and the Nearby Protocluster Environment}
A growing number of recent \textit{JWST} studies have highlighted the close connection between Ly$\alpha$ transmission and large-scale environment in the reionisation era. These works have identified both strong Ly$\alpha$ emitters and galaxy overdensities at $z\sim7.1$--$7.3$, and in several cases argued for the presence of a large ionized bubble extending across this region \citep{Saxena2023, Helton2023, Witstok2024, Meyer2024, Tang2023}. In this section, we examine the spatial and physical relationship between the \(z \sim 7\) protocluster and the previously identified strong Lyman-\(\alpha\) emitter JADES-GS-z7-LA. 

JADES-GS-z7-LA, identified at \(z = 7.2782\), is a remarkable LAE discovered in the JADES field. It exhibits an extremely high rest-frame equivalent width of \(EW_0(\mathrm{Ly}\alpha) = 388.0 \pm 88.8\)~\AA\ and a faint UV magnitude of \(M_{\text{UV}} = -17.0\), placing it among the most extreme LAEs at this epoch \citep{Saxena2023}. Spectroscopic observations with JWST/NIRSpec revealed strong Ly\(\alpha\) emission accompanied by prominent [O\,\textsc{iii}] and H\(\beta\) lines, indicative of a young, metal-poor star-forming system with a high ionization parameter and efficient ionizing photon escape. The Ly\(\alpha\) velocity offset from the systemic redshift is only \(113.3 \pm 80.0\) km s\(^{-1}\), implying the absence of strong galactic outflows. The Ly\(\alpha\) escape fraction exceeds \(70\%\), suggesting that the galaxy resides in a highly ionized region. The galaxy also shows extreme nebular line diagnostics, including a high [O\,\textsc{iii}]/[O\,\textsc{ii}] ratio \(O_{32} = 11.1 \pm 2.2\) and a strong \(R_{23} = 11.2 \pm 2.6\), consistent with a low-metallicity, highly ionized interstellar medium, potentially indicative of density-bounded nebulae\citep{Nakajima2014,Izotov2018}. Its estimated stellar mass of \(\sim 10^7\,M_\odot\) places it at the low-mass end of the galaxy population at \(z > 7\) \citep{Saxena2023}.

Meanwhile, our independent discovery of the \(z \sim 7\) protoclusters within GOODS-S presents a plausible possibility that JADES-GS-z7-LA might be dynamically or environmentally linked to this overdense region \citep{Qiong2024}. The protoclusters exhibit a significant enhancement in galaxy number density, with multiple spectroscopically confirmed members forming a connected structure.  Spectral energy distribution (SED) modeling shows that these galaxies have stellar masses ranging from $10^7$ to $10^9~M_{\odot}$, with most galaxies exhibiting blue rest-UV slopes indicative of young stellar populations.

A previously reported overdensity at similar redshift in the GOODS–S field is 
JADES--GS--OD--7.265 \citep{Helton2023}, which contains seven spectroscopically 
confirmed galaxies with an average overdensity of 
$\delta_{\rm gal} \simeq 6.6$ and a physical extent of 
$\sim 4.4$\,cMpc. We find that six galaxies in our two identified structures 
are also members of JADES--GS--OD--7.265, indicating that our overdensities trace 
the same large-scale structure previously identified in the literature. Additional evidence for a coherent overdense region comes from 
\citet{Saxena2023}, who proposed that the reduced neutral fraction inferred 
around JADES–GS–z7–LA and its companion galaxy may be due to their location 
within this overdensity at $z \approx 7.3$, which could host locally ionized 
regions.

\subsubsection{Spatial Connection Between JADES-GS-z7-LA and the Protocluster}

Given the relatively small projected distance of 40 arcseconds (\(\sim 0.2\) pMpc at $z\sim7$) between JADES-GS-z7-LA and the protocluster center, a key question is whether JADES-GS-z7-LA is a satellite member of the protocluster or an independent system within a nearby ionized bubble. Given its separation of $\sim 0.2$pMpc, it is plausible that JADES-GS-z7-LA lies within the same reionized volume, benefiting from the collective ionizing output of the protocluster. 

To address this issue, we investigates the surface density distribution of galaxies surrounding one of spectroscopically confirmed protoclusters at $z \sim 7.3$. We select galaxies from the publicly available JADES deep-field spectroscopic catalog \citep{Eisenstein2023} and our generated photometric catalog to probe the spatial clustering properties of galaxies within a projected radius of $3$ arcmin from two locations: the lunimous LAE JADES-GS-z7-LA (NIRSpec ID: 10013682) and the peak of protocluster at $z\sim7.27$(near the galaxy of NIRSpec ID: 9425).

To quantify the surface density as a function of distance from the central galaxies, we select galaxies within the redshift range $7.15 \leq z \leq 7.45$ from both spectroscopic and photometric catalogs, 
{down to an absolute UV magnitude limit of $M_{\mathrm{UV}} \approx -21$ mag. The same magnitude cut is applied to the photometric sample to ensure a consistent comparison.}
The central galaxies are defined as physically motivated, spectroscopically confirmed members of the overdensities, including the Ly$\alpha$-emitting galaxy (JADES-GS-z7-LA) and a spectroscopically confirmed protocluster member located near the apparent core of the structure. For the photometric sample, the same sky positions are adopted as centers to ensure a consistent comparison between spectroscopic and photometric surface density profiles.
We compute the angular separation between each selected galaxy and the central galaxies, and calculate the surface density $\Sigma$ in the increasing separation bins (up to $3$ arcmin, $1$ arcmin $\approx 0.484$ pMpc at $z \sim 7.3$). Error bars are Poissonian uncertainties. 

\begin{figure}
    \centering
    \includegraphics[width=0.5\textwidth]{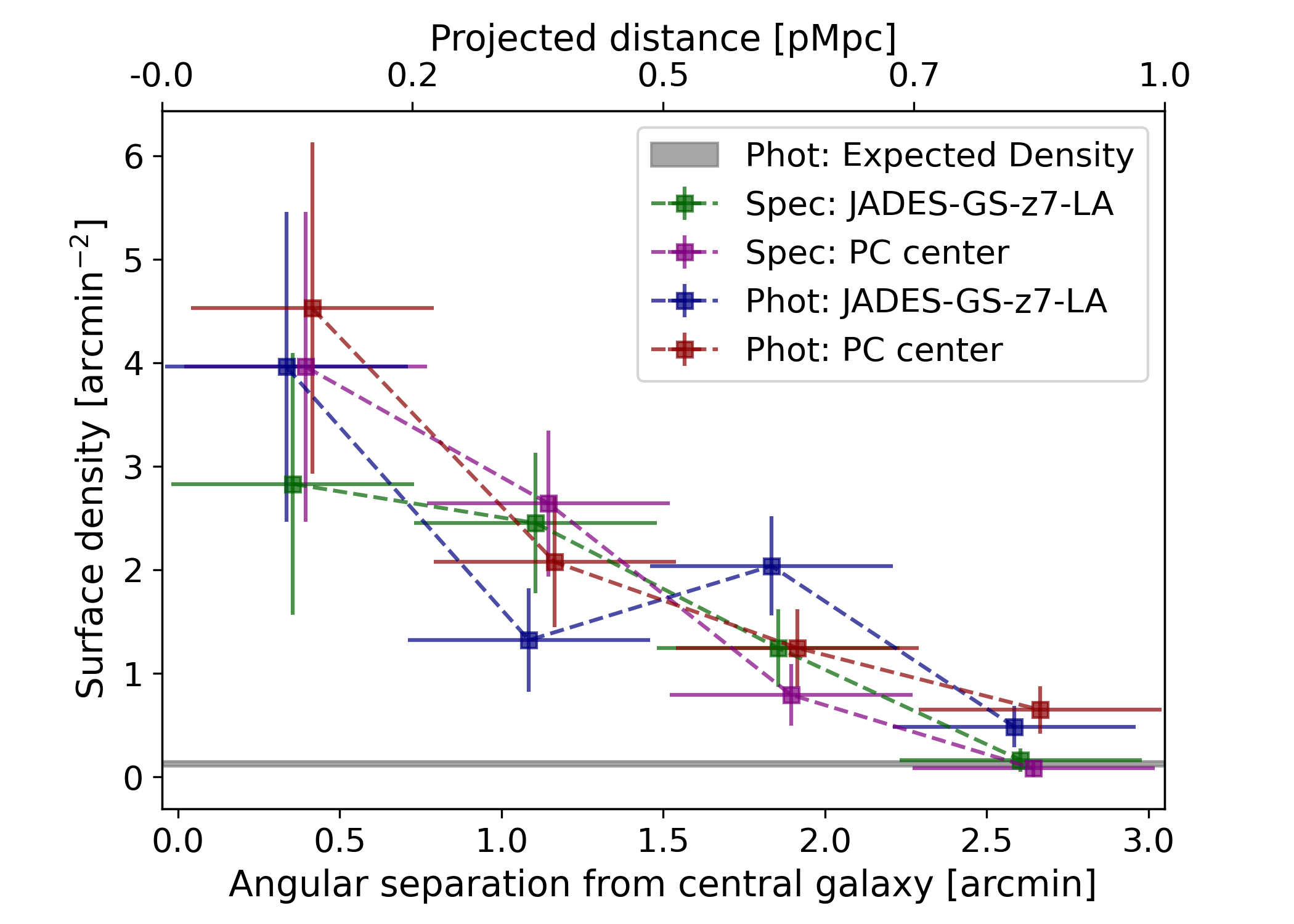}
    \caption{Surface density of galaxies as a function of angular separation from the central protocluster members. The shaded region represents the expected background galaxy density at $z\sim7$ estimated using the UV luminosity function from \citet{Adams2024}.}
    \label{fig:angular_separation}
\end{figure}

Figure~\ref{fig:angular_separation} shows the measured surface density profiles centered on the four selected galaxies. Both spectroscopic and photometric samples exhibit enhanced galaxy surface densities within $1$ arcmin, indicating a significant overdensity in the protocluster environment. 
As expected, the surface density derived from the photometric sample is systematically higher due to the inclusion of galaxies without spectroscopic confirmation, while the spectroscopic sample provides a lower bound owing to redshift incompleteness. Importantly, both samples show consistent radial trends, demonstrating that the detection of the overdensity is robust.

To assess the significance of the observed overdensity, we estimate the expected field galaxy surface density at $z \sim 7$ using the UV luminosity function (LF) from \citet{Adams2024}. 
We integrate the LF over the range $-22 \leq M_{\text{UV}} \leq -17$. The expected surface density is $\Sigma_{\text{expected}}(z \sim 7) \approx 0.1 - 0.15 \text{ arcmin}^{-2}$. This expected field surface density is shown as the gray shaded region in Figure~\ref{fig:angular_separation}. The observed overdensity exceeds this expected field density, confirming a significant enhancement of galaxy clustering in the protocluster environment. 
This suggests that JADES-GS-z7-LA resides within a highly overdense structure, consistent with expectations for early massive protoclusters.
It is possible that JADES-GS-z7-LA lies, at least in part, within the same ionized region as the protocluster, providing observational support for the idea that overdense regions may help start large-scale reionization earlier than the average field. The presence of strong LAEs within the protocluster suggests that ionized bubbles may extend over scales of up to $\sim 1$~pMpc in some directions, as predicted by models \citep{Mason2018, Jung2022}. However, the fact that many galaxies in the same structure do not show Ly$\alpha$ emission implies that these ionized regions are likely patchy or anisotropic, and not all galaxies within the overdensity are in fully ionized environments.

\subsubsection{Estimating the Ionized Bubble Size in the Protocluster Region}

To evaluate whether the protocluster galaxies can collectively or individually produce sufficiently large ionized regions to allow Ly$\alpha$ escape, we estimate the growth of ionized bubbles using the analytic formalism introduced by \citet{Haiman1997}. The time evolution of the H\,\textsc{ii} region radius around a galaxy is governed by

\begin{equation}
\frac{dR_{\mathrm{HII}}}{dt} = \frac{\langle \xi_{\mathrm{ion}} \rangle \langle f_{\mathrm{esc}} \rangle L_{\mathrm{UV}}}{4\pi R_{\mathrm{HII}}^2 \, \bar{n}_{\mathrm{HI}}(z)} + R_{\mathrm{HII}} H(z) - R_{\mathrm{HII}} \, \alpha_B \, \bar{n}_{\mathrm{HI}}(z) \frac{C_{\mathrm{HI}}}{3},
\end{equation}

where the first term describes ionization due to escaping UV photons, the second accounts for cosmological expansion, and the third represents recombinations in the clumpy intergalactic medium (IGM). Here, $\alpha_B = 2.59 \times 10^{-13}~\mathrm{cm}^3\,\mathrm{s}^{-1}$ is the case B recombination coefficient \citep{Osterbrock2006}, $C_{\mathrm{HI}} = 3$ is the clumping factor \citep{Robertson2013,Endsley2022}, and $\bar{n}_{\mathrm{HI}}(z)$ is the mean proper hydrogen number density. For each galaxy, the escape fraction is not treated as a fixed parameter. Instead, we estimate $f_{\mathrm{esc}}$ individually using an empirical relation between the UV continuum slope $\beta$ and $f_{\mathrm{esc}}$, motivated by studies of high-redshift, low-metallicity star-forming galaxies.

Using this equation, we compute $R_{\mathrm{HII}}$ for each of the ten spectroscopically confirmed protocluster members at $z \simeq 7.26$--7.30, taking into account their individual $M_{\mathrm{UV}}$, $\xi_{\mathrm{ion}}$, $\beta$, and $f_{\mathrm{esc}}$. The resulting H\,\textsc{ii} region sizes span a range of $\sim$0.09–0.61 comoving Mpc, with most galaxies producing bubbles smaller than $\sim$0.3 Mpc.

We then compute the physical separations between all galaxy pairs and assess whether their ionized regions overlap. We identify three pairs with potential bubble overlap: galaxies \#9425 and \#30141745 (separation = 0.50 Mpc, $R_1$ = 0.18 Mpc, $R_2$ = 0.36 Mpc), \#43252 and \#20046019 (separation = 0.49 Mpc, $R_1$ = 0.13 Mpc, $R_2$ = 0.61 Mpc), and \#20046019 and \#20046866 (separation = 0.34 Mpc, $R_1$ = 0.61 Mpc, $R_2$ = 0.16 Mpc). These results suggest that a subset of close galaxy pairs may reside in partially overlapping ionized regions, indicating ionized connectivity within the protocluster environment.

To estimate the maximum extent of a combined bubble assuming all ionizing output contributes collectively, we sum their total $\dot{N}_{\mathrm{ion}}$ and apply the similar formula:

\begin{equation}
R_{\mathrm{HII,\,total}} = \left( \frac{3 \sum_i \dot{N}_{\mathrm{ion},\,i}}{4\pi \, \alpha_B \, C_{\mathrm{HI}} \, \bar{n}_{\mathrm{H\,I}}^2(z)} \right)^{1/3},
\end{equation}

yielding a combined bubble radius of $R_{\mathrm{HII,\,total}} \simeq 0.69$ Mpc. However, this bubble is still not large enough to encompass the \lya-emitting galaxy (ID 10013682), located $\sim$1.38 Mpc from the nearest protocluster member. We additionally examine whether this LAE lies within the H\,\textsc{ii} region of any individual galaxy by comparing the pairwise distances to each $R_{\mathrm{HII},\,i}$ and find that it lies beyond all of them. 
These findings imply that either the LAE is producing its own ionized bubble (e.g., due to higher intrinsic $\dot{N}_{\mathrm{ion}}$ or $f_{\mathrm{esc}}$), or that the local IGM is pre-ionized by unseen nearby sources. Similar analyses at $z \sim 6.6$--6.9 by \citet{Endsley2022} suggest that LAE visibility depends not only on individual galaxy luminosity but also on the presence of nearby ionizing companions to form large enough bubbles ($R_{\mathrm{HII}} \gtrsim 0.8$ Mpc).

Future deep and more JWST/NIRSpec observations targeting additional members of the protocluster and their Ly$\alpha$ profiles will help confirm whether these galaxies contribute collectively to a shared ionized region. ALMA constraints on dust and [CII] emission will further elucidate the interplay between ionized and neutral gas in these environments. We note that these estimates include only spectroscopically confirmed protocluster members. A quantitative contribution from fainter galaxies is not explicitly included, as the faint-end population and its ionizing properties within the overdensity are currently unconstrained; the inferred bubble sizes should therefore be regarded as conservative lower limits. Fainter galaxies below current spectroscopic detection limits, which are expected to be more numerous in overdense regions, may contribute additional ionizing photons and further enlarge the effective ionized region, as suggested by luminosity-function-based studies at similar redshifts.

\subsection{Comparison with other high-\texorpdfstring{$z$}{z} LAE-rich overdensities}\label{sec:compare}

\begin{figure*}
    \centering
    \begin{subfigure}[t]{0.48\textwidth}
        \centering
        \includegraphics[width=\textwidth]{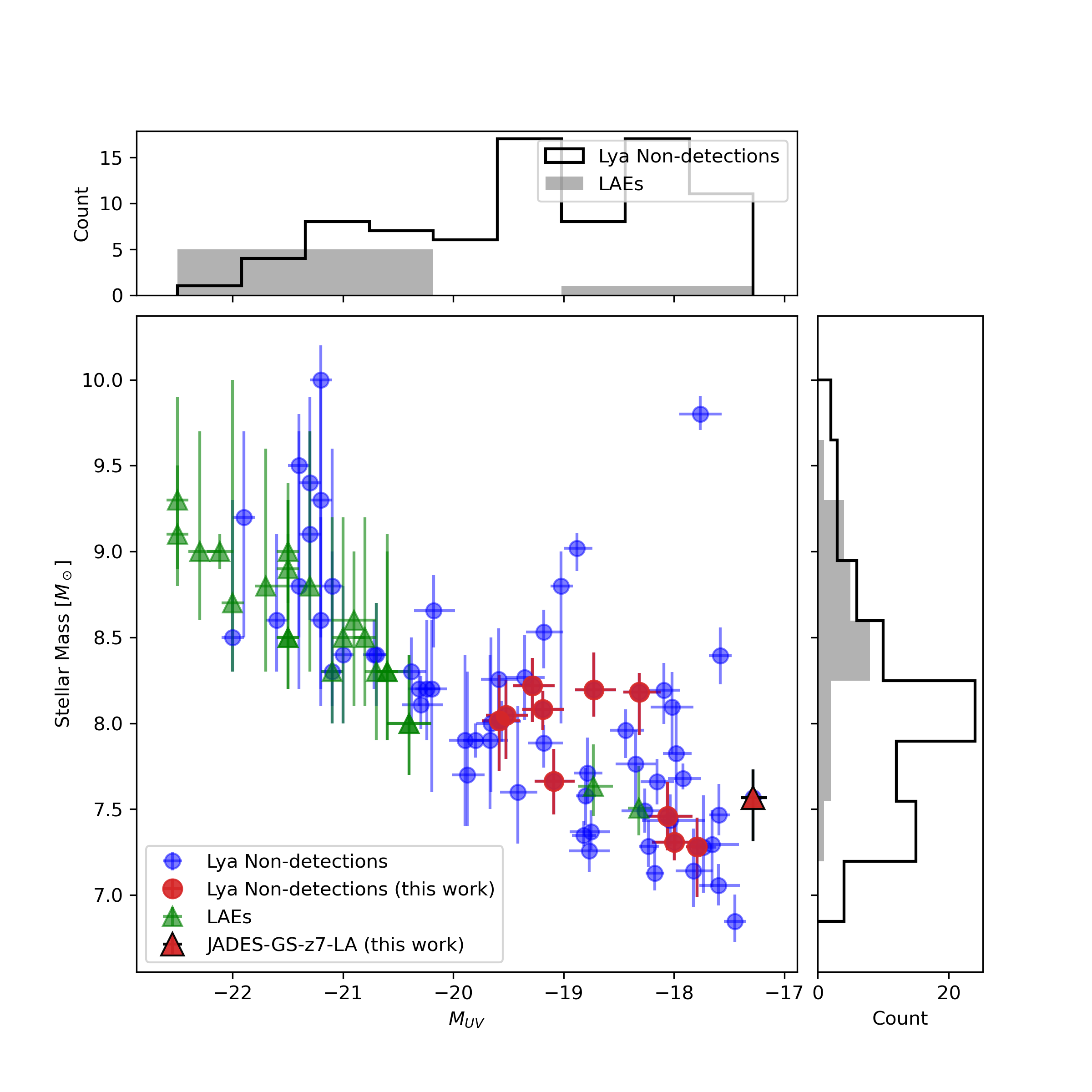}
        \caption{Stellar mass vs. $M_{\rm UV}$.}
    \end{subfigure}
    \begin{subfigure}[t]{0.48\textwidth}
        \centering
        \includegraphics[width=\textwidth]{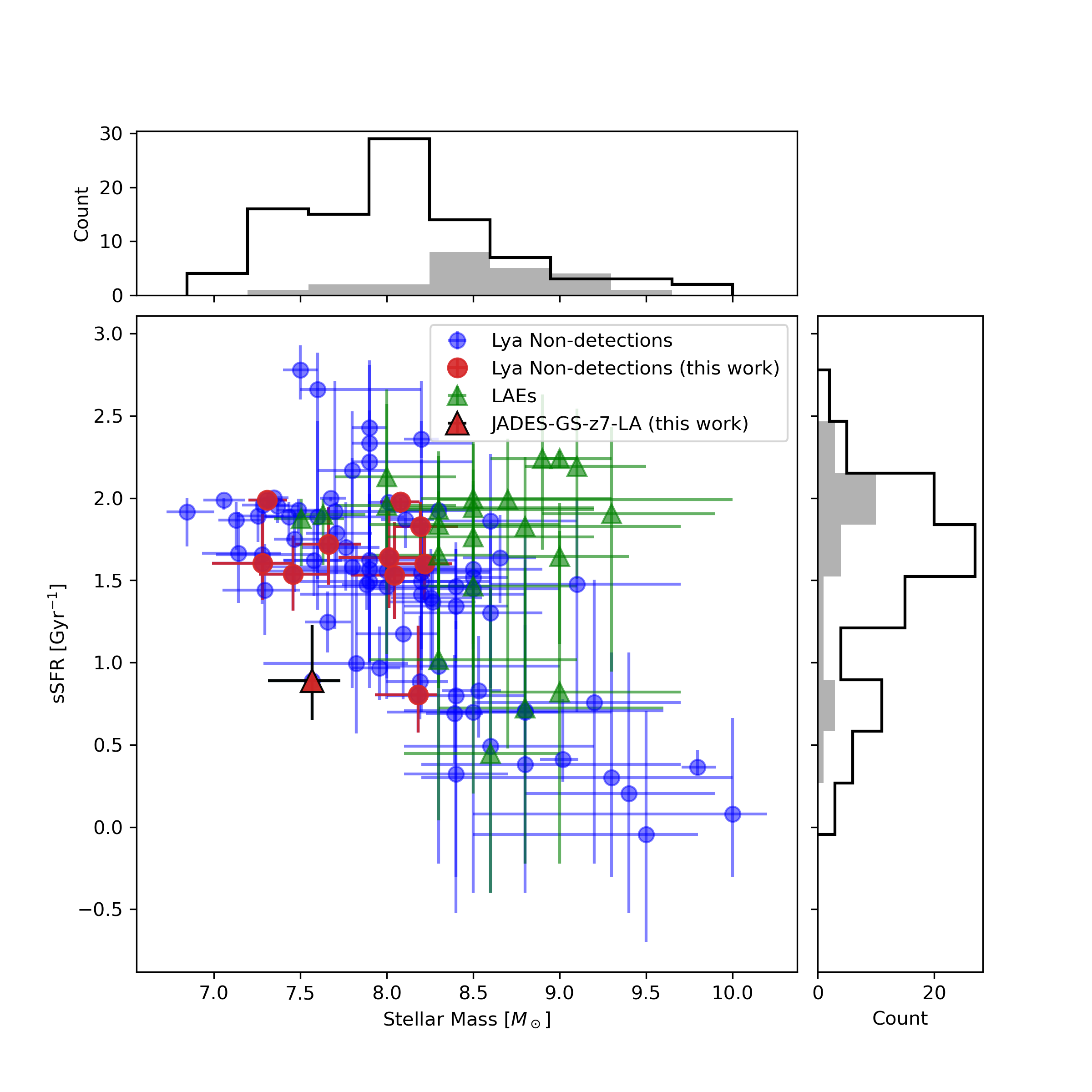}
        \caption{sSFR vs. stellar mass.}
    \end{subfigure}
    \begin{subfigure}[t]{0.48\textwidth}
        \centering
        \includegraphics[width=\textwidth]{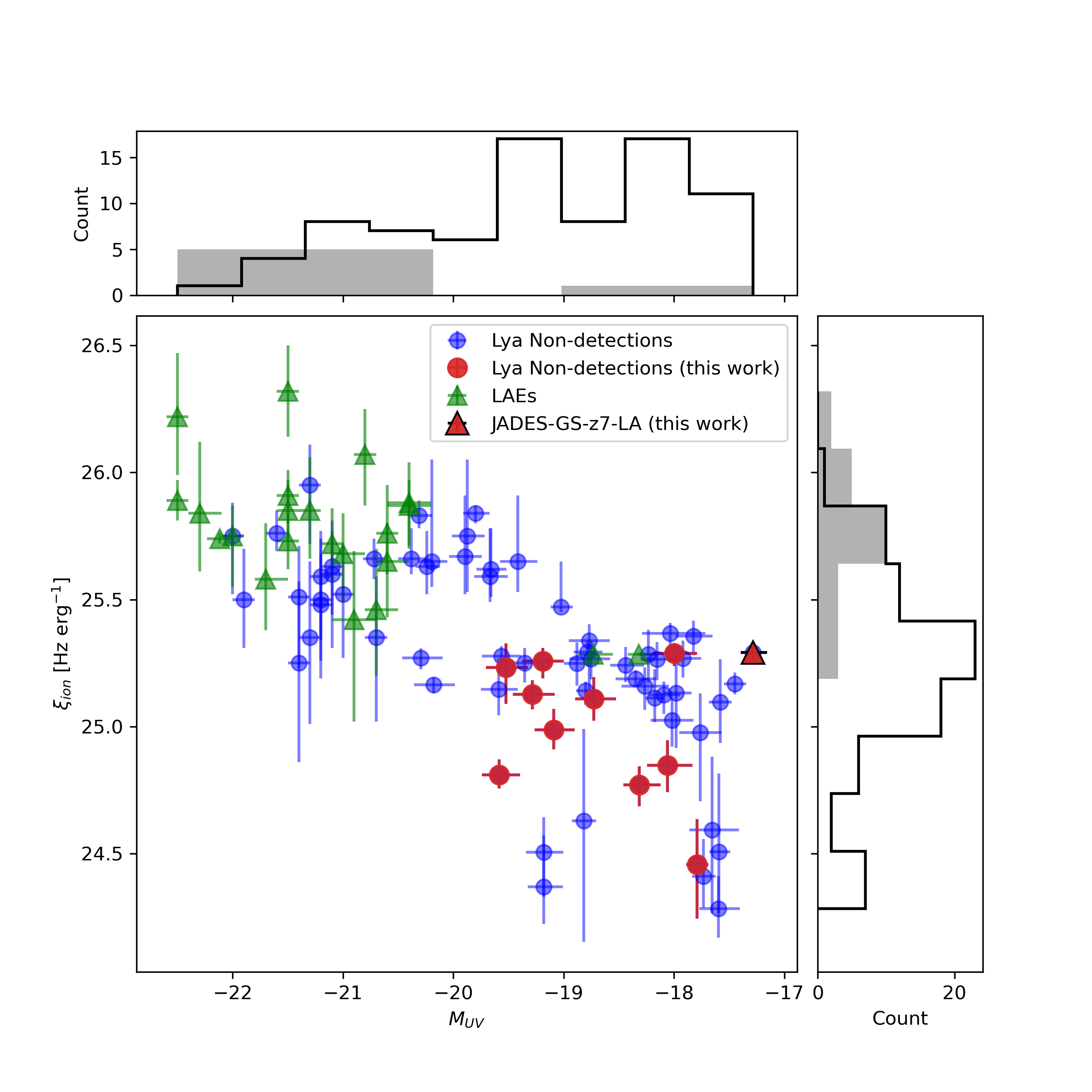}
        \caption{$\xi_{\rm ion}$ vs. $M_{\rm UV}$.}
    \end{subfigure}
    \begin{subfigure}[t]{0.48\textwidth}
        \centering
        \includegraphics[width=\textwidth]{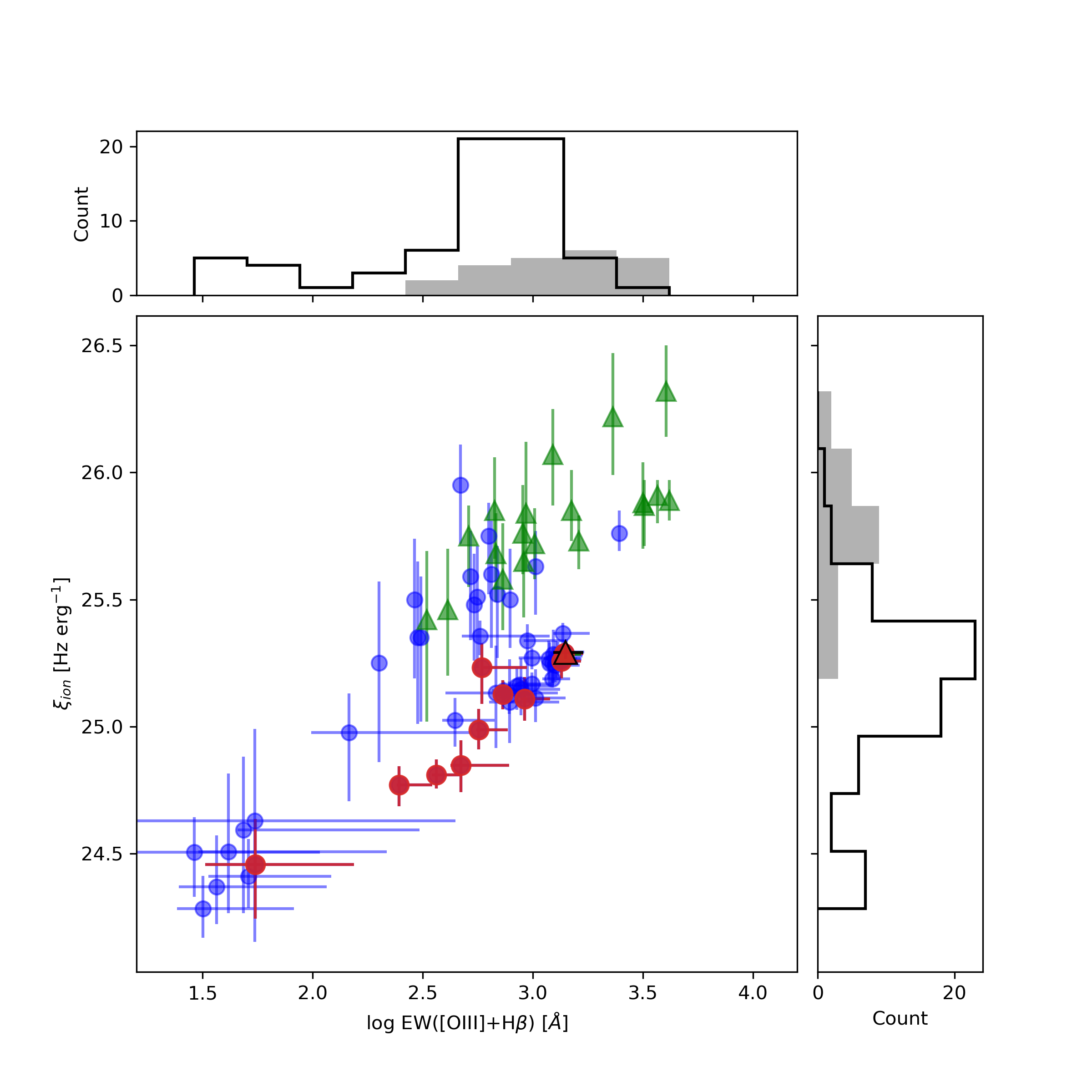}
        \caption{$\xi_{\rm ion}$ vs. EW([OIII]+H$\beta$).}
    \end{subfigure}
    \caption{Comparison of Lyman-alpha emitters (LAEs, red points) and non-detections (blue points) at $z\sim7$. LAEs exhibit higher masses, higher sSFRs, and enhanced ionizing efficiencies compared to non-detections.}
    \label{fig:lya_detections}
\end{figure*}

The contribution of faint galaxies to cosmic reionization remains a topic of active investigation and debate \citep{Bouwens2015, Stark2017}. Studies have shown that bright LAEs tend to be found in over-dense environments at $z>7$, suggesting that high-redshift LAE-rich systems could be key drivers of ionized bubbles in the early universe \citep{Endsley2022, Whitler2024}. However, we have identified a protocluster at $z\sim7.2$, primarily composed of faint galaxies, that lacks significant Ly$\alpha$ emission. To understand the nature of these galaxies, we compare their physical properties with those of LAE-rich systems from the literature. Specifically, we investigate key galaxy properties such as stellar mass, SFR, and UV magnitude ($M_{\text{UV}}$), to assess whether fundamental differences in these parameters could explain the absence of Ly$\alpha$ emission.

We utilize spectroscopically confirmed galaxy samples from multiple studies to compare our protocluster with LAE-rich environments at high redshift. The reference samples include LAE-rich overdensities at $z\sim6.8$   \citep{Endsley2022}, $z\sim7$ \citep{Endsley2021}, $z\sim9$ \citep{Whitler2024}, $z\sim7$ \citep{Hu2021,Martin2026} and $z\sim7-8$ \citep{Witstok2024}, drawn from existing literature. Throughout this work, we classify galaxies as LAEs if they exhibit a rest-frame Ly$\alpha$ equivalent width of ${\rm EW}_0 > 15,\text{\AA}$, consistent with previous studies. For all galaxies in the comparison samples, the stellar mass, SFR, UV magnitude, ionizing photon production efficiency $\xi_{\rm ion}$, UV slope are taken directly from published catalogs.

In Figure~\ref{fig:lya_detections} we present the comparison between Lyman-alpha emitters and non-detections across key physical properties, to investigate the physical conditions that influence cosmic reionization in the early universe. Panel (a) shows the relationship between absolute UV magnitude and stellar mass. At fixed $M_{\rm UV}$, the LAEs tend to have higher stellar masses, while the non-detections exhibit a broader and more scattered stellar mass distribution. The mass distribution of non-detections spans $7.2 \lesssim \log(M_*/M_{\odot}) \lesssim 10$, with LAEs preferentially clustering around $\log(M_*/M_{\odot}) \sim 8.0-9.2$. This suggests that, at a given UV luminosity, galaxies with higher stellar masses are more likely to exhibit strong Lyman-alpha emission. This trend is consistent with studies that find that LAEs at high redshift often reside in more massive galaxies, potentially due to their higher SFRs and more intense ionizing radiation fields \citep{Stark2017,Endsley2022}. Our cluster galaxies (red points) lie on the faint end of the distribution (with \(M_{\mathrm{UV}} > \text{-19.5}\)) and also exhibit relatively low stellar masses (\(\log(M_*/M_{\odot}) \sim 7.2\text{--}8.5\)).
Figure~\ref{fig:lya_detections} panel (b) explores the sSFR as a function of stellar mass. LAEs exhibit systematically higher sSFRs compared to non-detections, suggesting that Lyman-alpha visibility is linked to bursty or highly efficient star formation. This is particularly evident at high mass end $\log(M_*/M_{\odot}) \sim 9$, where LAEs tend to have sSFR values exceeding $\sim100~{\rm Gyr}^{-1}$. Although LAEs in our sample tend to have slightly higher stellar masses at fixed UV luminosity, this does not imply lower sSFRs. LAEs are systematically offset toward higher star-formation rates at fixed stellar mass compared to Ly$\alpha$ non-detections, indicating that the enhanced instantaneous star formation dominates over the modest increase in stellar mass. The high sSFRs observed in LAEs are consistent with earlier studies showing that strong Ly$\alpha$ emitters often go through recent bursts of star formation, which can boost the production of ionizing photons needed for their strong nebular lines \citep{Mason2018,Jung2020}. The scatter in sSFR, especially at lower stellar masses, likely reflects differences in star formation histories or stages of galaxy evolution at high redshift.

Figure~\ref{fig:lya_detections} panel (c) presents the relationship between $M_{\rm UV}$ and the ionizing photon production efficiency, $\xi_{\rm ion}$. LAEs are found to exhibit systematically higher $\xi_{\rm ion}$ values compared to non-detections, supporting the idea that these galaxies are efficient producers of ionizing photons. The mean $\xi_{\rm ion}$ for LAEs is approximately $\log \xi_{\rm ion} \sim 25.6$, higher than the typical values inferred for Lyman-alpha non-detections. This is consistent with previous findings suggesting that LAEs have harder ionizing spectra, possibly driven by low metallicities, young stellar populations, or binary stellar evolution effects \citep{Tang2023}. The trend that fainter UV galaxies exhibit lower \(\xi_{\rm ion}\) suggests that the ionizing efficiency of low-luminosity systems at \(z \sim 7\) may be limited. Their individual contributions to reionization are modest unless they exhibit unusually high escape fractions or bursty star formation histories.
Figure~\ref{fig:lya_detections} panel (d) shows the relation between the rest-frame equivalent width (EW) of the [OIII]+H$\beta$ emission and $\xi_{\rm ion}$. We find a clear positive trend, that LAEs mostly fall in the high EW and high $\xi_{\rm ion}$ region. By contrast, galaxies without Ly$\alpha$ detections tend to have lower EW and $\xi_{\rm ion}$ values. This suggests that strong nebular emission is linked to more efficient production of ionizing photons. LAEs typically have $\log {\rm EW} \gtrsim 2.5$, which may indicate strong optical emitters have physical conditions that help Ly$\alpha$ escape, such as low dust or high ionization \citep{Endsley2022,Tang2023}. The observed trend is also consistent with photoionization models, where galaxies with extreme emission lines are expected to have higher ionizing photon escape fractions, potentially making them key contributors to the cosmic reionization process.

It is worth noting that the lower Ly$\alpha$ detection rate among our cluster galaxies may be partly due to their intrinsically faint UV luminosities and low stellar masses. However, the differences observed at fixed $M_{\rm UV}$ suggest that other physical conditions, such as ionizing photon production, also play an important role.

\subsection{Impact of Environment on Lyman-alpha Emission at \texorpdfstring{$z>7$}{z>7}}

The presence or absence of Ly$\alpha$ emission in galaxies at $z \sim 7$ is closely tied to the ionization state of the intergalactic medium. At this epoch, cosmic reionization is still incomplete, leading to spatial variations in the IGM's neutral fraction. Consequently, while some galaxies reside in locally ionized regions that facilitate Ly$\alpha$ transmission, others are embedded in more neutral environments that suppress Ly$\alpha$ photons due to resonant scattering and absorption by neutral hydrogen \citep{Stark2017, Mason2018, Jung2020}. The spatial inhomogeneity of reionization thus plays a critical role in determining whether Ly$\alpha$ emission is observable from galaxies at this redshift.

One possible explanation for the presence of Ly$\alpha$ emission in certain galaxies is the formation of ionized bubbles driven by clusters of star-forming galaxies. These bubbles enable Ly$\alpha$ photons to escape the immediate vicinity of their host galaxies before encountering the neutral IGM, thereby increasing their transmission probability \citep{Mason2018, Endsley2022}. The size and growth rate of these ionized regions depend on several factors, including the number density of ionizing sources, their ionizing photon production efficiencies, and the local IGM density. Galaxies situated within or near such ionized bubbles have a higher likelihood of exhibiting detectable Ly$\alpha$ emission compared to those in more neutral regions. Moreover, galaxies with higher SFRs and lower dust content further enhance Ly$\alpha$ escape due to their higher production and escape fractions of ionizing photons \citep{Tang2023}.

Despite these general trends, the absence of detectable Ly$\alpha$ emission within the discovered $z \sim 7$ protocluster presents a notable exception. Protoclusters, characterized by their high galaxy densities, are expected to contribute significantly to reionization due to the collective ionizing output of their member galaxies \citep{Castellano2018,Endsley2022}. However, several factors may suppress Ly$\alpha$ visibility within such dense environments. Firstly, the higher gas densities within protoclusters can lead to increased resonant scattering of Ly$\alpha$ photons, reducing their escape fraction even if the surrounding IGM is partially ionized. Additionally, enhanced metal enrichment and dust production in these regions can further attenuate Ly$\alpha$ emission. Moreover, gravitational interactions and mergers, which are more common in overdense environments, can increase turbulence and gas kinematics, broadening the Ly$\alpha$ line profile and increasing the likelihood of photon scattering by residual neutral hydrogen in the IGM \citep{Overzier2016}.

Another plausible explanation is that the reionization process within protoclusters may lag behind that in lower-density regions. While the higher galaxy density enhances local ionizing photon production, the denser IGM in protoclusters requires a greater number of ionizing photons to achieve the same level of ionization as in more diffuse regions \citep{Mason2018}. Consequently, even though protoclusters host numerous star-forming galaxies, the IGM within these regions may remain partially neutral longer, suppressing Ly$\alpha$ transmission. Furthermore, the complex gravitational potential wells of protoclusters can lead to deeper potential barriers that trap ionizing photons, slowing the growth of ionized bubbles compared to less dense environments.

In summary, the presence or absence of Ly$\alpha$ emission at $z \sim 7$ is governed by a complex interplay of local and large-scale environmental factors. While galaxies residing within or near ionized bubbles have a higher probability of exhibiting Ly$\alpha$ emission, those embedded in more neutral regions or denser environments, such as protoclusters, may experience reduced Ly$\alpha$ transmission despite their high SFRs and ionizing photon production. \citep{Stark2017, Mason2018, Endsley2022, Tang2023}.

\subsection{Ionizing Photon Budget in Protoclusters and the Field}
\subsubsection{The ultraviolet luminosity density, \( \rho_{\rm UV} \)}

To estimate the ultraviolet (UV) luminosity density, \( \rho_{\rm UV} \), we use a sample of galaxies in the redshift range \( z = 7.25 - 7.30 \). We use the absolute UV magnitudes, \( M_{\rm UV} \), to derive the luminosity density. 
The UV luminosity density was calculated as:
\begin{equation}
\rho_{\rm UV} = \frac{\sum L_{\rm UV}}{V_{\rm com}},
\end{equation}
where \(\sum L_{\rm UV}\) is the total UV luminosity of galaxies in the sample, and \(V_{\rm com}\) is the comoving volume, estimated as \(V_{\rm com} = A_{\rm eff} \times D_{\rm com,\,depth}\), where \(A_{\rm eff} = \pi R^2\) is the effective projected area and \(D_{\rm com,\,depth}\) is the comoving depth of the redshift slice.
For the JADES protocluster, we adopt the physical size of the confirmed overdensity (i.e., the radius enclosing its members). For the comparison fields, we define \(R\) as half the projected separation between the most distant spectroscopically confirmed LAEs, serving as an approximation of the survey footprint. The uncertainty in \(\rho_{\rm UV}\) was estimated by propagating the measurement errors in \(M_{\rm UV}\), which affect the derived luminosities \(L_{\rm UV}\), while assuming negligible uncertainties in the survey area and redshift depth.

\begin{figure*}
    \centering
    \includegraphics[width=0.8\textwidth]{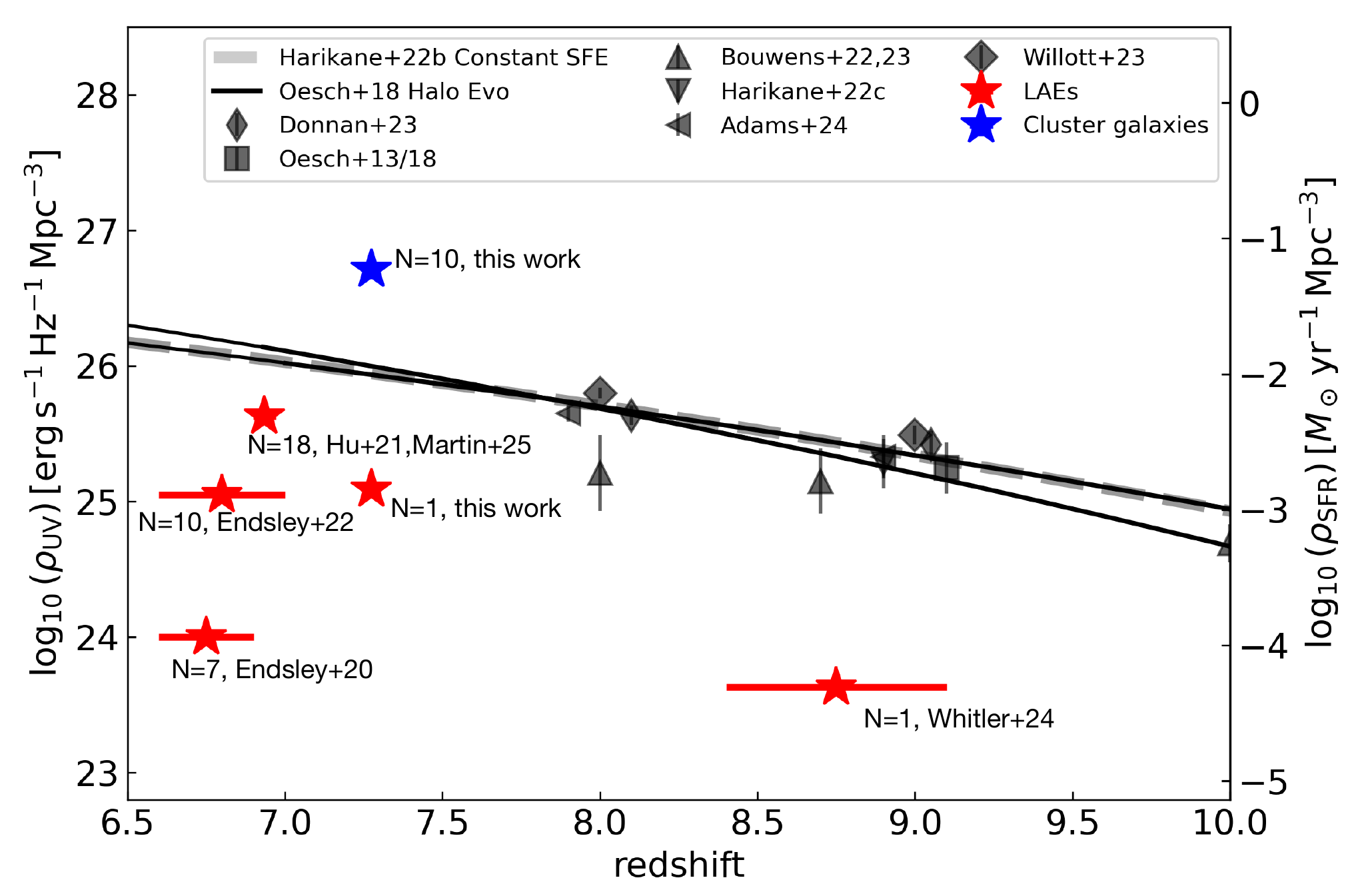}
    \caption{Comparison of the UV luminosity density \(\rho_{\mathrm{UV}}\) derived from our LAE and protocluster samples with results from previous studies \citep[e.g.,][]{Endsley2020, Endsley2022, Whitler2024,Hu2021,Martin2026}.} Red stars represent the contribution from LAEs, indicating that they account for only a small fraction of the total UV background across all redshifts. The blue star at \(z = 7.25-7.30\) shows the UV luminosity density from the compact protocluster region, with \(\log_{10}(\rho_{\mathrm{UV,\,cluster}}) = 26.75_{-0.08}^{+0.07}\). It is about one order of magnitude higher than the field-wide UV luminosity densities reported by previous  surveys at similar redshifts (see also Table~\ref{tab:comparison}).
    \label{fig:p_uv}
\end{figure*}

Table~\ref{tab:comparison} compares our derived \( \rho_{\rm UV} \) values with those from previous literature.
In Figure~\ref{fig:p_uv}, we compare the UV luminosity density \(\rho_{\mathrm{UV}}\) of our LAE and cluster samples with results from previous studies \citep[e.g.,][]{Oesch2018, Harikane2022b, donnan2023evolution}. The red stars indicate the \(\rho_{\mathrm{UV}}\) values contributed by LAEs. This implies that LAEs contribute only a small fraction of the total UV background at all redshifts.
In contrast, the blue star at \(z = 7.25\) marks the UV density from the compact protocluster region, with \(\log_{10}(\rho_{\mathrm{UV,\,cluster}}) = 26.71_{-0.08}^{+0.07}\). Compared to the other measurements reported by \citet{Endsley2022,Endsley2020,Hu2021,Martin2026} and \citet{Whitler2023}, our cluster region exhibits a significantly $\sim 1$ order higher $\rho_{\rm UV}$ (Table~\ref{tab:comparison}) than the total field-wide UV densities reported by previous deep surveys at similar redshifts. The result shows the impact of local galaxy overdensities, where clustered star formation can lead to locally enhanced ionizing photon production. In contrast, the LAEs from comparison fields show significantly lower values, typically 1 to 2 orders of magnitude below average. The comparison reveals that although individual galaxies have modest UV output, dense environments like protoclusters can host concentrated star formation activity, collectively contributing significantly to the reionization-era UV photon budget. This supports the idea that cosmic reionization may be driven not only by bright LAEs, but also by the spatial distribution and clustering of star-forming systems \citep{Endsley2023}.

The absence of Ly$\alpha$ emission in dense protocluster regions is not a rare phenomenon. For example, the overdensity identified at \(z \sim 7.88\) in the GLASS field \citep{Morishita2023} also lacks strong Ly$\alpha$ emitters, yet shows an elevated UV luminosity density of \(\log_{10}(\rho_{\rm UV}) = 28.65^{+0.02}_{-0.02}\) based on seven sources within a compact volume. In particular, the confirmed sources in \citet{Morishita2023} are located within a very compact region of radius $R \sim 60$~pkpc ($\sim 12''$) in the source plane, whereas the protocluster studied here extends over much larger scales of $\sim$0.5--1~pMpc.
These findings indicate that the high UV output from compact protoclusters without strong Ly$\alpha$ may be more common than previously appreciated, and their role in cosmic reionization may have been underestimated.

We also compute the fractional contribution of LAEs to the total UV luminosity density at four redshift samples at \(z \sim 6.7\), 6.85, 7.25, and 8.7 from COSMOS, CEERS and GOODS-S field (see Table~\ref{tab:comparison}). The resulting ratios \(\rho_{\mathrm{UV,\,LAE}} / \rho_{\mathrm{UV,\,total}}\) are \(0.003^{+0.001}_{-0.001}\), \(0.092^{+0.011}_{-0.008}\), \(0.016^{+0.002}_{-0.002}\) and \(0.140^{+0.017}_{-0.012}\). This suggests that LAEs contribute only a small fraction of the total UV output at all redshifts probed, typically below \(15\%\). This indicates that although LAEs are valuable tracers of ionizing galaxies during reionization, they represent only a subset of the UV-luminous population and do not dominate the ionizing photon budget. A noticeable peak at \(z \sim 7.25\) (\(14\%\)) may reflect favorable ionization conditions or local environmental effects, such as clustering in overdense regions where Ly$\alpha$ escape is enhanced. Conversely, the sharp decline in LAE contribution at \(z \sim 8.7\) (\(< 2\%\)) likely reflects increased IGM neutrality at earlier times, which strongly suppresses Ly$\alpha$ transmission. These trends are broadly consistent with theoretical predictions of reionization-era Ly$\alpha$ visibility \citep[e.g.,][]{Mason2018, Jung2022}.

\subsubsection{The escape ionizing photon production rate (\( \dot{N}_{\text{ion}}^{\text{esc}} \)) and the Lyman continuum escape fraction (\(f_{\mathrm{esc,\,LyC}}\))}

\begin{figure*} 
    \centering
    \includegraphics[width=0.48\textwidth]{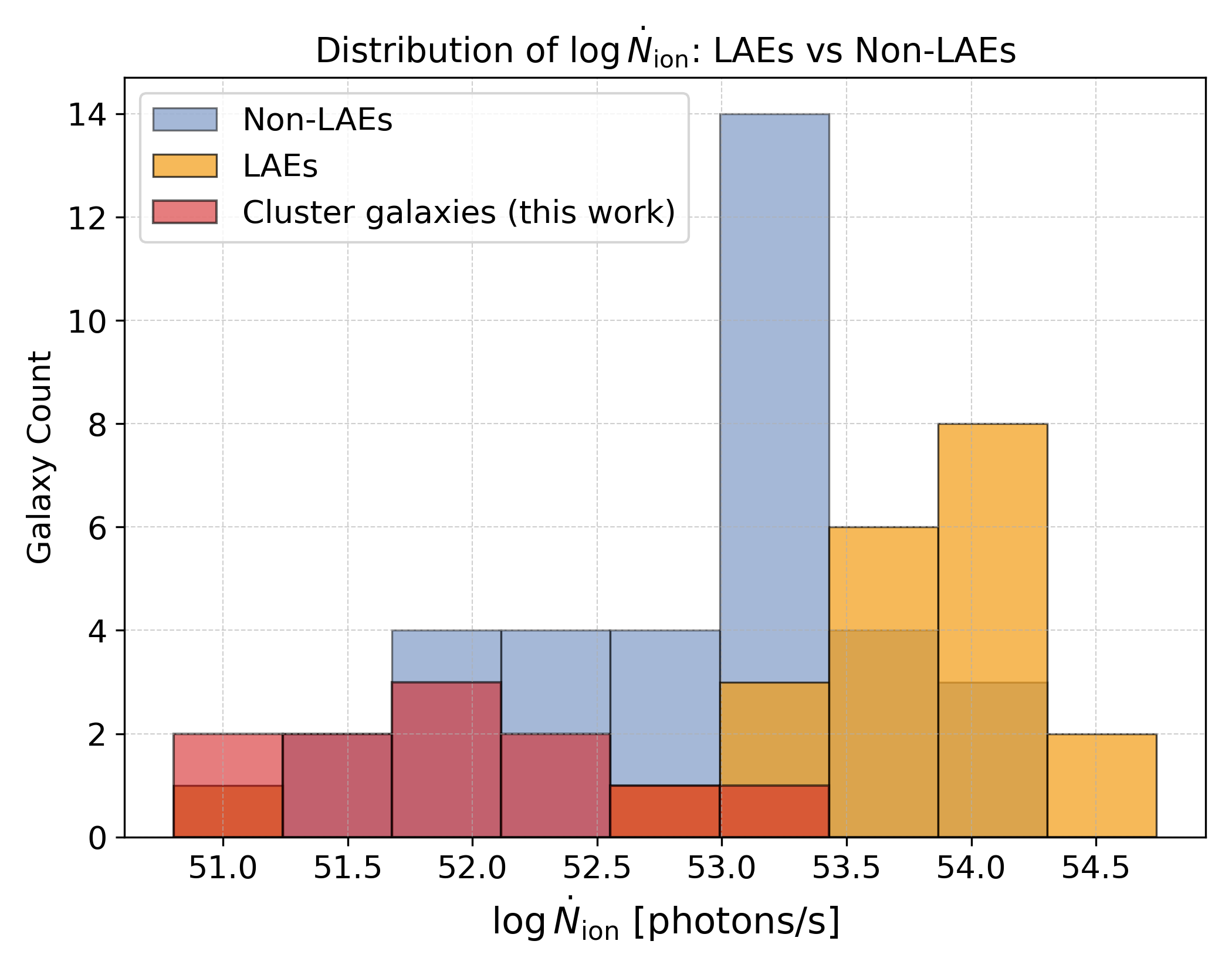}
    \includegraphics[width=0.48\textwidth]{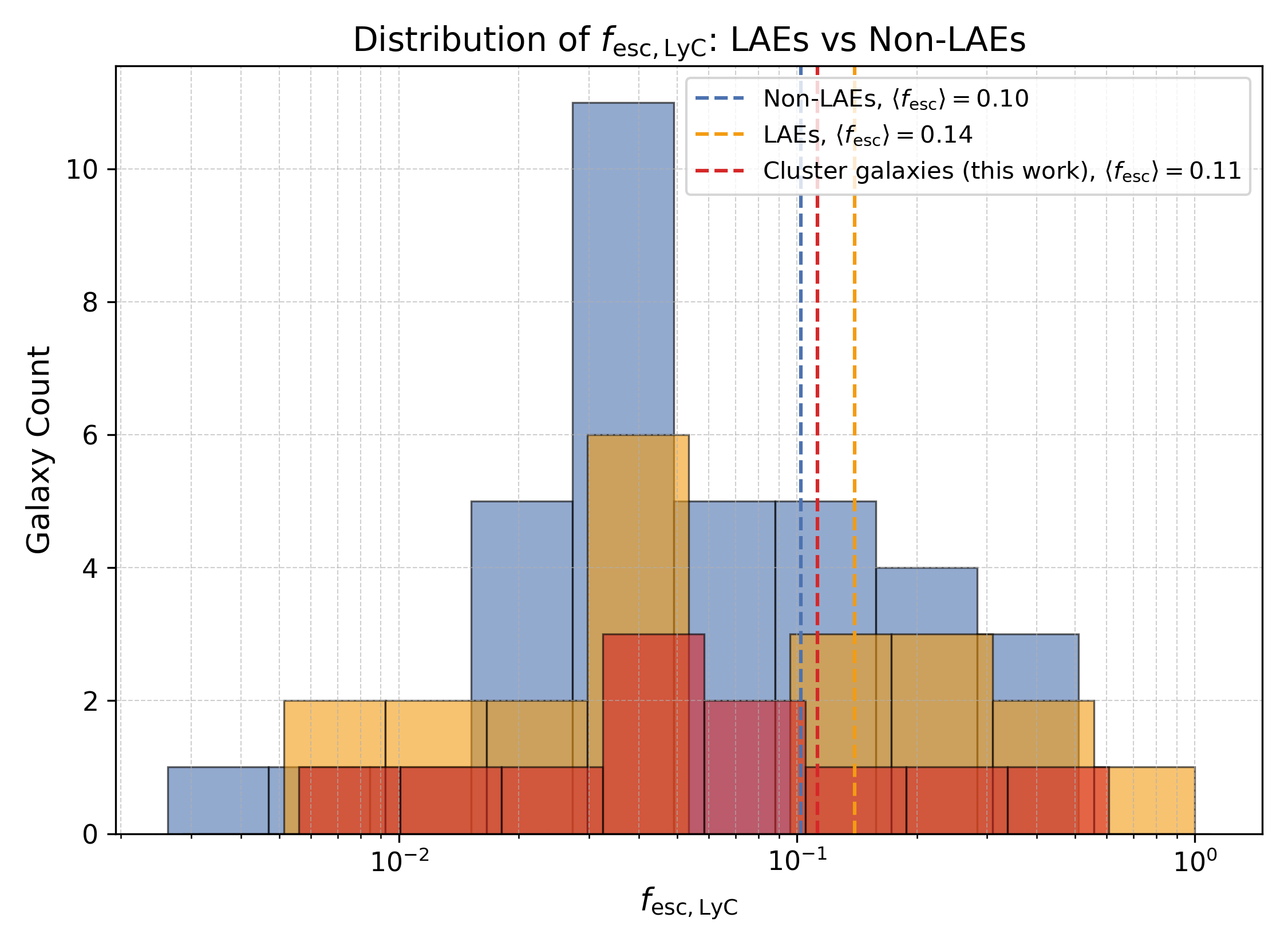}
    \caption{Distribution of ionizing photon production rates per galaxy, expressed as \(\log \dot{N}_{\text{ion}}\) [photons~s\(^{-1}\)]. The red histogram represents galaxies in the JADES protocluster field, while the blue histogram corresponds to non-LAEs galaxies in comparison fields. Although individual JADES galaxies exhibit lower ionizing output, their dense spatial configuration results in a significantly higher cumulative UV photon surface density.}
    \label{fig:log_Nion_distribution}
\end{figure*}

We estimated the escape ionizing photon production rate (\( \dot{N}_{\text{ion}}^{\text{esc}} \)) for individual galaxies, that is the number of photons that escape into the intergalactic medium (IGM), using the relation:

\begin{equation}
    \dot{N}_{\text{ion}}^{\text{esc}} = f_{\text{esc}} \times \xi_{\text{ion}} \times L_{\text{UV}}
\end{equation}

where \( f_{\text{esc}} \) is the Lyman continuum escape fraction, \( \xi_{\text{ion}} \) is the ionizing photon production efficiency in units of Hz\,erg\(^{-1}\), and \( L_{\text{UV}} \) is the monochromatic UV luminosity at 1500\,\AA. To estimate the Lyman continuum escape fraction (\(f_{\mathrm{esc,\,LyC}}\)) for our high-redshift galaxies, we adopt the empirical relation derived by \citet{Chisholm2022} from the Low-redshift Lyman Continuum Survey \citep[LzLCS;][]{Flury2022a, Flury2022b}, which links \(f_{\mathrm{esc,\,LyC}}\) to the UV continuum slope \(\beta_{1550}\). Their study shows a significant correlation between bluer UV slopes and higher escape fractions. Using hierarchical Bayesian regression (LINMIX; \citealt{Kelly2007}), they obtain the following relation:
\begin{equation}
f_{\mathrm{esc,\,LyC}} = (1.3 \pm 0.6) \times 10^{-4} \times 10^{(-1.22 \pm 0.1) \beta_{1550}}.
\end{equation}
We apply this relation to our high-redshift sample using the observed UV slopes, and find that the resulting escape fractions are broadly consistent with the commonly assumed value of \(f_{\mathrm{esc}} = 0.1\) in reionization studies \citep[e.g.,][]{Finkelstein2019} (see Figure~\ref{fig:log_Nion_distribution} Right).

Figure~\ref{fig:log_Nion_distribution} compares the ionizing photon production properties across different galaxy populations. The left panel shows the plot of the distribution of ionizing photon production rates per galaxy among LAEs, non-LAEs, and protocluster galaxies. For the galaxies in the JADES protocluster field, we found that their ionizing photon production spans the range of $\log \dot{N}_{\text{ion}} \in [51.0, 52.7]$ (photons s$^{-1}$), whereas LAEs in other comparison fields show higher photon output values of $\log \dot{N}_{\text{ion}} \in [51.7, 54.7]$ (photons s$^{-1}$).
Although the protocluster galaxies exhibit high \(\xi_{\mathrm{ion}}\) values and strong nebular emission (e.g., high [O\,\textsc{iii}]+H\(\beta\) equivalent widths), their individual \(\dot{N}_{\mathrm{ion}}\) values are lower than those of the LAEs. This is primarily due to their fainter UV magnitudes and lower stellar masses, which result in reduced intrinsic \(L_{\mathrm{UV}}\) and hence lower total ionizing photon output per galaxy. 

Nonetheless, the collective ionizing contribution from protocluster environments may still be significant. While individual field galaxies may appear more UV-luminous, the spatial concentration of galaxies in overdense regions like the JADES protocluster can result in a much higher total ionizing photon budget per unit volume (Figure~\ref{fig:p_uv}).

The right panel of Figure~\ref{fig:log_Nion_distribution} shows the distribution of the Lyman continuum escape fraction, \(f_{\mathrm{esc}}\), estimated from the observed UV continuum slope \(\beta\). The three populations have broadly similar median values around \(f_{\mathrm{esc}} \sim 0.1\), though LAEs show a mild enhancement. The cluster galaxies, despite their lower luminosities, do not exhibit significantly different escape fractions, implying their low \(\dot{N}_{\mathrm{ion}}\) values are primarily driven by their low UV luminosities rather than inefficient photon production.

Furthermore, our findings also show the limitations of using single-galaxy measurements (e.g., $\xi_{\text{ion}}$, SFR) alone, without considering their spatial environment. This aligns with recent studies that point to protocluster regions as key sites for early reionization \citep[e.g.,][]{Chiang2017,Endsley2022}. While parameters such as \(\xi_{\text{ion}}\) and \(f_{\text{esc}}\) reflect the ionizing efficiency of individual galaxies, they do not reflect the cumulative impact of local galaxy overdensities. In particular, our results show that galaxies with relatively modest ionizing output can collectively dominate the ionizing budget if they reside in dense environments. 

Moreover, measurements based solely on global galaxy properties may overlook the environmental effects of physical processes such as star formation, feedback, and gas accretion, all of which are known to vary with large-scale structure. Thus, reionization models that ignore spatial clustering or assume uniform source distributions may systematically underestimate the role of rare but highly concentrated regions. Future efforts that couple resolved galaxy properties with environmental effects — both observationally and in simulations — will be essential to fully understand reionization.

\subsubsection{The cosmic ionizing rate density ($\dot{n}_{\text{ion}}$)}

To assess the overall contribution of LAEs and protocluster galaxies to cosmic reionization, we estimate the cosmic ionizing photon production rate density, denoted as $\dot{n}_{\text{ion}}$. This represents the number of ionizing photons emitted per unit time per unit comoving volume. It is calculated by integrating the ultraviolet luminosity function (UVLF), $\phi(M_{\text{UV}})$, weighted by the product of the ionizing photon escape fraction ($f_{\text{esc}}$), the production efficiency of ionizing photons ($\xi_{\text{ion}}$), and the intrinsic UV luminosity ($L_{\text{UV}}$). The expression is given by:

\begin{equation}
\dot{n}_{\text{ion}} = \int \phi(M_{\text{UV}}) L_{\text{UV}} \, \xi_{\text{ion}} \, f_{\text{esc}} \, dM_{\text{UV}}.
\end{equation}

However, our sample is not complete across the full UV magnitude range, making this direct integration approach infeasible. To estimate the cosmic ionizing rate density contributed solely by our selected cluster galaxies or LAEs, rather than the full galaxy population, the integral can be simplified to:

\begin{equation}
\dot{n}_{\text{ion}} \approx \frac{1}{V} \sum_i \left( f_{\mathrm{esc}, i} \cdot \xi_{\mathrm{ion},i} \cdot L_{\mathrm{UV},i} \right)
.
\end{equation}

We find that the cosmic ionizing rate density near our $z \sim 7$ protocluster galaxies reaches a high value of $\log_{10}(\dot{n}_{\text{ion}}/\mathrm{s}^{-1}\,\mathrm{Mpc}^{-3}) \approx 50.81$, significantly exceeding that of the nearby strong LAE, which has $\log_{10}(\dot{n}_{\text{ion}}/\mathrm{s}^{-1}\,\mathrm{Mpc}^{-3}) \approx 48.40$. 
It may appear counterintuitive that the protocluster galaxies—despite their high ionizing photon output—exhibit no detectable Ly$\alpha$ emission. This discrepancy can be attributed to stronger attenuation by surrounding neutral hydrogen in the protocluster circumgalactic medium since the protocluster candidates also show on average large line-of-sight HI column densities.

It is challenging to explain how such high HI column densities can be observed within these dense environments where hard ionizing spectra from young, blue O/B-main sequence stars are expected to ionize their local ISM and surrounding CGM. One plausible explanation is that the LyC escape from protoclusters is characterized by ionization bounded nebulae with holes \citep{Gazagnes2020}. It is perhaps more typical of field galaxies to have density bounded nebulae \citep{Zackrisson2013, Nakajima2014} where environmental processes, which disrupt neutral gas to produce inhomogeneoous distributions, are less prominent. This inferred difference in LyC escape mechanism may prove crucial in distinguishing the contribution of protoclusters to reionization.


\begin{table*}
    \centering
    \caption{
        Comparison of UV luminosity density $\rho_{\rm UV}$ from this work and previous studies. 
        Values of $\log_{10}(\rho_{\rm UV})$ are given in units of erg\,s$^{-1}$\,Hz$^{-1}$\,Mpc$^{-3}$. 
        The sample area is listed in arcmin$^2$. The "Statistical Region" column distinguishes survey and cluster regions when applicable.
    }
    \label{tab:comparison}
    \begin{tabular}{lcccccc}
        \toprule
        Study & Redshift Range & $N_{\rm LAE}/N_{\rm total}$ & Area & Statistical Region & $\log_{10}(\rho_{\rm UV})$ & Field Name \\
              &                &     & (arcmin$^2$) &            & (erg\,s$^{-1}$\,Hz$^{-1}$\,Mpc$^{-3}$) & \\
        \midrule
        This Work (Cluster) & 7.25--7.30 & -/10 & 2.6   & cluster & $26.75^{+0.08}_{-0.07}$ & GOODS-S \\
        This Work (Total)   & 7.25--7.30 & 1/11 & 25.0  & total   & $25.76^{+0.08}_{-0.07}$ & GOODS-S \\
        This Work (LAE ID 10013682) & 7.28 & 1/- & 25.0 & total   & $24.11^{+0.05}_{-0.04}$ & GOODS-S \\
        Whitler et al. (2024)   & 8.4--9.1 & 1/27 & 530.7 & total & $23.63^{+0.05}_{-0.05}$ & CEERS \\
        Endsley et al. (2022)       & 6.6--7.0 & 10/11 & 5400  & total & $25.05^{+0.05}_{-0.04}$ & COSMOS \\
        Endsley et al. (2020)       & 6.6--6.9 & 7/20 & 165   & total & $24.00^{+0.06}_{-0.04}$ & COSMOS \\
        Hu et al. (2021) \& Martin et al. (2026)        & 6.899--6.971 & 18/- & 316.8   & cluster & $25.63^{+0.03}_{-0.03}$ &  LAGER COSMOS \\
        \bottomrule
    \end{tabular}
\end{table*}

\section{Conclusions}\label{sec:conclusion}

We have presented a detailed analysis of the reionization process for two protocluster structures at \( z \sim 7.2 \) identified in the GOODS-S field using deep \textit{JWST} imaging and spectroscopy from the JADES survey. Our main conclusions are as follows:

\begin{enumerate}
    \item Despite high local galaxy densities, the protocluster galaxies exhibit weak or undetectable Ly$\alpha$ emission. Stacked spectra show no significant Ly$\alpha$ above the \(3\sigma\) level. From individual spectra, we quantify this non-detection by deriving rest-frame $3\sigma$ upper limits on the Ly$\alpha$ equivalent width, finding typical limits of EW$_0^{3\sigma} \sim 10$--25~\AA\ (with a mean of $\sim$15~\AA) for the protocluster members. These limits fall below the commonly adopted LAE threshold of EW$_0 > 25$~\AA, indicating that Ly$\alpha$ emission is genuinely suppressed in this structure. This suggests the surrounding intergalactic medium (IGM) remains largely neutral or that ionizing photon production is insufficient to create fully ionized bubbles.

    \item Environment enhanced neutral hydrogen column densities. DLA modeling of stacked spectra indicates significantly higher neutral hydrogen column densities in protocluster galaxies, with \(\log(N_{\mathrm{HI}}/\mathrm{cm}^{-2}) = 21.55^{+0.52}_{-1.34}\), compared to \(19.3 - 20.5\) in field galaxies. These values suggest that gas-rich overdense environments may delay reionization locally and suppress Ly\(\alpha\) escape.
    
    \item We find that the luminous LAE JADES-GS-z7-LA, located $\sim$0.2 pMpc from the center of the $z \sim 7.3$ protocluster, exhibits exceptionally strong Ly$\alpha$ emission and high ionization conditions. This suggests that it resides within an extended ionized bubble likely sustained by the surrounding overdensity, further supporting the patchy reionization at $z \sim 7$.
    
    \item Ionizing efficiency is high but integrated output is modest. Protocluster galaxies show high ionizing photon production efficiencies, with typical values of \(\log(\xi_{\text{ion}}/\mathrm{Hz\,erg}^{-1}) \sim 25.5 - 25.9\), driven by strong nebular emission and low metallicities. However, their low UV luminosities (\(M_{\mathrm{UV}} > -19\)) and modest stellar masses (\(\log(M_*/M_\odot) \sim 8.5\)) lead to lower individual ionizing photon production rates compared to field LAEs.

    \item Spatial clustering enhances local ionizing budget. Although individual protocluster galaxies produce fewer ionizing photons, the spatial concentration of sources results in a UV luminosity density of \(\log(\rho_{\mathrm{UV}}/\mathrm{erg\,s^{-1}\,Hz^{-1}\,Mpc^{-3}}) = 26.75^{+0.08}_{-0.07}\), approximately 1 dex higher than field averages at the same redshift. This underscores the importance of galaxy clustering in driving early reionization.

    \item Lyman continuum escape fractions are consistent around broadly \(f_{\mathrm{esc}} \sim 0.1\), as inferred from the observed UV continuum slopes using the empirical \(\beta\)--\(f_{\mathrm{esc}}\) relation calibrated by \citet{Chisholm2022}. This suggests that even faint, clustered galaxies may contribute significantly to the ionizing background if their local environments permit photon escape. The protocluster galaxies likely host ionization-bounded nebulae with holes, limiting Ly$\alpha$ visibility despite high ionizing output. In contrast, field galaxies may exhibit density-bounded nebulae, promoting more uniform LyC escape.

\end{enumerate}

Taken together, while high-redshift overdensities enhances the integrated UV output, dense neutral gas reservoirs and low-luminosity stellar populations may locally suppress Ly$\alpha$ visibility. This implies the patchy and environmentally dependent nature of reionization, driven not only by bright LAEs but also by the collective contribution of faint galaxies in clustered environments. Future deep NIRSpec spectroscopy and ALMA observations will be crucial for resolving the internal ISM structure, escape pathways, and gas-phase conditions of these early protocluster systems.

\section*{Acknowledgements}

We acknowledge support from the ERC Advanced Investigator Grant EPOCHS (788113) and support from STFC studentships. 
This work is based on observations made with the NASA/ESA \textit{Hubble Space Telescope} (HST) and NASA/ESA/CSA \textit{James Webb Space Telescope} (JWST) obtained from the \texttt{Mikulski Archive for Space Telescopes} (\texttt{MAST}) at the \textit{Space Telescope Science Institute} (STScI), which is operated by the Association of Universities for Research in Astronomy, Inc., under NASA contract NAS 5-03127 for JWST, and NAS 5–26555 for HST. The observations used in this work are associated with JWST program 1345, 1180 1176, and 2738. The authors thank all involved in the construction and operations of the telescope as well as those who designed and executed these observations.

The authors thank Anthony Holloway and Sotirios Sanidas for their providing their expertise in high performance computing and other IT support throughout this work. This work makes use of {\tt astropy} \citep{Astropy2013,Astropy2018,Astropy2022}, {\tt matplotlib} \citep{Hunter:2007}, {\tt reproject}, {\tt DrizzlePac} \citep{Hoffmann2021}, {\tt SciPy} \citep{Virtanen2007} and {\tt photutils} \citep{Bradley2022}.

\section*{Data Availability}
The \textit{JWST} data used in this work are publicly available from the JWST Advanced Deep Extragalactic Survey \citep[JADES;][]{Rieke2023,Bunker2023, DEugenio2025} (PIs: Eisenstein, Lützgendorf; Program IDs: 1180, 1210). The \textit{HST} data are from the Hubble Legacy Fields project \citep{Illingworth2013, Whitaker2019}, and can be obtained upon reasonable request to that team. Additional data products generated in this study are available from the first author upon reasonable request.



\bibliographystyle{mnras}
\bibliography{example} 




\appendix

\section{Spectra of \texorpdfstring{$z\sim7$}{zsim7} Protocluster Members}\label{appendix:spectra}

\begin{figure*}
    \centering
    \includegraphics[width=\textwidth]{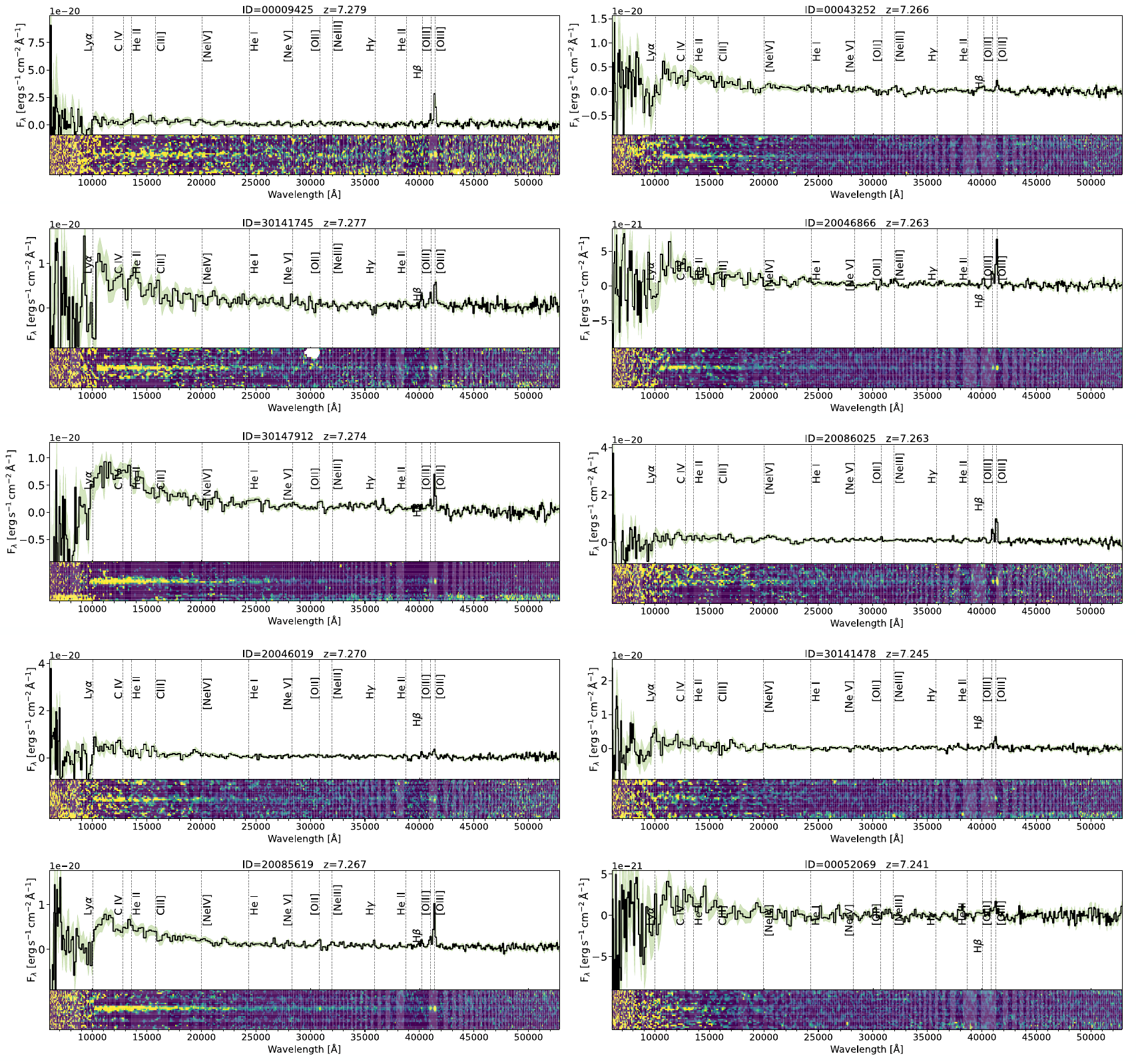}
    \caption{NIRSpec PRISM spectral observations of \(z \sim 7\) protocluster galaxy member. Key spectral features, including nebular emission lines and continuum breaks, are labeled.}
    \label{fig:protocluster_spectra}
\end{figure*}

In this appendix, we present the full set of 1D and 2D spectra for the spectroscopically confirmed galaxy members of the \(z \sim 7\) protocluster in Figure~\ref{fig:protocluster_spectra}. These spectra were obtained using JWST/NIRSpec PRISM observations.

Key emission lines are marked by vertical dashed lines. The black solid lines and green shaded regions indicate the spectral fluxes and their uncertainties, respectively. The source IDs and redshifts are adopted from the official JADES release and are labeled at the top of each panel.

\section{Stacked Spectra of Control Samples}

To compare with the protocluster sample, we also examine the stacked spectra of field galaxies. The stacked spectra for the two control samples are shown in Figure~\ref{fig:stacked_spectra_control} in the Appendix. The top panel shows galaxies in strictly underdense environments ($z=7.0-9.5$, lowest quartile of local density values $\Sigma_5$), selected by visual inspection to ensure isolation. The bottom panel shows the broader control sample ($z=7.1-7.4$), matched in redshift to the protocluster but excluding galaxies in highly overdense regions.

\begin{figure*} 
\centering 
\includegraphics[width=0.8\textwidth]{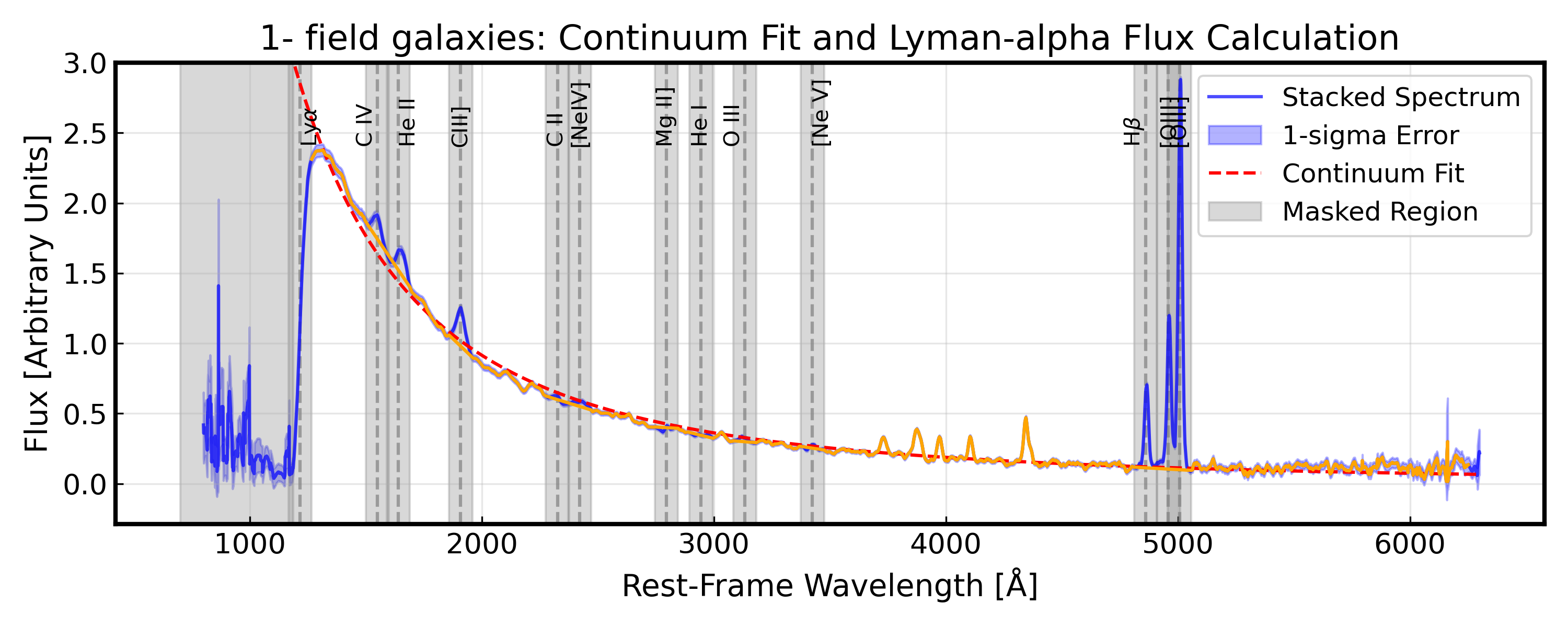} 
\vspace{0.5cm} 
\includegraphics[width=0.8\textwidth]{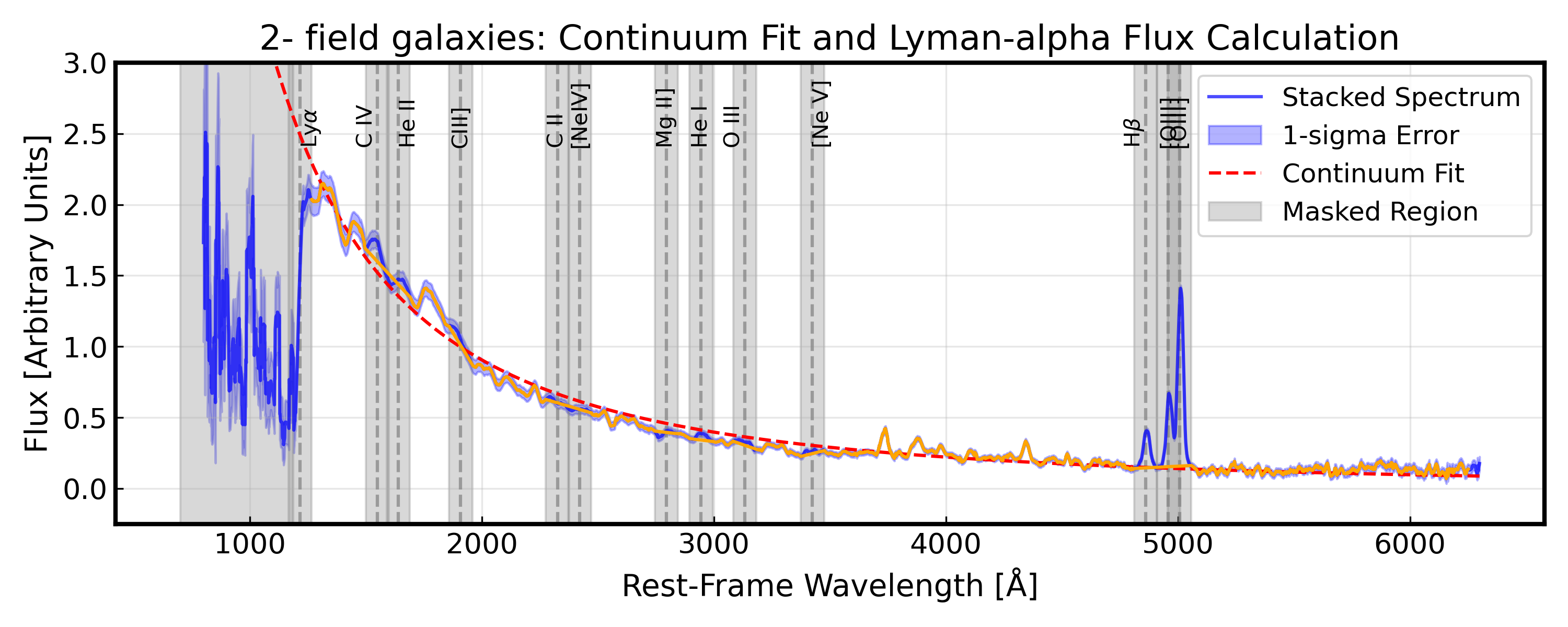} 
\caption{Stacked spectra of the two control samples. The top panel shows the strict field sample, while the bottom panel corresponds to the control sample at $z=7.1-7.4$. In both cases, Ly$\alpha$ remains undetected above the $3\sigma$ threshold.} 
\label{fig:stacked_spectra_control} 
\end{figure*}

\section{Physical Properties of Protocluster Members}\label{appendix:properties}

Table~\ref{tab:spec_z7_cluster} and Table~\ref{tab:bagpipes_z7_cluster} present the physical properties of galaxies that are spectroscopically confirmed members of the $z\sim7$ protocluster. The properties were derived using \texttt{Bagpipes} SED fitting with a continuity star formation history and dust attenuation model. {ID} is the object identifier from the JADES public catalog. {$z$} is the spectroscopic redshift. {$M_{\rm UV}$} is the rest-frame absolute UV magnitude at 1500\,\AA,. SFR and sSFR are the star formation rate and specific star formation rate, respectively. \textbf{$\log_{10}(\xi_{\mathrm{ion}})$} is the ionizing photon production efficiency (in units of Hz\,erg$^{-1}$). $\log_{10}(\mathrm{[OIII]}/\mathrm{H}\beta)$ is the logarithmic nebular emission line ratio of [OIII]$\lambda\lambda$4959,5007 to H$\beta$. $\log_{10}(M_\star/M_\odot)$ is the stellar mass. $\log_{10}(\dot{N}_{\mathrm{ion}})$ is the logarithmic ionizing photon production rate (in s$^{-1}$). EW([OIII]+H$\beta$) is the total rest-frame equivalent width of the [OIII]+H$\beta$ complex.

Table~\ref{tab:protocluster} presents the photometric properties of galaxies located within a projected radius of $30''$ (approximately 160 physical kpc at $z\sim7$) around the two most significant overdensity peaks identified in our surface density map. 

For each source, we provide its sky coordinates (RA and Dec in degrees), photometric redshift with 16th and 84th percentile uncertainties, stellar mass, SFR, and sSFR, all derived from \texttt{BAGPIPES} SED fitting. We also list the observed F444W apparent magnitude along with its lower and upper uncertainties.

All quantities are reported as median values. The uncertainties represent the 16th and 84th percentiles of the posterior distributions. The photometric magnitudes and their errors are measured using aperture-corrected fluxes (the specified circular apertures of 0.32 arcsec diameter) with local background subtraction.

\begin{table*}
\caption{Spectroscopically Confirmed Protocluster Member Galaxies.}
\label{tab:spec_z7_cluster}
\begin{tabular}{lccccccc}
\toprule
\midrule
ID & RA & Dec & $z_{\mathrm{spec}}$ & $M_{\mathrm{UV}}$ & $\log(M_*/M_\odot)$ & $f_{
\rm [OIII]+H\beta}$ & EW${\rm _{[OIII]+H\beta}}$ \\
-- & deg & deg & -- & mag & -- &  & Å \\
\hline
9425 & 53.179770 & -27.774649 & 7.279 & -18.66$_{-0.19}^{+0.25}$ & 8.61$_{-0.32}^{+0.39}$ & 8.45$_{-0.66}^{+0.58}$ & 1348.58$_{-109.17}^{+70.96}$ \\
43252 & 53.187141 & -27.801287 & 7.266 & -18.34$_{-0.09}^{+0.10}$ & 7.26$_{-0.11}^{+0.19}$ & 7.43$_{-1.68}^{+1.43}$ & 471.26$_{-115.12}^{+185.56}$ \\
30141745 & 53.180122 & -27.771437 & 7.277 & -19.22$_{-0.10}^{+0.10}$ & 8.35$_{-0.17}^{+0.14}$ & 7.82$_{-1.23}^{+1.09}$ & 567.54$_{-77.16}^{+81.87}$ \\
20046866 & 53.184041 & -27.797825 & 7.263 & -18.11$_{-0.13}^{+0.14}$ & 7.42$_{-0.12}^{+0.18}$ & 8.33$_{-0.62}^{+0.45}$ & 1390.23$_{-67.91}^{+52.09}$ \\
30147912 & 53.186276 & -27.779041 & 7.274 & -19.39$_{-0.05}^{+0.07}$ & 8.43$_{-0.10}^{+0.10}$ & 8.61$_{-0.92}^{+0.81}$ & 364.86$_{-36.77}^{+41.72}$ \\
20086025 & 53.183741 & -27.793902 & 7.263 & -17.97$_{-0.17}^{+0.23}$ & 8.46$_{-0.19}^{+0.21}$ & 8.72$_{-0.90}^{+0.54}$ & 919.66$_{-96.18}^{+107.93}$ \\
20046019 & 53.183934 & -27.799989 & 7.270 & -18.69$_{-0.09}^{+0.13}$ & 8.32$_{-0.16}^{+0.13}$ & 5.42$_{-1.81}^{+2.36}$ & 587.11$_{-136.73}^{+143.83}$ \\
30141478 & 53.186745 & -27.770636 & 7.245 & -17.26$_{-0.29}^{+0.48}$ & 7.38$_{-0.27}^{+0.37}$ & 8.64$_{-1.24}^{+0.59}$ & 1264.00$_{-114.46}^{+127.14}$ \\
20085619 & 53.191055 & -27.797314 & 7.267 & -19.13$_{-0.06}^{+0.07}$ & 8.53$_{-0.11}^{+0.09}$ & 8.39$_{-0.55}^{+0.52}$ & 729.48$_{-60.85}^{+74.55}$ \\
52069 & 53.182037 & -27.778048 & 7.241 & -16.95$_{-0.29}^{+0.31}$ & 7.03$_{-0.24}^{+0.31}$ & 8.23$_{-2.90}^{+1.10}$ & 783.20$_{-249.59}^{+326.05}$ \\
\bottomrule
\end{tabular}
\end{table*}

\begin{table*}
\caption{Physical properties of spectroscopically Confirmed $z\sim7$ protocluster members.}
\label{tab:bagpipes_z7_cluster}
\begin{tabular}{lccccc}
\toprule
\midrule
ID & SFR ($M_\odot$ yr$^{-1}$) & sSFR (Gyr$^{-1}$) & $\log_{10}(\xi_{\mathrm{ion}}^{0}\,/\,\mathrm{Hz}\,\mathrm{erg}^{-1})$ & $\log_{10}(\dot{N}_{\mathrm{ion}}^{\mathrm{esc}}\,/\,\mathrm{s}^{-1})$ & $R_{\mathrm{HII}}$ (Mpc) \\
\hline
9425 & 4.13$_{-2.10}^{+6.02}$ & -7.98$_{-0.02}^{+0.00}$ & 25.26$_{-0.07}^{+0.05}$ & 51.73$_{-0.70}^{+0.70}$ & 0.18$\pm0.39$ \\
43252 & 0.19$_{-0.04}^{+0.09}$ & -7.98$_{-0.05}^{+0.01}$ & 24.85$_{-0.11}^{+0.10}$ & 50.96$_{-0.53}^{+0.52}$ & 0.13$\pm0.32$ \\
30141745 & 1.52$_{-0.34}^{+0.52}$ & -8.13$_{-0.19}^{+0.15}$ & 24.99$_{-0.08}^{+0.08}$ & 51.84$_{-0.67}^{+0.67}$ & 0.36$\pm1.08$ \\
20046866 & 0.27$_{-0.06}^{+0.13}$ & -7.98$_{-0.01}^{+0.00}$ & 25.29$_{-0.05}^{+0.04}$ & 51.36$_{-0.77}^{+0.77}$ & 0.16$\pm0.38$ \\
30147912 & 2.06$_{-0.42}^{+0.47}$ & -8.10$_{-0.15}^{+0.12}$ & 24.81$_{-0.05}^{+0.06}$ & 50.80$_{-0.66}^{+0.66}$ & 0.30$\pm0.83$ \\
20086025 & 2.24$_{-0.76}^{+1.65}$ & -8.06$_{-0.16}^{+0.09}$ & 25.11$_{-0.08}^{+0.09}$ & 53.38$_{-0.73}^{+0.72}$ & 0.17$\pm0.38$ \\
20046019 & 0.94$_{-0.23}^{+0.42}$ & -8.33$_{-0.17}^{+0.18}$ & 25.23$_{-0.14}^{+0.10}$ & 51.66$_{-0.71}^{+0.71}$ & 0.61$\pm1.95$ \\
30141478 & 0.24$_{-0.11}^{+0.30}$ & -7.98$_{-0.06}^{+0.01}$ & 25.24$_{-0.07}^{+0.07}$ & 52.27$_{-0.69}^{+0.69}$ & 0.12$\pm0.02$ \\
20085619 & 1.80$_{-0.38}^{+0.61}$ & -8.26$_{-0.13}^{+0.13}$ & 25.13$_{-0.06}^{+0.05}$ & 51.69$_{-0.68}^{+0.68}$ & 0.26$\pm0.65$ \\
52069 & 0.09$_{-0.04}^{+0.06}$ & -8.03$_{-0.18}^{+0.07}$ & 25.10$_{-0.16}^{+0.17}$ & 52.70$_{-0.70}^{+0.70}$ & 0.08$\pm0.02$ \\
\bottomrule
\end{tabular}
\end{table*}

\begin{table*}
\caption{Photometric galaxies near the two overdensity peaks.}
\label{tab:protocluster}
\begin{tabular}{llcccccc}
\toprule
RA & Dec & $z_{\rm phot}$ & logM$_{\star}$ & SFR & sSFR & F444W$_{\rm mag}$ & Cluster \\
\midrule
53.185354 & -27.773188 & 7.313$_{-0.084}^{+0.100}$ & 7.38$_{-0.12}^{+0.15}$ & 1.78$_{-0.15}^{+0.28}$ & -7.10$_{-0.20}^{+0.09}$ & 28.20$_{-0.11}^{+0.10}$ & Protocluster 1 \\
53.179755 & -27.774646 & 7.219$_{-0.074}^{+0.137}$ & 8.08$_{-0.12}^{+0.11}$ & 10.64$_{-2.03}^{+2.74}$ & -7.02$_{-0.07}^{+0.03}$ & 26.68$_{-0.11}^{+0.10}$ & Protocluster 1 \\
53.179543 & -27.774437 & 7.318$_{-0.112}^{+0.129}$ & 8.38$_{-0.19}^{+0.14}$ & 1.56$_{-0.56}^{+1.13}$ & -8.20$_{-0.26}^{+0.38}$ & 28.06$_{-0.11}^{+0.10}$ & Protocluster 1 \\
53.180665 & -27.772396 & 7.124$_{-0.072}^{+0.104}$ & 7.01$_{-0.08}^{+0.12}$ & 0.89$_{-0.11}^{+0.17}$ & -7.03$_{-0.10}^{+0.04}$ & 28.78$_{-0.11}^{+0.10}$ & Protocluster 1 \\
53.179537 & -27.773949 & 7.212$_{-0.064}^{+0.096}$ & 7.22$_{-0.10}^{+0.11}$ & 1.67$_{-0.35}^{+0.47}$ & -7.00$_{-0.00}^{+0.00}$ & 28.14$_{-0.11}^{+0.10}$ & Protocluster 1 \\
53.181486 & -27.769501 & 7.316$_{-0.100}^{+0.097}$ & 7.26$_{-0.12}^{+0.15}$ & 1.33$_{-0.15}^{+0.23}$ & -7.11$_{-0.16}^{+0.10}$ & 28.42$_{-0.11}^{+0.10}$ & Protocluster 1 \\
53.180119 & -27.771435 & 7.361$_{-0.103}^{+0.106}$ & 7.66$_{-0.19}^{+0.19}$ & 2.38$_{-0.45}^{+0.48}$ & -7.28$_{-0.25}^{+0.22}$ & 27.88$_{-0.11}^{+0.10}$ & Protocluster 1 \\
53.191057 & -27.797317 & 7.350$_{-0.108}^{+0.118}$ & 8.22$_{-0.21}^{+0.16}$ & 6.69$_{-1.91}^{+1.60}$ & -7.40$_{-0.22}^{+0.25}$ & 27.25$_{-0.11}^{+0.10}$ & Protocluster 2 \\
53.187147 & -27.801285 & 7.312$_{-0.114}^{+0.121}$ & 7.46$_{-0.20}^{+0.20}$ & 0.99$_{-0.14}^{+0.19}$ & -7.46$_{-0.22}^{+0.24}$ & 28.95$_{-0.11}^{+0.10}$ & Protocluster 2 \\
53.183937 & -27.799989 & 7.387$_{-0.102}^{+0.105}$ & 8.05$_{-0.25}^{+0.20}$ & 4.05$_{-1.33}^{+1.02}$ & -7.47$_{-0.27}^{+0.32}$ & 27.55$_{-0.11}^{+0.10}$ & Protocluster 2 \\
53.186312 & -27.795612 & 7.407$_{-0.143}^{+0.168}$ & 8.71$_{-0.11}^{+0.07}$ & 1.87$_{-0.40}^{+0.74}$ & -8.44$_{-0.15}^{+0.24}$ & 27.86$_{-0.11}^{+0.10}$ & Protocluster 2 \\
53.184046 & -27.797826 & 7.265$_{-0.108}^{+0.129}$ & 7.31$_{-0.11}^{+0.11}$ & 1.89$_{-0.37}^{+0.48}$ & -7.01$_{-0.05}^{+0.02}$ & 28.16$_{-0.11}^{+0.10}$ & Protocluster 2 \\
53.183754 & -27.793889 & 7.324$_{-0.114}^{+0.128}$ & 8.20$_{-0.16}^{+0.22}$ & 9.78$_{-1.90}^{+2.56}$ & -7.17$_{-0.29}^{+0.15}$ & 27.23$_{-0.11}^{+0.10}$ & Protocluster 2 \\
53.183747 & -27.793701 & 7.346$_{-0.112}^{+0.094}$ & 7.31$_{-0.11}^{+0.14}$ & 1.86$_{-0.32}^{+0.46}$ & -7.02$_{-0.08}^{+0.02}$ & 28.28$_{-0.11}^{+0.10}$ & Protocluster 2 \\
53.183008 & -27.789453 & 7.193$_{-0.131}^{+0.187}$ & 7.76$_{-0.26}^{+0.20}$ & 2.75$_{-0.61}^{+0.79}$ & -7.30$_{-0.26}^{+0.27}$ & 28.23$_{-0.11}^{+0.10}$ & Protocluster 2 \\
\bottomrule
\end{tabular}
\end{table*}

\section{DLA Model Fit with Fixed IGM Neutral Fraction}\label{appendix:dla}

To test the robustness of our neutral hydrogen column density estimates, we explore an alternative scenario in which the IGM is assumed to be fully neutral (\(x_{\mathrm{HI}} = 1\)). In this case, the absorption is entirely attributed to the intergalactic medium.

\begin{figure*}
    \centering
    \begin{minipage}{0.43\textwidth}
        \centering
        \includegraphics[width=\textwidth]{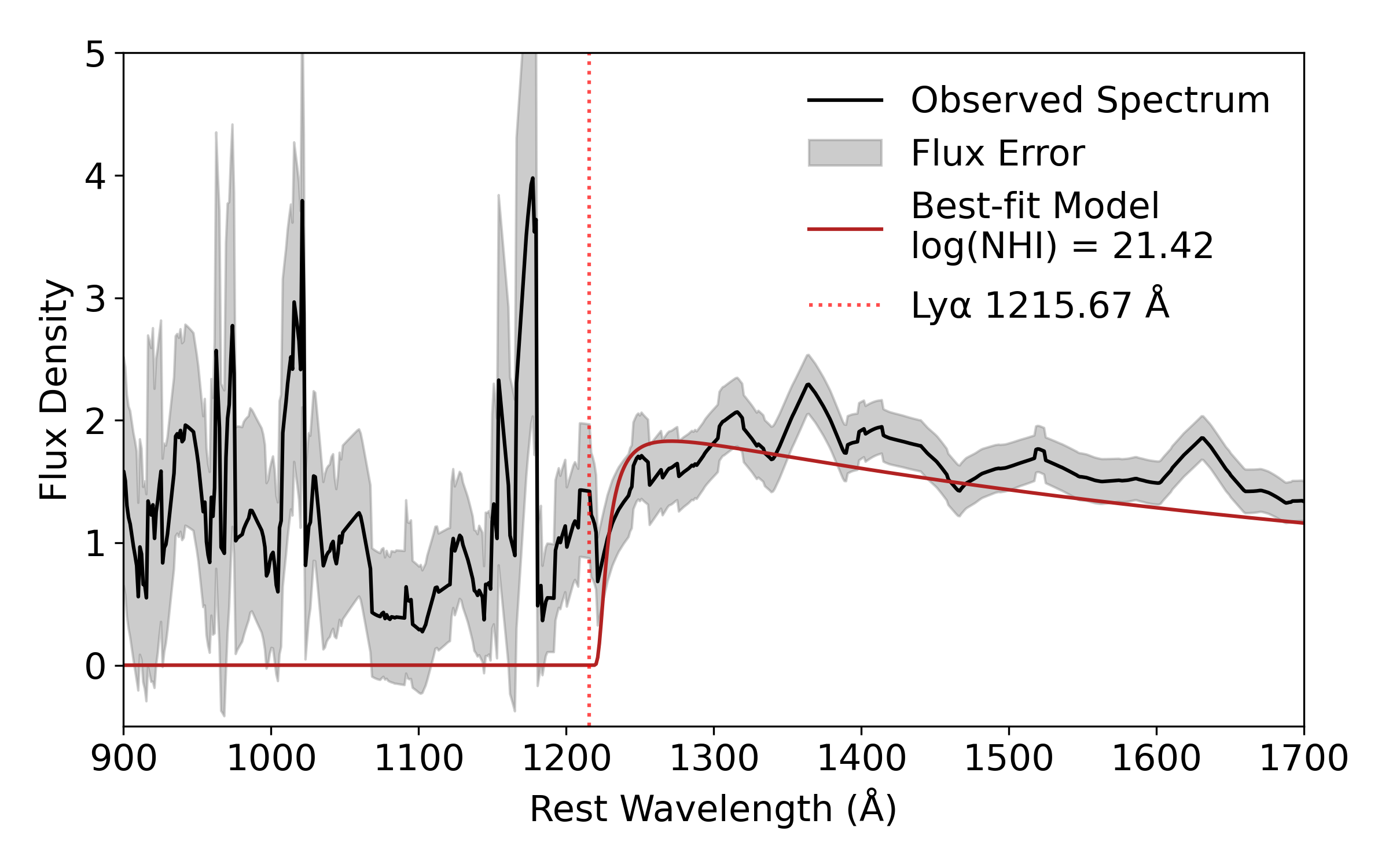}
    \end{minipage}
    \hfill
    \begin{minipage}{0.55\textwidth}
        \centering
        \includegraphics[width=\textwidth]{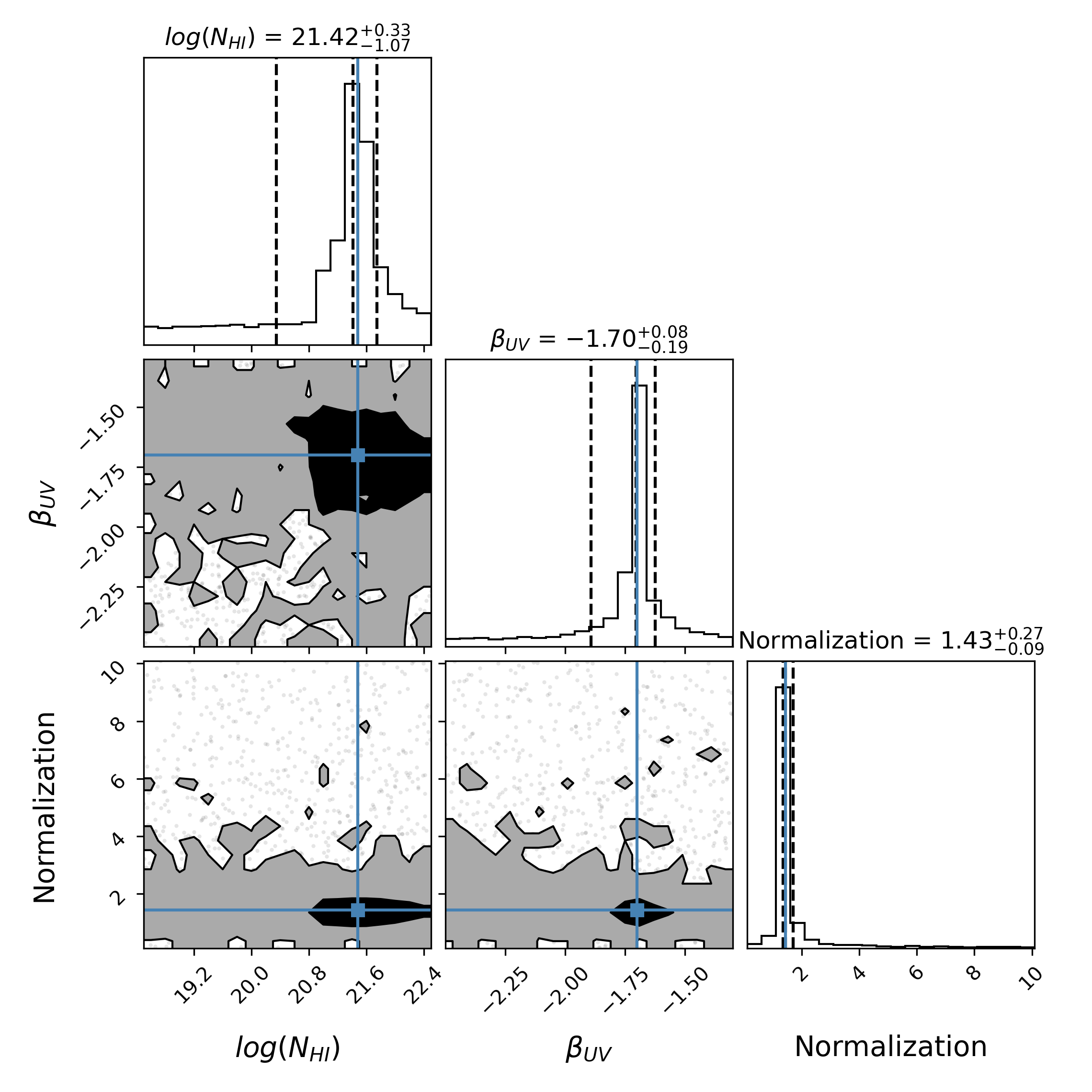}
    \end{minipage}
    \caption{
    {\it Left:} Best-fit DLA model assuming a fully neutral IGM (\(x_{\mathrm{HI}} = 1\)) for the protocluster galaxy stack. The red curve shows the model, while the observed spectrum is plotted in black with the shaded region indicating flux uncertainties. The vertical dotted line marks the rest-frame Ly$\alpha$ wavelength. 
    {\it Right:} Posterior distributions of the fitted parameters assuming \(x_{\mathrm{HI}} = 1\). The inferred \(\log(N_{\mathrm{HI}})\) and \(\beta_{\mathrm{UV}}\) values are consistent with those from the default fit in which \(x_{\mathrm{HI}}\) is a free parameter, indicating the robustness of the results.
    }
    \label{fig:dla_combined_fixedxhi}
\end{figure*}

Figure~\ref{fig:dla_combined_fixedxhi} left panel presents the best-fit DLA model under this assumption, overlaid on the observed rest-frame spectrum. The red curve shows the best-fit absorption + continuum model, yielding \(\log(N_{\mathrm{HI}}/\mathrm{cm}^{-2}) = 21.42\). Despite the simplification, the model reproduces the observed flux depression blueward of Ly$\alpha$ reasonably well. The posterior distributions for the remaining parameters are shown in Figure~\ref{fig:dla_combined_fixedxhi} right panel. We obtain 
\(\log(N_{\mathrm{HI}}/\mathrm{cm}^{-2}) = 21.42^{+0.33}_{-1.07}\), 
\(\beta_{\mathrm{UV}} = -1.70^{+0.08}_{-0.19}\), and 
a continuum normalization of \(1.43^{+0.27}_{-0.09}\). These results are consistent with those derived using the free-\(x_{\mathrm{HI}}\) fit, suggesting that the data do not strongly constrain the exact contribution of the IGM to the total neutral hydrogen absorption.

\bsp	
\label{lastpage}
\end{document}